\documentclass{aa}  

\usepackage{graphicx}
\usepackage{txfonts}
\usepackage{xcolor}
\newcommand{\gaia}{\textit{Gaia}}
\providecommand{\ww}{\mbox{$W(G,G_{BP}-G_{RP})$}}
\newcommand{\sos}{SOS Cep\&RRL}
\providecommand{\grp}{\ensuremath{G_\mathrm{RP}}}
\providecommand{\gbp}{\ensuremath{G_\mathrm{BP}}}

\begin{document}

   \title{Gaia DR3: Specific processing and validation of all-sky RR Lyrae and Cepheid stars - The Cepheid sample\thanks{Tables~\ref{tab:multi},~\ref{tab:reclassification},~\ref{tab:noUseTheseMetallicities},~\ref{tab:association}, and ~\ref{tab:contamination} are only available in electronic form at the CDS via anonymous ftp to cdsarc.u-strasbg.fr (130.79.128.5) or via http://cdsweb.u-strasbg.fr/cgi-bin/qcat?J/A+A/}}


   \author{V. Ripepi
          \inst{1}
          \and
          G. Clementini \inst{2}
          \and 
          R. Molinaro \inst{1}
          \and    
          S. Leccia \inst{1}
          \and 
         E. Plachy \inst{3,4,5}
        \and
       L. Moln\'ar \inst{3,4,5}
          \and
        L. Rimoldini\inst{6}      
        \and
          I. Musella\inst{1}
        \and  
            M. Marconi\inst{1}
          \and
        A. Garofalo\inst{2}
          \and
          M. Audard\inst{6,7}
          \and
          B. Holl\inst{6,7}
          \and
          D.~W. Evans\inst{8} 
          \and
          G. Jevardat de Fombelle\inst{6}
          \and
          I. Lecoeur-Taibi\inst{6}
          \and
          O. Marchal\inst{9}
          \and
          N. Mowlavi\inst{6}
           \and
          T. Muraveva\inst{2}
          \and
          K. Nienartowicz\inst{10}
          \and
         P. Sartoretti\inst{11}
          \and
          L. Szabados\inst{3,4}
          \and
          L. Eyer\inst{7}
          }

\institute{ INAF-Osservatorio Astronomico di Capodimonte, Salita Moiariello 16, 80131, Naples, Italy\\  \email{vincenzo.ripepi@inaf.it}
\and
INAF-Osservatorio di Astrofisica e Scienza dello Spazio, Via Gobetti 93/3, I-40129 Bologna, Italy 
\and
Konkoly Observatory, Research Centre for Astronomy and Earth Sciences, E\"otv\"os Lor\'and Research Network, H-1121 Budapest, Konkoly Thege M. \'ut 15-17, Hungary
\and
MTA CSFK Lend\"ulet Near-Field Cosmology Research Group, Konkoly Thege Mikl\'os \'ut 15-17, 1121 Budapest, Hungary
            \and
ELTE E\"otv\"os Lor\'and University, Institute of Physics, 1117, P\'azm\'any P\'eter s\'et\'any 1/A, Budapest, Hungary
             \and
Department of Astronomy, University of Geneva, Ch. d'Ecogia 16, 1290 Versoix, Switzerland
\and
Department of Astronomy, University of Geneva, Chemin Pegasi 51, 1290 Versoix, Switzerland
\and
Institute of Astronomy, University of Cambridge, Madingley Road, Cambridge CB3 0HA, UK
\and
Observatoire Astronomique de Strasbourg, Universit\'{e} de Strasbourg, CNRS, UMR 7550, 11 rue de l'Universit\'{e}, 67000 Strasbourg, France
\and
Sednai S\`arl, Geneva, Switzerland
\and
GEPI, Observatoire de Paris, Universit\'e PSL, CNRS, 5 Pla Jules Janssen, 92190 Meudon, France\\
             }

   \date{}

 
  \abstract
   {Cepheids are pulsating stars that play a crucial role 
   in several astrophysical contexts. 
   Among the different types, the Classical Cepheids are  fundametal tools for the calibration of the extragalactic distance ladder. 
   They are also powerful stellar population tracers in the context of Galactic studies. The \gaia\ Third Data Release (DR3) publishes improved data on Cepheids collected during the initial 34 months of operations.
   }
   {We present the \gaia\ DR3 catalogue of Cepheids of all types, obtained through the analysis carried out with the Specific Object Study (SOS) 
   Cep\&RRL pipeline.  }
   {We discuss the procedures adopted to clean the Cepheid sample from spurious objects, 
   to validate the results, and 
   to re-classify sources with a wrong outcome from the \sos\ pipeline.} 
   {The \gaia\ DR3 includes  multi-band time-series photometry and characterisation by the \sos\ pipeline for a sample of 15\,006 Cepheids of all types. 
   The sample includes  4\,663,  4\,616,  321 and 185 pulsators, distributed in the  LMC, SMC, M31 and M33, respectively, as well as 5\,221 objects in the remaining All Sky sub-region which includes stars in the MW field/clusters 
   and in a number of 
   small satellites of our Galaxy. Among this sample, 327 objects were known as variable stars in the literature but with a different classification, while, to the best of our knowledge, 474 stars have not been reported before to be 
   variable stars and therefore they likely are new Cepheids  discovered by {\it Gaia}. 
   }
   {}

   \keywords{Stars: distances -- Stars: variables: Cepheids --
                Magellanic Clouds --
                Galaxy: disc -- 
                Surveys -- Methods: data analysis
               }

   \maketitle
%

\section{Introduction}

Cepheids made their appearance 
on the scene, when Edward Pigott discovered their first representative, $\eta$~Aql, in 1784; thus opening a field of astrophysical research that is still fully active nowadays. The variable stars that are collectively called Cepheids are actually 
an ensemble of different types which we now separate into three groups: Classical Cepheids (DCEPs, whose prototype is $\delta$ Cep), type II Cepheids (T2CEPs) and anomalous Cepheids (ACEPs). 

The 
crucial role played by DECPs resides in their Period-Luminosity ($PL$) and Period-Wesenheit ($PW$) relations 
that represent fundamental tools at the basis of the extra-galactic distance ladder  \citep[e.g.][]{Leavitt1912,Madore1982,Caputo2000,Riess2016}. However, DCEPs are important astrophysical objects also 
for stellar evolution and Galactic studies. Indeed, since their pulsational properties (mainly periods) are linked to the intrinsic stellar parameters (effective temperature, mass, luminosity) DCEPs can be used as an independent test for 
stellar evolution models. Moreover, given their young age ($\sim$50-500 Myr) they are preferentially located in the
 Milky Way (MW) thin disc, 
 and, thanks to 
 precise distances that can be derived from their 
 $PL$ and $PW$ relations, DCEPs can be used to model the disc and 
 trace their birthplaces in the spiral arms, where star formation is most active \citep[e.g.][and references therein]{Skowron2019,Poggio2021}. Furthermore, if the chemical abundance of the DCEPs is available, they can be used to trace the metallicity gradient of the MW \citep[e.g.][and references therein]{Genovali2014,Luck2011,Luck2018,Ripepi2022a}.

While DCEPs are luminous, young and massive ($M\sim$3--11 $M_{\odot}$) stars, T2CEPs are more evolved objects, older than 10 Gyr, more luminous and slightly less massive than RR Lyrae variables \citep[$M\sim$0.55--0.7 $M_{\odot}$, see e.g.][for a more extended description of T2CEPs properties]{Caputo1998,Sandage2006}. They are preferentially metal poor objects and, as the RR Lyrae variables, populate the main Galactic components, that is disc, bulge and halo. T2CEPs pulsate with periods from $\sim$ 1
to $\sim$ 24 d and are separated into BL Herculis stars (BLHER; periods between 1 and 4 d) and W Virginis stars (WVIR; periods between 4 and 24 d). Historically, a third class of variables is considered as an additional subgroup of the T2CEP class, namely the RV Tauri (RVTAU) stars \citep[see e.g.][and references therein]{Feast2008}, with  periods from about 20 to 150 d and often less regular light curves. They are post-asymptotic giant branch stars in their path to become planetary nebulae. This evolutionary phase corresponds to the latest evolution of intermediate mass stars and therefore the link of RVTAU with the low-mass WVIR stars should be
considered with caution. T2CEPs follow very tight $PL$ and $PW$ relations, especially in the near infra-red \citep[NIR, see e.g.][and references therein]{Matsunaga2011,Ripepi2015} and are therefore excellent distance indicators. 

The third Cepheid-like class of pulsating stars is represented by the anomalous Cepheids (ACEPs). They have periods approximately between 0.4 d and 2.5 d and absolute magnitudes brighter than RR Lyrae stars by 0.3 mag to 2 mag \citep[][and references therein]{Caputo2004}. ACEP variables are thought to be in their central He burning evolutionary phase and to have masses approximately between 1.3 to 2.1 $M_{\odot}$, as well as metallicities lower than $Z$=0.0004 \citep[corresponding to an iron abundance lower than $\sim-1.6$ dex, for $Z_{\odot}$ = 0.0152, see][for details]{Caputo1998,Marconi2004}. Similarly to RR Lyrae stars, the ACEPs can pulsate in the fundamental or first overtone modes and in both cases show well defined $PL$ and $PW$ relations, especially 
in the Large Magellanic Cloud \citep[LMC;][]{Ripepi2014,Sos2015}.  

Great advances in the study of variable sources 
have been obtained thanks to the \gaia\ mission  \citep[][]{Gaia2016} and its subsequent data releases \citep[DR1, DR2 and EDR3,][]{Gaia2016Brown,Gaia2018,Gaia2021,Riello2021}.   
Indeed, the multi-epoch nature of \gaia\ observations makes the satellite a very
powerful tool to identify, characterise and classify many different classes of variable stars across the 
whole Hertzsprung-Russell (HR) diagram \citep[see][]{Eyer2019}. In \gaia\ DR1 
time-series photometry in the $G$-band, and 
parameters derived from the $G$ light curves, were released for a small number of objects in and around the LMC, including 599 Cepheids of all types and 2595 RR Lyrae stars \citep[][hereafter Paper I]{Clementini2016}. 
\gaia\ DR2 
in 2018, released 
more than 550\,000 variable sources belonging to a variety of different classes \citep[see][]{Holl2018}, including about 9\,500 Cepheids of all types and about 140\,000 RR Lyrae stars \citep[][hereafter Paper II]{Clementini2019}.  
{\it Gaia} Early Data Release 3 \citep[EDR3;][]{Gaia2021} in December 2020 published 
average photometry, parallaxes and proper motions but no time-series data. Epoch data 
are now made available with {\it Gaia} 
Data Release 3 \citep[DR3, see][]{GaiaVallenari}, that provides  multiband 
time series photometry for nearly 12 million variable sources \citep[see][]{DR3-DPACP-162}.  

The Specific Objects Study (SOS) Cep\&RRL  pipeline (\sos\ pipeline hereafter) 
was developed to validate and 
characterise  
Cepheids and RR Lyrae stars observed by \gaia. 
The pipeline has been described in detail in Papers I and II, to
which we refer the interest reader. 
The general properties of the entire sample of variable objects released in \gaia\ DR3  are discussed in \citet{DR3-DPACP-162}, that 
also describes the chain of subsequent steps carried out in 
 the general variability analysis before the \sos\ processing of the data \citep[see also][]{Holl2018}.

In this paper we 
describe the properties of the Cepheids for which time-series data are released in DR3 and their characteristic parameters, that populate the {\tt vari\_cepheid} catalogue which 
is part of the data release. In more detail, we: i) illustrate the changes 
we implemented in the \sos\ pipeline
 to process the DR3 photometric and radial velocity (RV) time-series of candidate Cepheids 
 provided by the general variable star classification pipelines \citep[][]{DR3-DPACP-162,DR3-DPACP-165}; ii) discuss the procedures adopted to clean the sample of Cepheids that are released in the \gaia\ DR3; iii) present the ensemble properties of the DR3 Cepheids; iv)  describe the validation procedures adopted to estimate the completeness and contamination of the sample. 

 A complementary paper \citep[][]{DR3-DPACP-168} describes the \sos\ pipeline and the relative results for the RR Lyrae variables.

\section{SOS Cep\&RRL pipeline: changes from DR2 to DR3}
\label{sect:2}

The main steps of the \sos\ pipeline for candidate Cepheids are  shown in figs. 1 and 3 of Papers I and II.
The procedures used for DR3 are basically the same as for DR1 and DR2, but with some important changes that we describe  
below: 

\begin{enumerate}
    \item {\bf Sub-regions in the sky:} For the processing of the DR2 data, the SOS pipeline subdivided the sky in three regions, two around the LMC and the Small Magellanic Cloud (SMC), respectively, and a third one, called All-sky, including all the remaining stars, which 
    mainly belonged to the MW. This subdivision was needed because of the different observational properties of Cepheids and RR Lyrae stars in the Magellanic Clouds (MCs) and in the MW. In fact, while 
    Cepheids (and RR Lyrae) in the LMC and SMC are all more or less at the same distance from us within each galaxy, so that we can just use their apparent magnitudes to define their position in/around the reference $PL$ or $PW$ relations, for the MW we need absolute magnitudes calculated from 
    \gaia\ parallaxes to place these  stars on the $PL$ and $PW$ diagrams. These differences, in turn, required different steps in the SOS pipeline. 
    In DR3 we have enlarged the regions around the LMC and SMC and 
    have introduced two new sub-regions encircling the Andromeda (M31) and Triangulum (M33) galaxies, 
     whose brightest Cepheids are within reach of the \gaia\ mission. These four sub-regions are listed in Table~\ref{tab:subregions}. The fifth sub-region is composed by all the remaining sky after excluding the four sub-regions defined above,  that, 
     for continuity with Paper II, 
     we called  
     All-Sky. This sub-region contains in large majority MW stars, with a small fraction of objects belonging to dwarf galaxies that are satellites of our Galaxy. 
    
    \item{\bf Treatment of multi-mode DCEPs:} To avoid spurious detection of multi-mode DCEPs we have searched for more than one pulsation mode 
    only the time-series of stars with a number of epochs greater or equal to 40. In addition, we introduced an analysis of the residuals after the fit of the $G$ light curve with just one pulsation mode, and retained  as potential multi-mode only objects showing a dispersion larger or equal to 0.025 mag \citep[a similar procedure, although with a larger scatter, is adopted for RR Lyrae stars, see][]{DR3-DPACP-168}.
    
    \item {\bf RV curves treatment:} Since in DR3 RV time-series are published for a small sample of Cepheids and RR Lyrae stars \citep[see also][]{Sartoretti}, a  new module of the pipeline analyses the RV curves, providing average RV values, peak-to-peak amplitudes and epoch of minimum RVs\citep[see][for more details]{DR3-DPACP-168}. 
    
    \item {\bf Update of the $PL$ and $PW$ relations:} We have updated the $PL$ and $PW$ relations that are used in the pipeline, adding those needed to deal with M31 and M33 data. As all these relations are significantly changed with respect to DR2, we describe them in detail in Sect.~\ref{sect:newPL}. 
    
    \item{\bf Errors with bootstrap:} To estimate the uncertainties on all Cepheid parameters published in DR3,  we have applied a bootstrap technique. Specifically, to estimate the uncertainties on the Fourier fit parameters (period, amplitudes and phases), as well as, on all the other quantities characterising the light and RV curves (e.g. mean magnitudes, mean RV, peak-to-peak amplitudes etc.), the input data have been randomly re-sampled (allowing data point repetitions) and all 
    parameters have been recalculated on each simulated sample. This procedure has been repeated 100 times and, for each parameter, the respective uncertainty was estimated by considering the robust standard deviation ($1.486\cdot MAD$) of the distributions obtained with the bootstrap method.
    A similar procedure has been applied to estimate the uncertainties on all other released quantities, such as the metallicity and the Fourier parameters.
    
    \item{\bf Fine tuning of the ROFABO outlier rejection operator:} The photometric and RV time-series are inserted in the  
    \sos\ pipeline after 
    undergoing a chain of routines which elaborate the observations to obtain a standard time, magnitudes, RVs and relative uncertainties, that  constitute the input time-series data. 
    Among these operators standard outlier rejection techniques are applied 
    to remove as many bad points as possible, without affecting the scientific information contained in the time-series. 
    To improve the rejection of 
    outliers from the time-series of Cepheids and RR Lyrae stars, the \sos\ pipeline adopted a customised configuration set of parameters for the 
    ROFABO \citealp[Remove Outliers on both FAint and Bright sides Operator, see][]{DR3-DPACP-162} routine. To determine the best configuration parameters of ROFABO allowing 
    to maximise, specifically for Cepheids and RR Lyrae stars,  the rejection of bad points preserving good measures, 
    we processed with the SOS 
    pipeline a sample of hundreds time-series affected by different kinds of outliers, together with time-series not presenting obvious bad measures. 
    A specific ROFABO function for the \sos\ pipeline with configuration parameters fine tuned as described above 
    was then added to the whole operator chain.
\end{enumerate}

\begin{table}
\caption{Sky sub-regions considered by the \sos\ pipeline. The fifth sub-region, called All Sky comprises all the sky except the four sub-regions listed in the table.}             
\label{tab:subregions}      
\centering                          
\begin{tabular}{c c c c c}        
\hline\hline                 
galaxy & $\alpha_{min}$ & $\alpha_{max}$ & $\delta_{min}$ & $\delta_{max}$ \\    
       &    J(2000)       &   J(2000)        &     J(2000)      & J(2000)        \\
            &   (deg)           &     (deg)        &     (deg)        &    (deg)       \\
\hline                        
LMC &   67.50 & 97.50 & $-75.000$ &$-62.000$ \\
SMC &   0.00   &  30.00 & $-76.000$ &$-70.000$   \\
M31 &   8.75 & 12.75 & 39.667 & 42.833 \\
M33 &   22.90 & 24.00 & 30.000 & 31.300 \\
\hline                                   
\end{tabular}
\end{table}

\begin{table*}[]
\caption{Coefficients and scatter values of the $PL$ and $PW$ relations used for the sky regions including the LMC and SMC. All relations are of the form $mag = \alpha + \beta\cdot \log(P)$. The relations for M31 and M33 are not shown because are the same as for the LMC, but scaling the zero points according to the distance moduli and adopting 24.40 mag for M31, 24.57 mag for M33, and 18.49 mag for the LMC (see text for details).}
    \label{tab-plpw-coeffs}
    \centering
    \begin{tabular}{lcccccc}
    \hline
    Type & Mode & $\alpha$ & $\beta$ & $\sigma$ & Band & N\\
        \hline
        \hline
          \multicolumn{7}{c}{LMC} \\
        DCEP & F & 17.333$\pm$0.010 & $-$2.793$\pm$0.015 &  0.169 & \mbox{$G$} & 2477\\
         DCEP & 1O & 16.890$\pm$0.007 & $-$3.280$\pm$0.020 & 0.197 & \mbox{$G$}& 1775\\
         DCEP & F & 15.998$\pm$0.005 & $-$3.317$\pm$0.007 & 0.075 & \ww  & 2447\\
         DCEP & 1O & 15.521$\pm$0.003 & $-$3.476$\pm$0.009 & 0.081 & \ww  & 1745\\
         ACEP & F & 18.022$\pm$0.026 & $-$2.930$\pm$0.17 & 0.238 & \mbox{$G$}& 102\\
         ACEP & 1O & 17.248$\pm$0.062 &$-$3.340$\pm$0.290 & 0.227 & \mbox{$G$}& 44\\
         ACEP & F & 16.790$\pm$0.021 & $-$3.080$\pm$0.140 & 0.180 & \ww & 97\\
         ACEP & 1O & 16.201$\pm$0.051 & $-$3.500$\pm$0.240 & 0.184 & \ww & 43\\
         T2CEP & -- & 18.731$\pm$0.033 & $-$1.905$\pm$0.036 & 0.275 & \mbox{$G$}& 205\\
         T2CEP & -- & 17.516$\pm$0.019 & $-$2.577$\pm$0.019 &0.138 & \ww & 197\\
         \hline
         \hline
          \multicolumn{7}{c}{SMC} \\
          DCEP & F (P<2.95) & 17.935$\pm$0.012 & $-$3.155$\pm$0.046 & 0.226 & \mbox{$G$} & 1911\\ 
          DCEP & F (P>2.95) & 17.757$\pm$0.026 & $-$2.830$\pm$0.033 & 0.251 & \mbox{$G$} & 843\\ 
          DCEP & 1O & 17.260$\pm$0.007 & $-$3.185$\pm$0.029 & 0.256 & \mbox{$G$} & 1790\\
          DCEP & F (P<2.95) & 16.711$\pm$0.009 & $-$3.627$\pm$0.037 & 0.172 & \ww & 1880\\ 
          DCEP & F (P>2.95) & 16.592$\pm$0.017 & $-$3.382$\pm$0.021 & 0.156 & \ww & 839\\ 
          DCEP & 1O & 16.133$\pm$0.051 & $-$3.595$\pm$0.021 & 0.177 & \ww & 1755\\
          ACEP & F & 18.255$\pm$0.024 & $-$2.430$\pm$0.160 & 0.171 & \mbox{$G$}& 79\\
          ACEP & 1O & 17.633$\pm$0.050 & $-$3.450$\pm$0.290 & 0.186 & \mbox{$G$}& 43 \\
          ACEP & F & 17.161$\pm$0.025 & $-$3.020$\pm$0.160 & 0.169 & \ww & 77\\
          ACEP & 1O & 16.712$\pm$0.054 & $-$3.540$\pm$0.310 & 0.198 & \ww & 40\\
          T2CEP & -- & 19.105$\pm$0.098 & $-$2.140$\pm$0.100 & 0.372 & \mbox{$G$}& 42\\
          T2CEP & -- & 17.843$\pm$0.052 & $-$2.505$\pm$0.054 & 0.190 & \ww & 42\\
                  \hline
        \hline
    \end{tabular}
    
\end{table*}

\subsection{New $PL$ and $PW$ relations employed in the \sos\ pipeline}
\label{sect:newPL}

A number of significant changes, with respect to DR2, were introduced in the branch of the \sos\ pipeline that processes the candidate Cepheids. Specifically:  1) we adopted new $PL$ and $PW$ relations 
directly derived from the \gaia\ data, while in DR2 we had used photometry in the Johnson system transformed into the \gaia\ bands (see sect. 3.2 of Paper II); 2) for DR3 we used the new Wesenheit magnitudes defined by  \citet{Ripepi2019}, that is  \ww=$G-1.90$(\gbp-\grp), which replaced the $W(G,G-G_{RP})$ magnitudes used in DR2 (see eq. 5 in Paper II). 

To calculate the $PL$ and $PW$ relations we gathered Cepheids of all types known from the literature and used the 
SOS pipeline to analyse their light curves in the \gaia\ bands to obtain periods and  intensity-averaged magnitudes in the $G$, \gbp, \grp\ bands \citep[see sect. 2.1 of][for details on how the \sos\ pipeline determine these quantities]{Clementini2016}.  

The calculation of the $PL$ and $PW$ relations required different approaches for the different sub-regions as specified in the following:

\begin{enumerate}
    \item {\bf LMC and SMC:} For both galaxies, we adopted as reference  
    9\,649  DCEPs 
    and 262 ACEPs 
    from \citet{Sos2017}, while the T2CEPs (338 objects) were taken from \citet{Sos2018}.  
    We retrieved the DR3 
    time-series photometry of these stars and used the \sos\ pipeline to derive periods and intensity-averaged magnitudes in the $G$, \gbp, \grp\ bands for the objects with more than 20 epochs (we only wanted good light curves to build the reference 
    $PL$ and $PW$ relations). We discarded all 
    objects for which the SOS and the literature periods did not agree with 
    within 1\%. After all these steps we remained with the number of stars listed in the last column of  Table~\ref{tab-plpw-coeffs}. 
    Linear $PL$ and $PW$ relations were derived from them using the {\tt python LtsFit} package \citep{Cappellari2013}, which has a robust outlier removal procedure.
    \item {\bf M31 and M33:} Given the faint apparent magnitude of the Cepheids in these distant galaxies, we adopted as reference the $PL$ and $PW$ relations we calculated for the LMC (see above), which have the lowest scatter, and simply re-scaled the zero points to take into account the difference in distance moduli between the LMC and M31/M33. For the latter we adopted: $\mu_{M31}=24.40$ mag \citep[the typical value for the M31 globular clusters, see][]{Perina2009} and $\mu_{M31}=24.57$ mag \citep{Conn2012}. 
    However, a different choice for the 
    distance moduli of M31 and M33 would not affect our analysis and results,   
   as we used rather large magnitude intervals (up to 0.6--0.8 mag) around the $PL$ and $PW$ relations to select the candidate Cepheids. 
    \item{\bf All Sky:} The first step consisted in collecting a reliable sample of Cepheids of all types in the MW. To this aim, we adopted the most updated lists of Cepheids  available as of October 2020, namely: \citet[][all types]{Ripepi2019}, \citet[][only DCEPs]{Skowron2019}, \citet[][including DCEPs, ACEPs and T2CEPs]{Sos2020} and \citet[][only DCEPs and T2CEPs, the former not classified according to the pulsation mode and the latter in the  different T2CEPs sub-types]{Chen2020}. After removing overlaps between catalogues, we filtered the resulting list of objects adopting the {\it Gaia} EDR3 astrometry. In particular, we retained only objects with relative error on parallax better than 20\% and RUWE$<$1.4\footnote{Section 14.1.2 of "Gaia Data Release 2 Documentation release 1.2"; https://gea.esac.esa.int/archive/documentation/GDR2/}. This choice was driven by the need of cleaning the sample from contaminants, particularly from binaries, which are easily spotted in the $PW$ diagram as they are usually significantly sub-luminous compared to Cepheids. At the end of this procedure, we were left with a "clean" sample  of All-sky Cepheids, for which numbers divided into  
    various types and/or modes are provided  in the last column of Table~\ref{tab-plpw-allSky-coeffs}. Note that the T2CEP sample only includes BLHER and WVIR stars, because the physical connection with RVTAU stars is questioned (see Introduction). Table ~\ref{tab-plpw-allSky-coeffs} shows that for ACEPs we have only 4 stars in each pulsation mode. Therefore, 
    for ACEPs we 
    adopted the slope of the LMC $PW$ relation 
    and fitted only the zero point.
    For DCEPs and T2CEP the number of objects is instead sufficient to obtain good $PWs$. In fitting the relations 
    to preserve as much as possible the symmetry of the uncertainties on the parallax, we adopted the Astrometric Based Luminosity \citep[ABL][]{Feast1997,Arenou1999}:

\begin{equation}
{\rm ABL}=10^{0.2 W}=10^{0.2(\alpha+\beta\log P)}=\varpi10^{0.2w-2}
\label{eqABLZ}
\end{equation}

\noindent
where, $W$ and $w$ are the absolute and apparent Wesenheit magnitudes and $\varpi$ is the parallax. The fitting procedure is similar to that adopted in \citet{Ripepi2019} and \citet{Ripepi2022a} and will not be repeated here. The resulting coefficients for the $PW$ relations of All-Sky Cepheids with different  type/mode are summarised in Table~\ref{tab-plpw-allSky-coeffs}.  
\end{enumerate}

The $PL$ and $PW$ relations described above represent a fundamental tool of the Cepheid branch in the \sos\ pipeline, as we use them for a first 
classification of the candidate Cepheids 
in different types/modes (see Paper~I and II, for full details on the pipeline). 
In practice, we define a band across each $PL$ and $PW$ relation, 
as $\pm n \times \sigma$, where $\sigma$ is the dispersion of each relation. For DR3, we used 1 $\sigma$ for the ACEPs, 4 $\sigma$ (10 $\sigma$ for the ABL formalism) for the DCEPs and 2$\sigma$ for the T2CEPs. These values have been calibrated using the LMC, SMC and All Sky samples of known Cepheids defined 
above, as to minimise the overlap between contiguous variable types/modes, at the same time maximising the number of correct classifications.

            \begin{table*}[]
    \caption{Same as in  Table~\ref{tab-plpw-coeffs}, but for All-Sky Cepheids. The reported values of the scatter refer to the residuals around the fit in the ABL formalism.}

    \centering
    \begin{tabular}{lcccccc}
    \hline
    Type & Mode & $\alpha$ & $\beta$ & $\sigma_{ABL}$ & Band & N\\
        \hline
        \hline
          \multicolumn{6}{c}{All Sky} \\
          DCEP & F & $-$2.744$\pm$0.045 & $-$3.391$\pm$0.052 & 0.015 & \ww & 898\\
          DCEP & 1O & $-$3.224$\pm$0.028 & $-$3.588$\pm$0.065 & 0.021 & \ww & 416\\
          ACEP & F & $-$1.717$\pm$ & $-$4.061$\pm$ & 0.010 & \ww & 4\\
          ACEP & 1O & $-$2.220$\pm$ & $-$4.057$\pm$ & 0.013 & \ww & 4\\
          T2CEP & -- & $-$1.224$\pm$0.039 & $-$2.542$\pm$0.088 & 0.041 & \ww & 264\\
         \hline
    \end{tabular}
    \label{tab-plpw-allSky-coeffs}
\end{table*}

\section{Application of the SOS Cep\&RRL pipeline to the DR3 data: cleaning of the sample}

The \gaia\ DR3 data analysed by the \sos\ pipeline consist in $G$ and integrated \gbp\ and \grp\ time-series photometry collected between 25 July 2014 and 28 May 2017, spanning a period of 34 months (for reference, DR2 was based on 22 months of observations). In addition to the time-series photometry, for DR3 we also analysed the RV time-series \citep[see][for the general procedures used to measure RV in {\it Gaia}]{Sartoretti} for a selected sample of 799 Cepheids of all types. Among these, 798 are Cepheids present in the {\tt vari\_cepheid} catalogue, while one objects, previously classified as RR Lyrae,  resulted to be a DCEP\_MULTI variable (source\_id=5861856101075703552) and is present in the {\tt vari\_rrlyrae} catalogue \citep[see][for full details]{DR3-DPACP-168}. 

The general treatment of the light and RV curves and the processing steps that precede the \sos\ pipeline are schematically summarised by \citet{Holl2018} and \citet{DR3-DPACP-162}. In particular, the \sos\ pipeline processed candidate Cepheids (and RR Lyrae stars)\footnote{We recall that the RR Lyrae stars are discussed in a companion paper \citep[][]{DR3-DPACP-168}} identified as such by the supervised classification of the general variability pipeline \citep[see][for details]{DR3-DPACP-162,DR3-DPACP-165}  with various probability levels. In order to maximise the number of DCEPs known from the literature that are recovered, 
we considered classification candidates that 
have also low probability levels.
Among the Cepheid candidates, the \sos\ pipeline retained for analysis 
only objects with at least 12 measurements in the $G$-band, while the RV time-series were processed 
only for sources with a number of RV  measurements larger than/equal to 7. 

At the end of this first processing we obtained a sample of about 1 million Cepheid candidates of all types. Among them, only about 5\,000 
were in M31 and M33. To reduce the huge number of candidate Cepheids in the LMC, SMC and, in particular, in the All Sky sample, to more manageable numbers, we applied the following series of filters:

\begin{enumerate}
\item {\bf Separation of known or suspected Cepheids in the literature:} We separated from the whole sample of Cepheid candidates, sources that are known or suspected Cepheids of all type in the literature. This was done for  each of the five sub-regions defined in Table~\ref{tab:subregions}. This first step was necessary to avoid filtering out possible good objects in the following cutting steps. The majority of the literature Cepheids were then  validated by visual inspection, as described in Sect.~\ref{sect:visualInspection}. 
For the known Cepheids 
in the LMC, and SMC we retained those mentioned in Sect.~\ref{sect:newPL}, while to the All-Sky known Cepheids we added all the objects classified as Cepheids as of February 2021 in the {\sl SIMBAD} database \citep[available at CDS, Centre de Donn\'es astronomiques de Strasbourg,][]{Wenger2000}. For M31 and M33 we adopted the samples by \citet{Kodric2018} and \cite{Pellerin2011}, respectively. 
After eliminating 
overlaps, the overall literature sample 
within 
the one million candidates, adds to 
about 16,000 objects.
These literature Cepeheids were elected for visual inspection, with the exclusion of 
about 9\,000 Cepheids in the MCs, for which the OGLE classification 
is already reliable.    

\item {\bf Goodness of the light curves:} We filtered the remaining sample based on uncertainties on the light curve parameters. More in detail, we applied the cuts listed in Table~\ref{tab:cutsLC}. This allowed us to filter out about 10\% of the sources
and, in particular, to reduce 
the All-Sky sample to approximately 667\,000 sources. 

\item{\bf Probability of the classifiers (LMC and SMC samples):} Since the number of candidates remaining from the previous steps was still too large for the LMC and SMC, we reconsidered the probability adopted in selecting Cepheid candidates from the classifiers of the general variability pipeline. Adopting again the highly reliable sample of literature objects in the MCs, we calculated for each classifier the probability which returned us the 95\% of the known Cepheids. This procedure was very effective, leaving us with only about 2\,500 new Cepheid candidates in the two MCs.

\item{\bf Filtering of aliasing periods (M31 and M33 samples):} As discussed in \citet{DR3-DPACP-164}, instrumental effects produce false variable sources with periods packed at specific values that are strictly correlated with the position 
on the sky. These effects are particularly disturbing in the case of M31 and M33, given that for these galaxies we can count only on the $G$-band photometry. 

Luckily, as the range in coordinates spanned by the M31 and M33 data is rather small, the aliases correlated with the position on the sky produce narrow peaks in period. A histogram of the periods provides five and seven narrow period peaks in M31 and M33, respectively. Filtering the stars in those intervals left us with 1\,923 candidate Cepheids in M31 and 1\,332 stars in M33,  to be further checked.

\item{\bf Filtering on number of epochs, limiting magnitude, amplitude and period (All-Sky sample):} Since the All-Sky sample resulting from the previous filtering was still too large 
we applied the following further filtering: $G<19.0$ mag, amp($G$)$>$0.15 mag, maximum period $P_{max}$=100 days and number of epochs in the $G$-band $>$30. The selection on the number of epochs was motivated by the need of measuring accurate periods, while the limits in magnitude and amplitude allowed us to reduce significantly the number of spurious variability detections caused by instrumental effects \citep[see e.g.][]{DR3-DPACP-164} which are more likely among faint sources, whose 
\gbp\ and \grp\ magnitudes are also in  most cases 
not accurate. The cut in period is justified because 
very few Cepheids, both DCEPs and RVTAU, are expected to exceed a period of 100 days.
In the end, the above filtering left us with 166k candidates for further analysis.

\item{\bf Machine Learning filtering:} While the sample in the MCs was small enough to be checked visually, the All Sky sample was still too large. We therefore applied  an additional filtering, based on machine learning techniques. We adopted a supervised classification method based on 
a reliable training set. To build the 
training set, we adopted a sample of Cepheids of all types similar to that described in Sect.~\ref{sect:newPL}, but not limited in relative parallax error, to increase the statistics, thus including in total about 4\,100 objects. To this sample we added about 2\,250 contaminants of different types, including RR~Lyrae stars, long period variables, eclipsing binaries, etc. taken from objects for which the general classification pipeline assigns very high probability of belonging to the given class. In addition, we verified that the vast majority of the contaminants were also known in the literature with a classification in agreement with that assigned by the classification pipeline. 

After establishing  
the training set, we defined the input attributes for the machine learning algorithm. Based on parameters that are already used by the \sos\ pipeline, we adopted: the first periodicity, the second periodicity (if any), the absolute magnitudes in all bands, the absolute Wesenheit magnitudes (in $G$, \gbp$-$\grp), the amplitudes in all bands, the amplitude ratios (amp(\gbp)/amp(\grp); amp(\gbp)/amp($G$); amp($G$)/amp(\grp)), colours (\gbp$-$\grp; \gbp$-G$; $G-$\grp) and the Fourier parameters ($R_{21}; \,R_{31};\, \phi_{21};\, \phi_{31}$). The classes feed to the algorithm were: ACEP\_F,  ACEP\_1O, DCEP\_F, DCEP\_1O, DCEP\_MULTI, BLHER, WVIR, RVTAU and OTHER, where the last tag included all the non-Cepheid  objects. To execute the machine learning procedure we used the {\tt H2O} platform\footnote{h2o.ai}. After ingesting the training set, we divided it into training/validation sets in proportion of 85\%/15\%. We then carried out several tests to find the best model for our case amongst those offered by the {\tt H2O} package. The model that returned the largest percentage of precision in detecting the right classes/modes resulted to be the {\tt XGBOOST} algorithm.

We applied this model to the sample of 166k All-Sky candidate Cepheids returned by the selection described in point 5, obtaining a probability of belonging to one of the classes mentioned above for each candidate. A quick visual examination of samples of light curves for objects with probability to be Cepheids of all types larger than 50\% revealed that there were no reliable  candidates with probability $<90\%$. We therefore considered only candidates  with probability larger than 90\%, for a total of 10\,273 sources. Finally, to further restrict the number of stars for visual inspection, we adopted the peak-to-peak amplitudes, requiring that:  1.3$\leq$amp(\gbp)/amp(\grp)$\leq$2.0; 1.1$\leq$amp($G$)/amp(\grp)$\leq$1.5 and 100$\times \sigma G$/amp($G$)$\leq2.0$. These broad limits include the large majority of bona fine Cepheids 
according to tests carried out on the training set adopted for the machine learning procedure.  
After applying this last filtering, we were left with 7\,349 stars to inspect visually. 
\end{enumerate}
To conclude, at the end of the whole filtering procedure described in this section, we were left with about 20\,100 Cepheids to visually inspect, in order to validate the classification provided by the \sos\ pipeline. 

\begin{table}[]
\caption{Constraints on the results from the light curve fitting.}
\label{tab:cutsLC}
\centering
    \begin{tabular}{l}
\hline
    \hline
    Parameter\\
        \hline
$0.0 < G \leq 22.0$ mag \\
$0.0 <$ \gbp\ $\leq 22.0$ mag\\
$0.0 <$ \grp\ $\leq 22.0$ mag\\
$0.0 <\sigma G \leq 0.5 $ mag\\
$0.0 < \sigma$\gbp\  $\leq 0.5 $ mag\\
$0.0 < \sigma $\grp\ $\leq 0.5 $ mag\\
$0.0 < $ amp($G$) $\leq 2.5 $ mag\\
$0.0 < $ amp(\gbp) $\leq 2.5 $ mag\\
$0.0 < $ amp(\grp) $ \leq 2.5 $  mag\\
$0.0 < \sigma$amp($G$)/ amp($G$)$\leq 1.0 $ \\
$0.0 < \sigma$amp(\gbp)/ amp(\gbp) $\leq 1.0 $ \\
$0.0 < \sigma$amp(\grp)/ amp(\grp) $\leq 1.0 $ \\
$0.0 < R_{21} \leq 2.0 $ mag\\
$0.0 < R_{31} \leq 2.0 $ mag\\
$0.0 < \sigma R_{21}/R_{21} \leq 1.0 $ \\
$0.0 < \sigma R_{31}/R_{31} \leq 1.0 $ \\
$0.0 < \sigma \phi_{21}/\phi_{21} \leq 1.0 $ \\
$0.0 < \sigma \phi_{31}/\phi_{31} \leq 1.0 $ \\
\hline
\end{tabular}

\end{table}

\section{Correction of the \sos\ pipeline classification}
\label{sect:visualInspection}

As 
mentioned in the previous section, a number of sources were selected for further inspection to verify their classification. Different procedures were adopted for the 
LMC/SMC, M31/M33 and the All Sky samples due to the different characteristics of the available data. More in detail, for the LMC and SMC, the literature samples have a solid classification and we already knew from DR2 that the \sos\ pipeline provides reliable classifications for the Cepheids  
in these two galaxies. For this reason we did not check visually the known Cepheids in the MCs, 
but only the new candidates. Differently, 
the classification of Cepheids 
in M31 and M33 required a careful validation, due to the low signal to noise ratio of the \gaia\ data 
and the much less established literature for Cepheids in these galaxies.
Concerning the All-Sky sample, the literature sample is likely contaminated both by non-Cepheids and by wrong classifications (i.e. wrong Cepheid types/pulsation modes) because they do not rely on solid distances, but mainly on the analysis of the light curve shapes.
For this reason, we checked all the known Cepheids in addition to the new candidates for the All Sky sample.  

\subsection{Visual inspection of the Gaia DR3 light curves}

In general, for each star we evaluated the shape of the light curves in all  \gaia\ bands, the position in the period-Fourier parameters diagrams ($P-R_{21};\,P-R_{31}; \,P-\phi_{21}; \, P-\phi_{31})$, the position on the $PL$ and $PW$ relations as well as the amplitude ratios amp(\gbp)/amp(\grp) and amp($G$)/amp(\grp). In case of negative parallax values, the ABL function was used. We adopted the very useful "OGLE atlas of light curves"\footnote{http://ogle.astrouw.edu.pl/atlas/index.html} as reference for the shapes of the light curves of Cepheids of all types. 
For Cepheids in 
M31 and M33, 
we could rely only on the $G$-band light curves and the position in the $PL$ relation, as 
for 
$G \sim$ 20-21 mag the \grp\ and \gbp\ magnitudes are often missing or totally  
unreliable, hence the Wesenheit relation was not usable in most cases. 

In the All-Sky sample, major difficulties were to distinguish DCEP\_1O from first overtone RR Lyrae stars with periods smaller than 0.4 days, in case of light curves not very well defined and parallaxes with relative errors larger than 10--20\%. Similarly, in some cases it was difficult to separate DCEP\_1O and ACEP\_1O with periods $\sim$0.7-0.8 days. Even more challenging was to distinguish ACEP\_F from ab-type RR Lyrae for periods around 0.6 days and from DCEP\_F in the period range 1.0--1.4 days. These difficulties arose mainly from the very similar shape of the light curves for these 
types of variable stars, which can be distinguished 
only based on 
fine details of the light curves, such as humps and bumps, not always clearly visible. Also WVIR and DCEP\_F can be confused in presence of noisy light curves and inaccurate parallaxes. In all those cases, the Fourier parameters provided as well ambiguous results, since they 
stem directly from the light curve shape.  

A main source of contamination is given by contact binary stars, whose light curves mimic that of the overtone Cepheids and, to some extent, also those of the WVIR variables. To mitigate this problem, we always inspected the light curves folded according to once and twice the  period provided by the \sos\ pipeline. 
In this way it was often possible to identify stars for which there was a small, but detectable difference between the light curve minima. This check, in conjunction with the amplitude ratios  amp(\gbp)/amp(\grp) and amp($G$)/amp(\grp), which for binaries tend to assume values close to unity (see Sect. 4 in Paper~II), while is much larger for pulsating stars \citep[see e.g. table 4 in][]{Ripepi2019}, allowed us to detect and reject the large majority of
potential contaminants that are contact binaries. 

During visual inspection, many objects classified as DCEP\_1O variables by the \sos\ pipeline were found to show a  larger scatter than other sources of same magnitude, leading us to suspect they might be missed multi-mode objects. As discussed in detail in Sect.~\ref{multi}, we searched for secondary periodicities in the light curves of the stars in this sample, finding that many of them are actually multi-mode pulsators. A 
large fraction of them were missed simply because of the too strict constraint on the number of epochs and scatter in the light curves, introduced in the \sos\ pipeline (see point 2 of Sect.~\ref{sect:2}), which allowed us to minimise the number of spurious detection, but at the same time also  prevented us from detecting many genuine multi-mode pulsators. 

\begin{table*}
\caption{Re-processing of the \gaia\ data for DCEP\_MULTI objects not detected as such by the \sos\ pipeline. The different columns show: source identification, longest and shortest pulsation periods with relative errors, period ratio, classification (for brevity we use F, 1O, 2O to indicate the fundamental, first and second overtones, respectively).   
Only the first ten lines are shown
to guide the reader about the table content. The entire version of the
table will be published at CDS (Centre de Données astronomiques de
Strasbourg, https://cds.u-strasbg.fr/).}             
\label{tab:multi}     
\begin{tabular}{rrrrrrr}
\hline
  \multicolumn{1}{c}{source\_id} &
  \multicolumn{1}{c}{$P_L$} &
  \multicolumn{1}{c}{$\sigma P_L$} &
  \multicolumn{1}{c}{$P_S$} &
  \multicolumn{1}{c}{$\sigma P_S$} &
  \multicolumn{1}{c}{$P_S$/$P_L$} &
  \multicolumn{1}{c}{Class} \\
  \multicolumn{1}{c}{} &
  \multicolumn{1}{c}{(days)} &
  \multicolumn{1}{c}{(days)} &
  \multicolumn{1}{c}{(days)} &
  \multicolumn{1}{c}{(days)} &
  \multicolumn{1}{c}{} &
  \multicolumn{1}{c}{} \\
\hline
           5864639514713019392  &  0.216830   & 1.25e-07   &   0.172544   & 4.73e-07    &   0.796   &   1O/2O   \\    
           5853820767014992128  &  0.236592   & 1.08e-04   &   0.188584   & 2.80e-05    &   0.797   &   1O/2O   \\
           5423800601092727168  &  0.238321   & 3.74e-04   &   0.190348   & 3.63e-05    &   0.799   &   1O/2O   \\
           5601418217705666560  &  0.239664   & 3.99e-07   &   0.191709   & 8.04e-05    &   0.800   &   1O/2O   \\
           4313476032410287104  &  0.242839   & 1.10e-02   &   0.193298   & 8.99e-05    &   0.796   &   1O/2O   \\
           5409512756735301120  &  0.247992   & 3.75e-07   &   0.197926   & 1.27e-06    &   0.798   &   1O/2O   \\
           5939019827046790272  &  0.249715   & 8.01e-04   &   0.198971   & 4.43e-06    &   0.797   &   1O/2O   \\
           5941658375763435648  &  0.254389   & 6.52e-07   &   0.202699   & 2.43e-06    &   0.797   &   1O/2O   \\
           5254261818006768512  &  0.262113   & 1.61e-06   &   0.209564   & 4.74e-06    &   0.800   &   1O/2O   \\
           3314887198215151104  &  0.263087   & 6.39e-07   &   0.199706   & 2.39e-06    &   0.759   &   F/1O    \\
  \hline\end{tabular}
\end{table*}

\begin{table*}
\caption{Reclassification of objects 
incorrectly classified by the  \sos\ pipeline. The column "class" provides  the correct Cepheid type/mode; "comment" describes if the class is wrong or 
the pulsation mode; "region" shows the sky region to which the particular star belongs.  
Only the first ten lines are shown
to guide the reader about the table content. The entire version of the
table will be published at CDS (Centre de Données astronomiques de
Strasbourg, https://cds.u-strasbg.fr/).}             
\label{tab:reclassification}     
\begin{tabular}{rrrlll}
\hline
  \multicolumn{1}{c}{sourceid} &
  \multicolumn{1}{c}{RA} &
  \multicolumn{1}{c}{Dec} &
  \multicolumn{1}{c}{class} &
  \multicolumn{1}{c}{comment} &
  \multicolumn{1}{c}{Region} \\
  \multicolumn{1}{c}{} &
  \multicolumn{1}{c}{(deg)} &
  \multicolumn{1}{c}{(deg)} &
  \multicolumn{1}{c}{} &
  \multicolumn{1}{c}{} &
  \multicolumn{1}{c}{} \\
\hline
  2422853521974230400 &0.073909 & $-$10.221463 & ACEP\_F & WRONG\_CLASS & All Sky\\
  565137161224290944 & 1.836580 & 80.297101 & DCEP\_F & WRONG\_CLASS & All Sky\\
  419703349473530240 & 4.678347 & 54.039237 & WVIR & WRONG\_CLASS & All Sky\\
  2367033515654862720 & 4.817180 & $-$18.075243 & ACEP\_F & WRONG\_CLASS & All Sky\\
  430629986799994880 & 5.630325 & 63.033041 & DCEP\_1O & WRONG\_MODE & All Sky\\
  431184518613946112 & 6.407203 & 64.229891 & DCEP\_MULTI & WRONG\_MODE & All Sky\\
  382372112206462336 & 7.230740 & 43.033510 & BLHER & WRONG\_CLASS & All Sky\\
  4906654274849806592 & 7.296960 & $-$57.939912 & ACEP\_1O & WRONG\_CLASS & All Sky\\
  4980356188527065472 & 7.672437 & $-$44.272952 & ACEP\_F & WRONG\_CLASS & All Sky\\
  375318264077848448 & 11.680155 & 42.092860 & DCEP\_1O & WRONG\_MODE  & M31\\
  \hline\end{tabular}
\end{table*}

\subsection{Multi-mode Cepheids}\label{multi}

\begin{figure}
   \centering
         \includegraphics[width=\hsize]{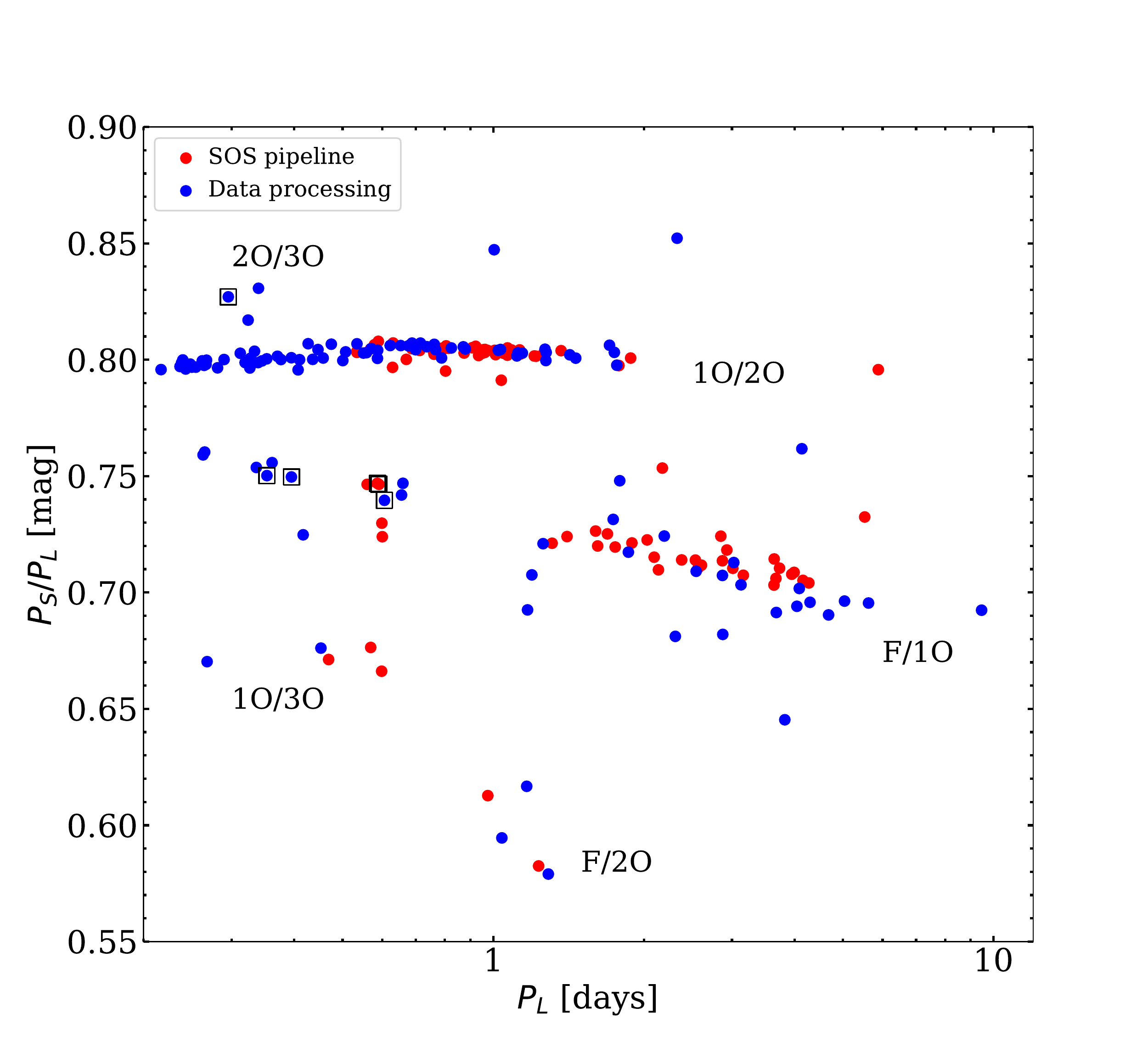}
      \caption{Petersen diagram for  confirmed DCEP\_MULTI published in the \gaia\ DR3 catalogue (red filled circles) and for  additional DCEP\_MULTI detected in the re-processing of the data (blue filled circles). $P_L$ and $P_S$ represent the longest and shortest pulsation periods 
      of the multi-mode object.
      Labels show the typical location of the different multi-mode pulsation combinations identified
      in these sources. Black squares mark six objects known in the literature as ARRDs (see Sect.~\ref{sect:arrd}).
              }
         \label{fig:multi}
   \end{figure}

DCEPs in the All-Sky sample that during visual inspection were 
suspected to be multi-mode pulsators,  were further investigated, 
by analysing their light curves with software external to the \sos\ pipeline. In particular, we used the {\tt Period04} package \citep{Lenz2005} for a first selection of most promising candidates and to determine their  periodicities. Then we used a custom program written in {\tt python} to carry out the non-linear fitting with truncated Fourier series, the prewhitening of the first periodicity and then the fitting of all periodicities together. In close similarity  with the \sos\ pipeline, we finally determined the period uncertainties with a  bootstrap procedure. The re-processing led to identify 109 DCEP\_MULTI variables, 
in addition to the 86 DCEP\_MULTI for which the \sos\ pipeline provided the correct classification. The list of additional DCEP\_MULTI variables and their 
periods,  with the relative errors, are  provided in Table~\ref{tab:multi}. The Petersen diagram (period ratios vs longer period) for the DCEP\_MULTI 
in the All-Sky sample is shown in Fig.~\ref{fig:multi}, where the loci occupied by the different period ratios have been 
 taken from \citet{Sos2020}.  

\begin{table}
\caption{Number and type/mode classification of Cepheids confirmed by the \sos\ pipeline that are published in {\it Gaia} DR3. The classification corrections discussed in Sect.~\ref{sect:visualInspection} have been taken into account in the calculations. The number of objects is provided for each of the five regions in the sky adopted in this work. The meaning of the last three column is the following: OTHER= stars present in the {\tt vari\_cepheid} table which after visual inspection resulted in variable stars of type other than Cepheid; Reclassified=objects classified as Cepheids in the {\tt vari\_cepheid} table which are known in the literature with different variability types; New=Cepheid variables present in the {\tt vari\_cepheid} table which, as far as we know, were not reported before in the literature.}             
\label{table:numberResult}     
\begin{tabular}{l r r r r r}       
\hline\hline                 
\noalign{\smallskip}
Type & All-sky & LMC  & SMC  & M31 & M33 \\ 
\hline                          
\hline                        
\noalign{\smallskip}
DCEP F          & 2.008  & 2,357    & 2,487   & 309 & 173 \\ 
DCEP 1O         & 1,101  & 1,931    & 1,803   & 10  & 12  \\ 
DCEP MULTI      & 195    &    58    &   110   & --  &  -- \\ 

\hline                          
\noalign{\smallskip}
DCEP Total      & 3,304  &  4,346  &  4,400  & 319  &  185 \\ 
\hline                          
\noalign{\smallskip}
ACEP F          & 150    &    69   &  87   &  --  & -- \\  
ACEP 1O         & 132    &    32   &  80   &  --  & -- \\  
\hline                          
\noalign{\smallskip}
ACEP Total      &  282   &   101   &  167  & -- & -- \\ 
\hline                       
\noalign{\smallskip}
T2CEP BLHER     &  579  &     66   &  16    & -- & -- \\  
T2CEP WVIR      &  795  &     120  &  20    & -- & -- \\   
T2CEP RVTAU     &  261  &     30   &  13    & 2  & -- \\   
\hline                          
\noalign{\smallskip}
T2CEP Total     & 1,635 &     216  &  49    & 2  & -- \\ 
\hline
\hline
\noalign{\smallskip}
Cepheid Total & 5,221   & 4,663   & 4,616   & 321 & 185 \\
\hline
\hline
\noalign{\smallskip}
OTHER           & 15     &    --    &  --     & --  &  -- \\
\hline  
\noalign{\smallskip}
Reclassified    & 327    &    15    &  1     & 18  &  5 \\
\hline  
\noalign{\smallskip}
New             & 472     &    3    &  11    & 22  &  57 \\
\hline  
\noalign{\smallskip}
\end{tabular}
\end{table}

\subsection{Final classification}
The processing of the SOS pipeline along with the validation, cleaning and re-classification procedures described in the previous sections, produced a final catalogue of 15\,021 Cepheids 
of all types, that populate the {\tt vari\_cepheid} table in the \gaia\ DR3 archive.
Notwithstanding our efforts to clean the sample from spurious objects, after a deeper analysis, 15 sources  
turned out to be non-Cepheid variables (these objects are listed with the new classification in Table~\ref{tab:reclassification}), thus bringing the total number of bona-fide  Cepheids of all types in \gaia\ DR3 to 15\,006.  

In total, we changed the \sos\ 
classification of 1\,160 stars. This corresponds to 
about 
8\% of the total sample. The new classifications are given in Table~\ref{tab:reclassification}. 
Taking into account all 
re-classifications, we report in Table~\ref{table:numberResult} the breakdown of the DR3 Cepheids 
by types in the different sub-regions in which we divided our sample.

The comparison with the literature, which will be discussed in more detail in  Sect.~\ref{sect:validationLiterature}, along with a cross match with the SIMBAD database\footnote{http://simbad.u-strasbg.fr/simbad/} \citep[][]{Wenger2000} allowed us to calculate the number of Cepheids of any type 
already known in the literature, how many 
are classified as variables but of non-Cepheid type and,  the number of new discoveries. The result of this exercise is reported in the last line of Table~\ref{table:numberResult}. The largest number of new or reclassified objects belongs to the All-Sky sample, but we note that many new Cepheids were  discovered also in M31 and M33.  

\begin{table*}
\tiny
\setlength\tabcolsep{5pt}
\caption{
Links to {\it Gaia} archive table to retrieve the 
pulsation characteristics: period(s), epochs of maximum light and minimum radial velocity (E), peak-to-peak amplitudes, intensity-averaged mean magnitudes, mean radial velocity, $\phi_{21}$, $R_{21}$, $\phi_{31}$, $R_{31}$ Fourier parameters with related uncertainties and metallicity computed by the SOS Cep\&RRL pipeline for the 15\,021 objects (15\,006 Cepheids and 15 stars of different type) released in {\textit{Gaia}} DR3.}           
\label{tab:vari-cep}     
\centering                         
\begin{tabular}{ll}       
\hline\hline                 
\noalign{\smallskip}
Table URL & \texttt{http://archives.esac.esa.int/gaia/}\\
\noalign{\smallskip}
\hline
\noalign{\smallskip}
\multicolumn{2}{c}{Cepheids 
main parameters computed by the SOS Cep\&RRL pipeline}\\
\hline
\noalign{\smallskip}
Table Name&  \texttt{\textbf{gaiadr3.vari\_cepheid}}\\
Source ID & \texttt{source\_id}\\
Type& \texttt{type\_best\_classification} (one of \texttt{T2CEP}, \texttt{DCEP} or \texttt{ACEP})\\
Type2& \texttt{type2\_best\_classification} (for type-II Cepheids, one of \texttt{BL\_HER}, \texttt{W\_WVIR} or \texttt{RV\_TAU})\\
Mode& \texttt{mode\_best\_classification} (one of \texttt{FUNDAMENTAL}, \texttt{FIRST\_OVERTONE}, \texttt{SECOND\_OVERTONE}\\
{}	& \texttt{MULTI}, \texttt{UNDEFINED}, or \texttt{NOT\_APPLICABLE})\\
Multi-mode& \texttt{multi\_mode\_best\_classification} (for multi-mode $\delta$ Cepheids, one of \texttt{F/1O}, \texttt{F/2O}, \texttt{1O/2O}, \\
{}	& \texttt{1O/3O}, \texttt{2O/3O}, \texttt{F/1O/2O}, or \texttt{1O/2O/3O})\\
$Pf, P1O, P2O, P3O$
&\texttt{p\_f, p1\_o, p2\_o, p3\_o} \\
$\sigma (Pf, P1O, P2O, P3O)$
&\texttt{pf\_error, p1\_o\_error, p2\_o\_error, p3\_o\_error}
\\
E\tablefootmark{\rm (a)}($G$,  $G_{\rm BP}$, $G_{\rm RP}$,  RV)&\texttt{epoch\_g, epoch\_bp, epoch\_rp, epoch\_rv}\\
$\sigma {\rm E}(G, G_{\rm BP}, G_{\rm RP}, {\rm RV})$
&\texttt{epoch\_g\_error, epoch\_bp\_error, epoch\_rp\_error,  epoch\_rv\_error}\\
$\langle G \rangle, \langle G_{\rm BP} \rangle, \langle G_{\rm RP} \rangle$, $\langle {\rm RV} \rangle$ &\texttt{int\_average\_g, int\_average\_bp, int\_average\_rp, average\_rv}\\
$\sigma \langle G \rangle$ , $\sigma \langle G_{\rm BP} \rangle$ , $\sigma \langle G_{\rm RP} \rangle$, $\sigma \langle {\rm RV} \rangle$ &\texttt{int\_average\_g\_error, int\_average\_bp\_error, int\_average\_rp\_error,} \\
{} & \texttt{average\_rv\_error}\\
Amp$(G, G_{\rm BP}, G_{\rm RP}, RV)$&\texttt{peak\_to\_peak\_g, peak\_to\_peak\_bp, peak\_to\_peak\_rp, peak\_to\_peak\_rv}\\
$\sigma [{\rm Amp}(G)], \sigma [{\rm Amp}(G_{\rm BP}], \sigma [{\rm Amp}(G_{\rm RP}], \sigma [{\rm Amp}({\rm RV})]$ &\texttt{peak\_to\_peak\_g\_error, peak\_to\_peak\_bp\_error, peak\_to\_peak\_rp\_error,}\\
{} & \texttt{peak\_to\_peak\_rv\_error}\\
$\phi_{21}(G)$ &\texttt{phi21\_g}\\
$\sigma [\phi_{21}(G)]$ &\texttt{phi21\_\linebreak g\_error} \\
$R_{21}(G)$ &\texttt{r21\_g}\\
$\sigma [R_{21}(G)]$&\texttt{r21\_g\_error}\\ 
$\phi_{31}(G)$ &\texttt{phi31\_g}\\
$\sigma [\phi_{31}(G)]$ &\texttt{phi31\_\linebreak g\_error} \\
$R_{31}$ &\texttt{r31\_g}\\
$\sigma [R_{31}(G)]$&\texttt{r31\_g\_error} \\ 
${\rm [Fe/H]}\tablefootmark{\rm (b)}$&\texttt{metallicity}\\
$\sigma$ ([Fe/H]) &\texttt{metallicity\_error}\\
$N_{\rm obs}$($G$ band) & \texttt{num\_clean\_epochs\_g}\\
$N_{\rm obs}$($G_{\rm BP}$ band) & \texttt{num\_clean\_epochs\_bp}\\
$N_{\rm obs}$($G_{\rm RP}$ band) & \texttt{num\_clean\_epochs\_rp}\\
$N_{\rm obs}$(RV)  & \texttt{num\_clean\_epochs\_rv}\\
\noalign{\smallskip}
\hline                                  
\end{tabular}
\tablefoot{To ease table access, we also provide the correspondence between parameter [period(s), E, etc.] and the name of the parameter in the \textit{Gaia}  archive table. 
$^{\rm (a)}$E corresponds to the time of   maximum in the light curve and the time of minimum in the RV curve. The BJD of all epochs 
is offset by JD 2455197.5 d (= J2010.0).}  
\end{table*}

\section{Properties of the Cepheids in the \gaia\ DR3}

A summary of the parameters provided by the \sos\ pipeline which form the entries of the {\tt vari\_cepheid} table is provided in Table~\ref{tab:vari-cep}.

In the following subsections we describe the main properties of the Cepheids in \gaia\ DR3. 

Examples of light and RV curves for DCEPs of different pulsations modes are shown in Fig.~\ref{fig:lc_dceps}. Similarly, Fig.~\ref{fig:lc_t2ceps} displays the \gaia\ time-series for the prototypes of the T2CEP classes, namely BL\,Her, W\,Vir and RV\,Tau. Finally, Fig.~\ref{fig:lc_aceps} shows the light and RV curves for ACEP\_F and ACEP\_1O variables.   

\subsection{Number of epochs}

\begin{figure}
   \centering
      \includegraphics[width=\hsize]{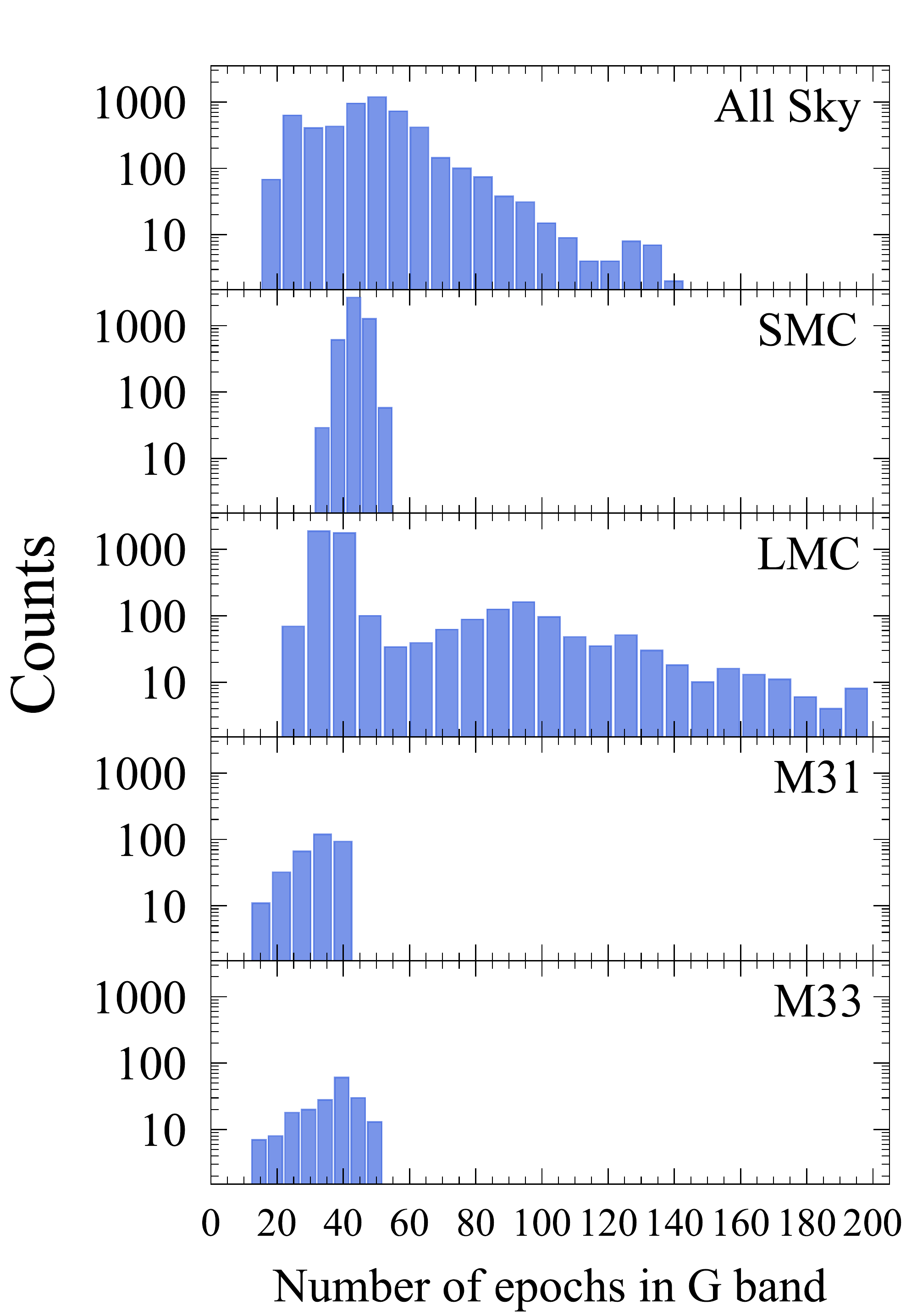}
            \caption{Number of epochs in the $G$-band time series. From top to bottom the different panels show the data for the different sub-samples corresponding to the five regions of the sky defined in Sect.~\ref{sect:2}. 
            }
         \label{fig:nEpochs}
   \end{figure}

An important quantity affecting the quality of the results is the number of epochs in the light and RV curves. This feature is strictly depending on the position of a specific object in the sky, as the \gaia\ scanning law is extremely non-uniform \citep[see][]{Gaia2016}. Obviously, more epochs are available for the analysis of the time-series, more precise will be the determination of the periods, amplitudes, etc. Figure~\ref{fig:nEpochs} shows histograms with the number of epochs in the $G$ band for each sub-sample (the number of epochs in \gbp\ and \grp\ provide similar distributions). Restricted regions in the sky such as the SMC, M31 and M33 show narrower intervals of epochs than  both the All-Sky and LMC samples. The latter shows an extended tail with many DCEPs having more than 140 epochs because it is located in the region interested by the EPSL (Ecliptic Pole Scanning Law) which was covered continuously during the first 28 days of the \gaia\ mission \citep[see][]{Gaia2016Brown}. Unfortunately, for M31 and M33, which are the most difficult sub-regions, due to the large distance, the number of epochs is small (less than 40 on average for M31), making it difficult to study the Cepheids in these systems. 

\begin{figure}
   \centering
         \includegraphics[width=\hsize]{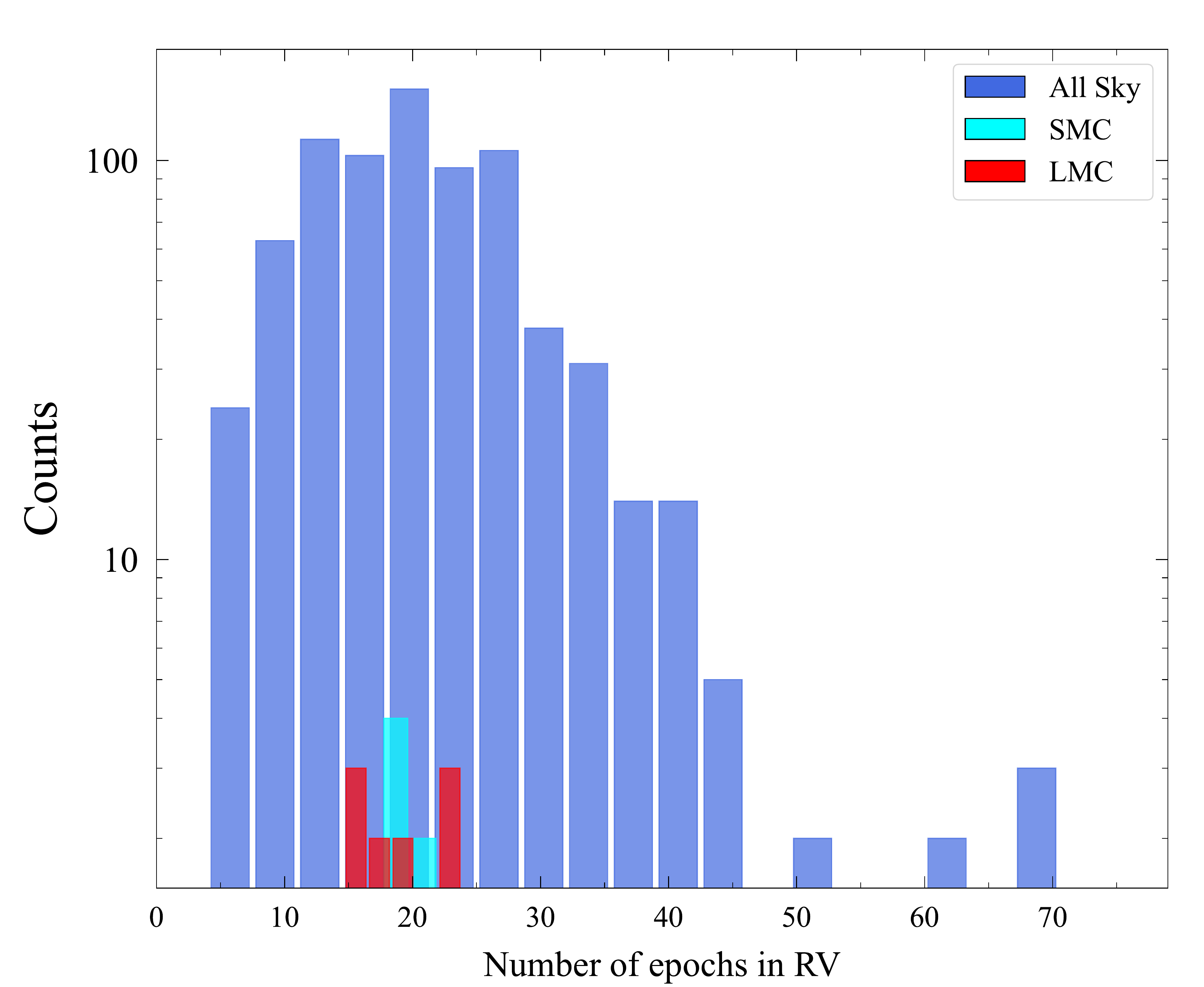}
      \caption{Number of epochs in the RV time series for the labelled sub-samples. 
              }
         \label{fig:nEpochsRV}
   \end{figure}

Concerning the RVs, the number of useful epochs for the Cepheids with RV time-series published in DR3 is displayed in Fig.~\ref{fig:nEpochsRV}. There are 15 and 9 DCEPs with RV time series in the LMC and SMC, respectively. The rest of the objects belong to the All-Sky sample.  

\subsection{Spatial distribution}

\begin{figure}
   \centering
   \includegraphics[width=\hsize]{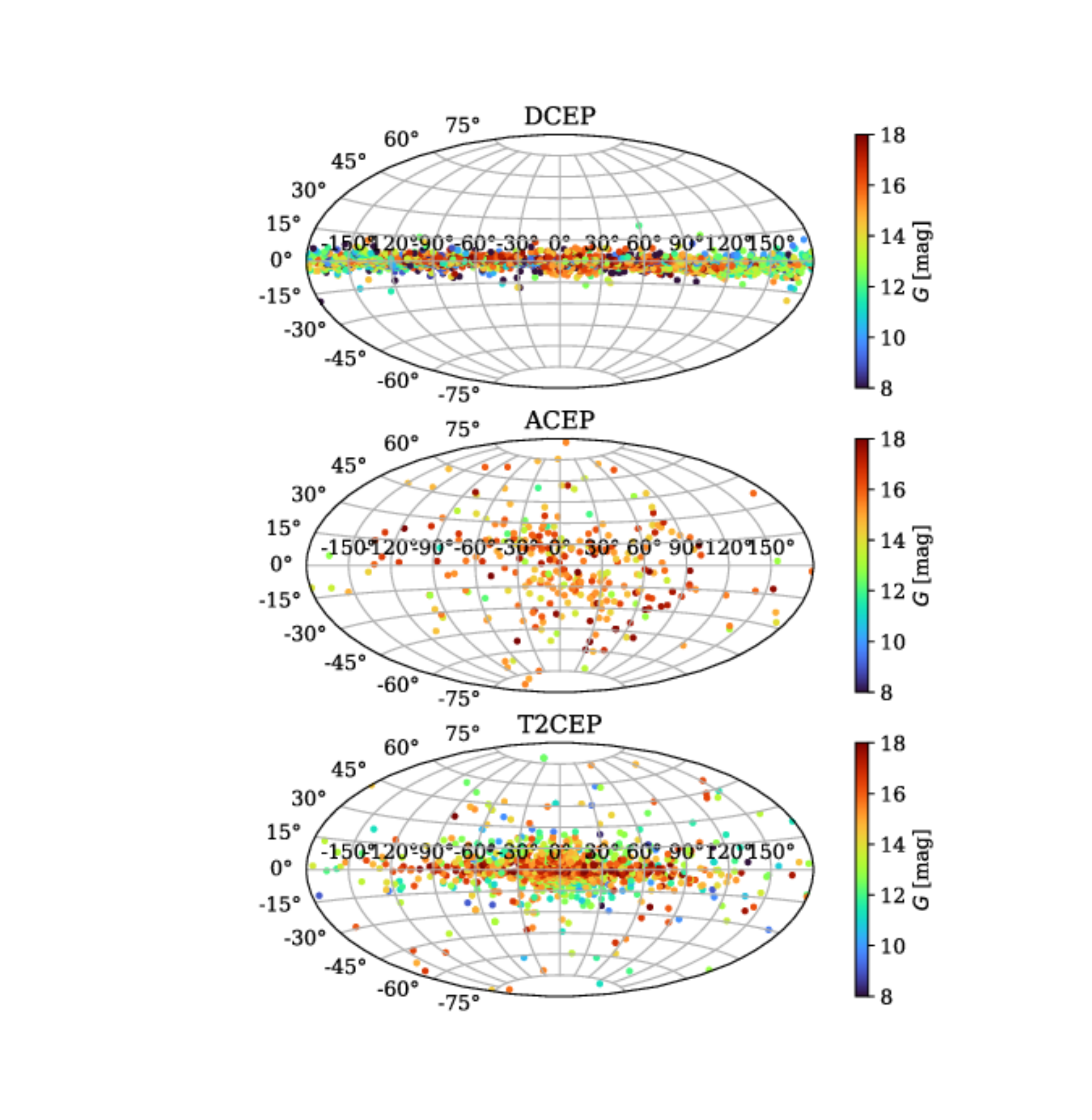}
      \caption{Map in galactic coordinates of the different Cepheid types in the MW. The objects are colour-coded according to their apparent $G$ magnitude.
              }
         \label{fig:aitoffMapMW}
   \end{figure}

\begin{figure}
   \centering
   \vbox{
   \includegraphics*[width=\hsize, trim=1cm 10.5cm 1cm 2cm]{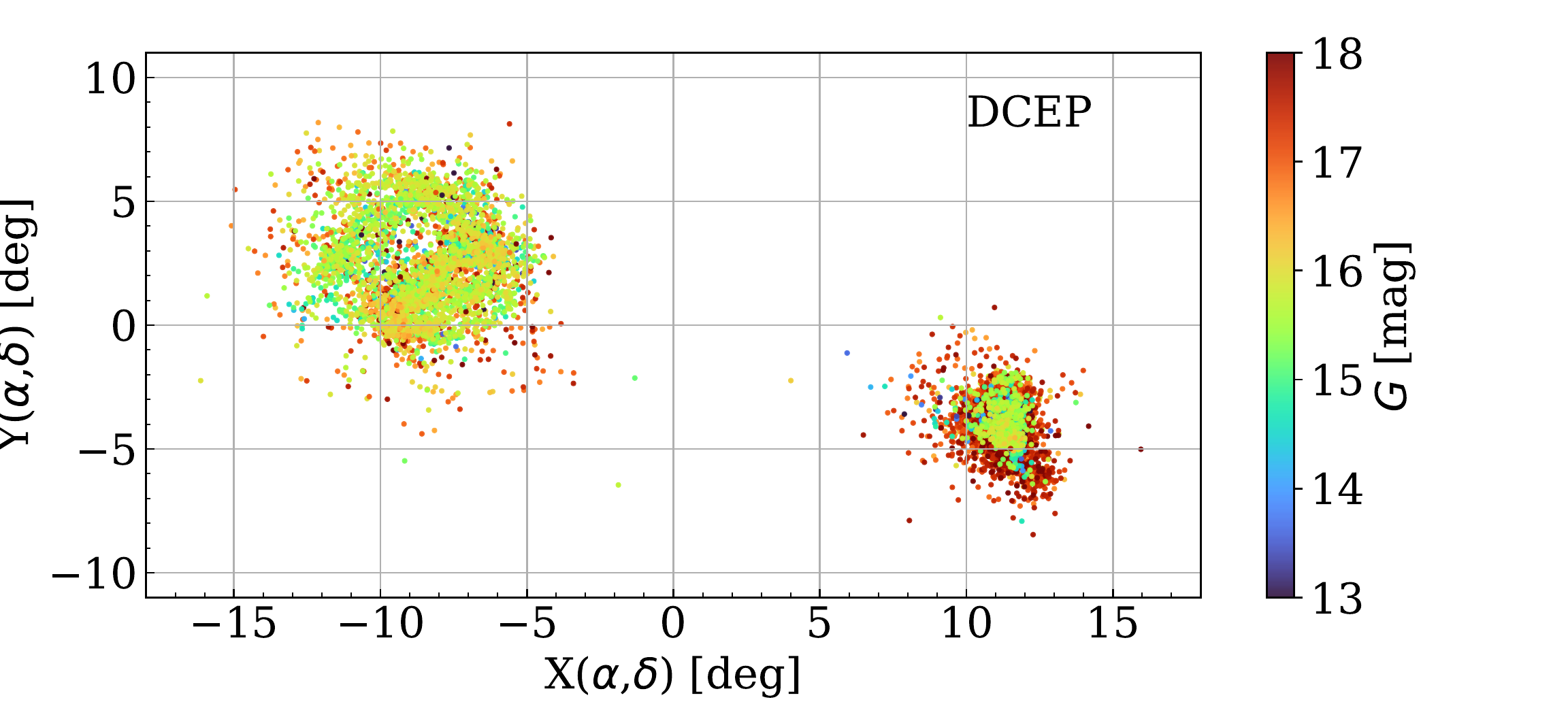}
      \includegraphics*[width=\hsize, trim=1cm 10.5cm 1cm 2cm]{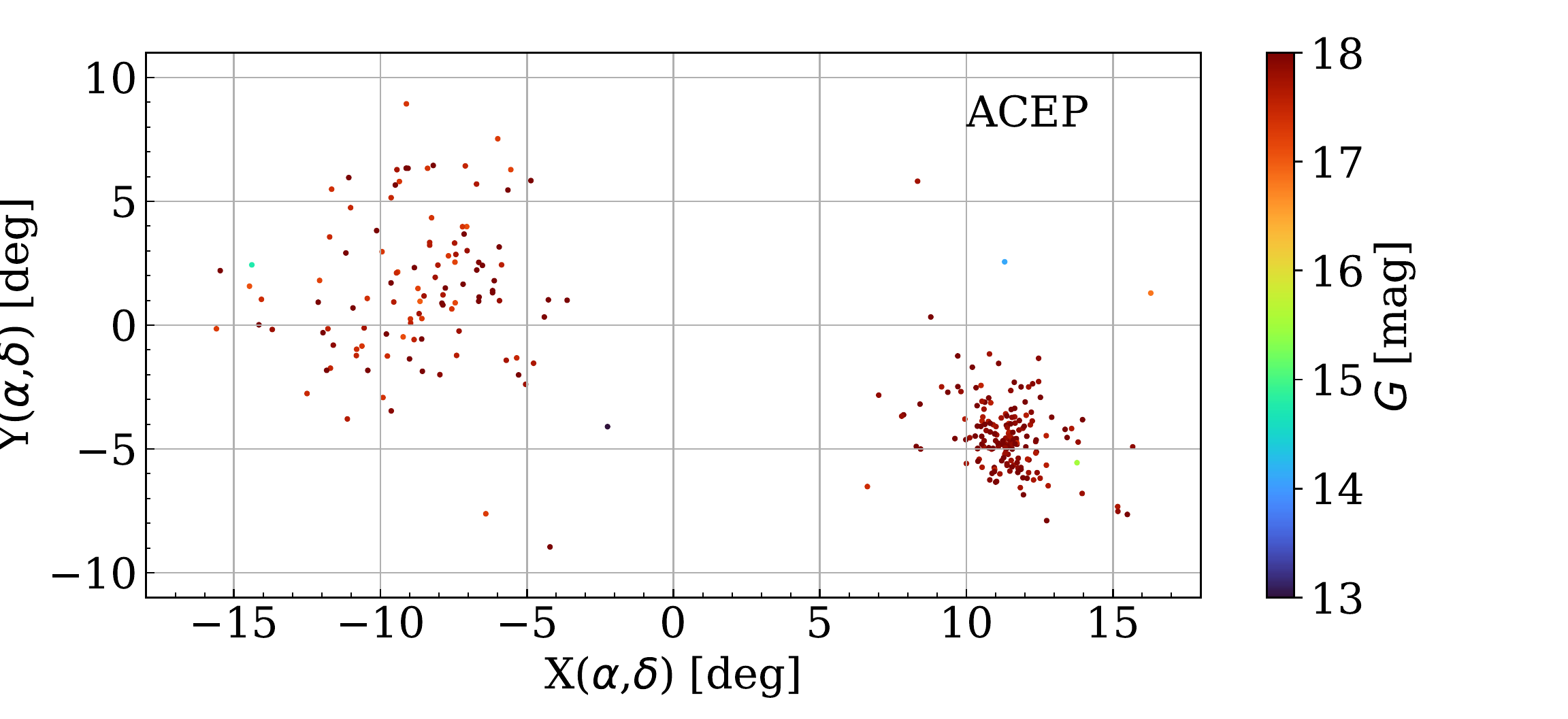}
         \includegraphics*[width=\hsize, trim=1cm 10.5cm 1cm 2cm]{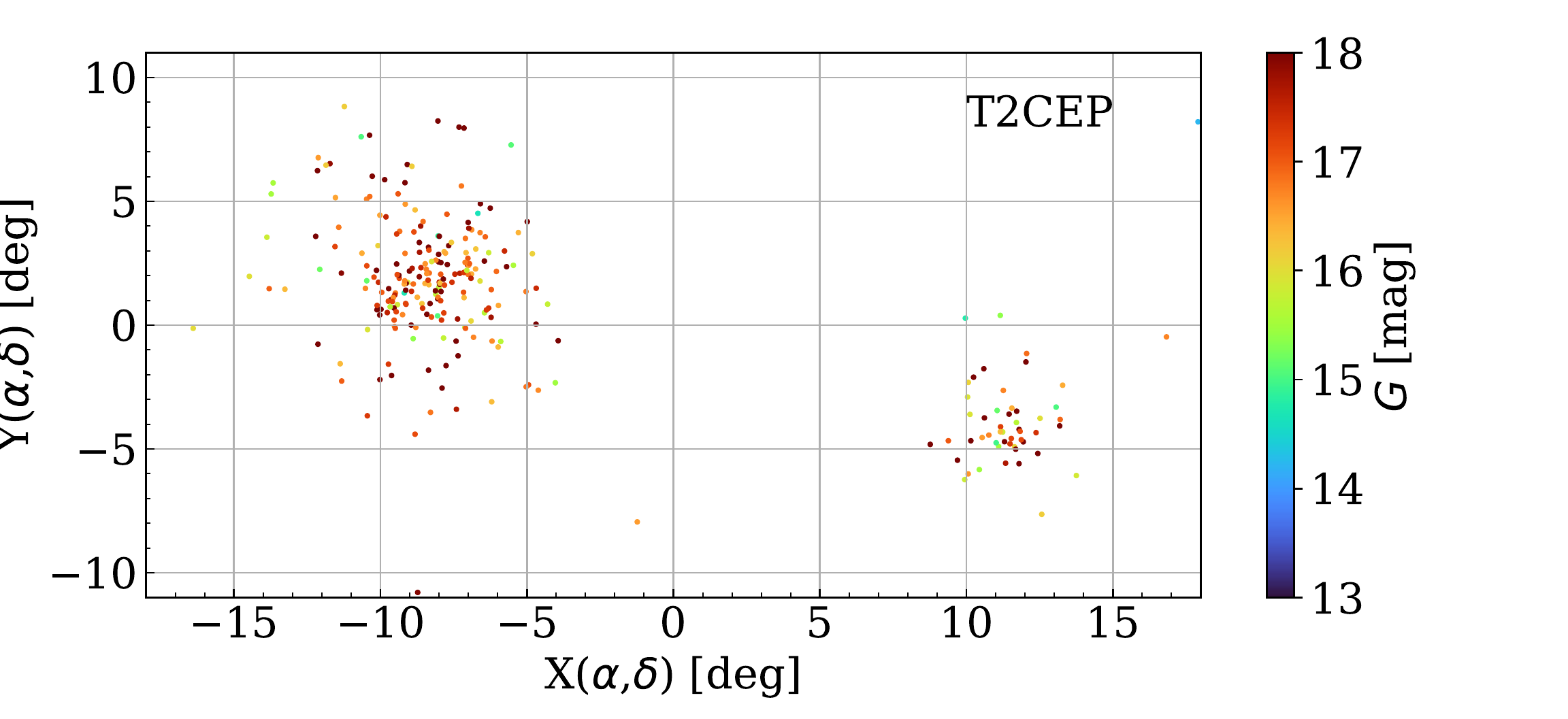}
   
      }
      \caption{Map of the different Cepheid types in the MCs. The objects are colour-coded according to their apparent $G$ magnitude. The map is a zenithal equidistant projection centred at equatorial coordinates $\alpha$,$\delta$=56.0\,$-$73.0 deg (J2000).
              }
         \label{fig:aitoffMapMCs}
   \end{figure}

The spatial distribution of Cepheids of different types in the All-Sky sample  is shown in Fig.~\ref{fig:aitoffMapMW}. The different distributions reflect the progenitor stellar populations of the different types: DCEPs are concentrated in the Galactic disc, as expected for a young population\footnote{In the figure we have removed from the All-Sky sample objects physically bound to the LMC and SMC (see Sect.~\ref{sect:associationWithStellarSystems}).}; ACEPs, which are intermediate-age objects are preferentially located in the Galactic halo; T2CEPs are present in almost all Galactic components, disc, thick disc, halo and bulge, where they are more concentrated.    
The spatial distributions of the LMC and SMC Cepheids are shown in Fig.~\ref{fig:aitoffMapMCs}. Also in these galaxies the DCEPs trace the young populations inhabiting the LMC bar and the spiral arms \citep[see e.g.][and references therein]{Ripepi2022} as well as the body and the wing of the SMC \citep[see e.g.][and references therein]{Ripepi2017}. The spatial distributions of ACEPs and T2CEPs are more sparse and connected with the spheroids describing the intermediate-old populations in both galaxies \citep[see e.g.][]{GaiaLuri2021}.

\begin{figure}
   \centering
   \includegraphics[width=\hsize]{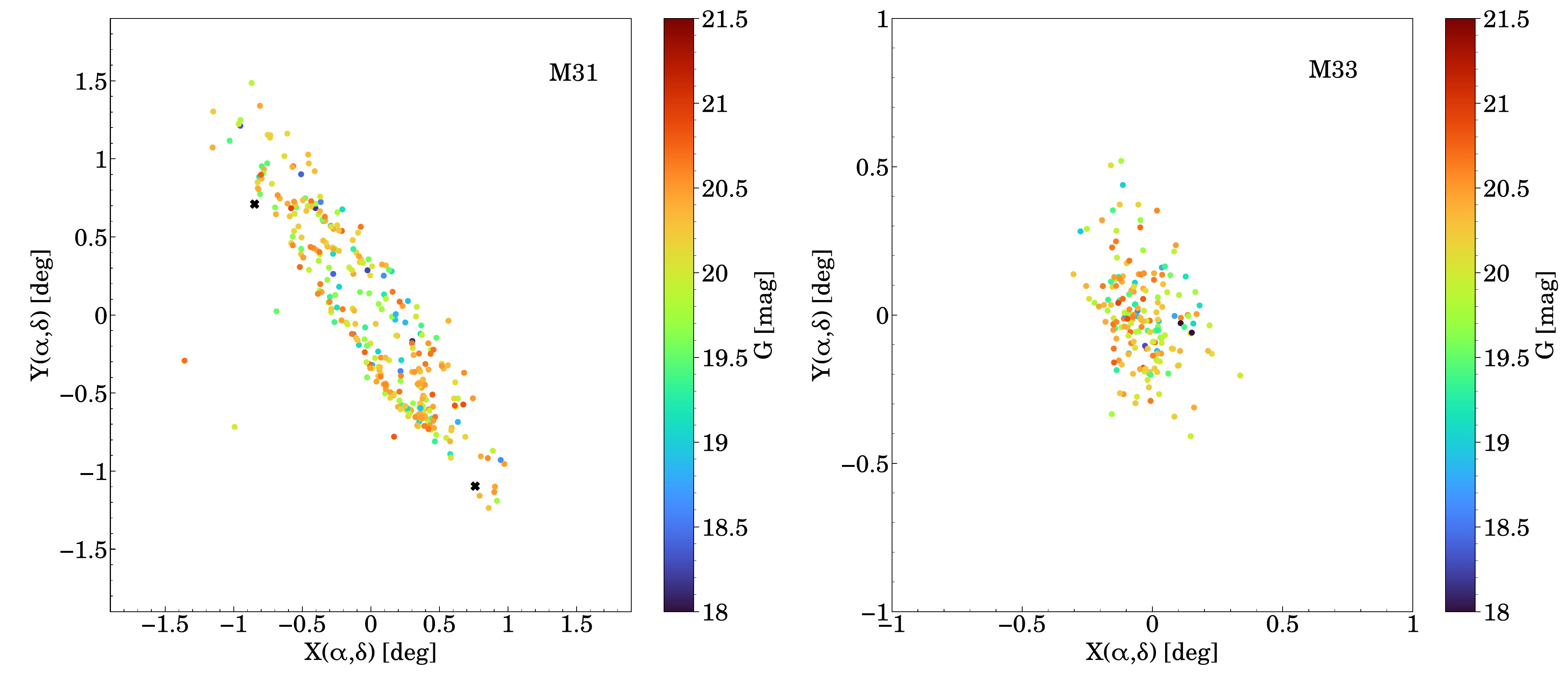}
      \caption{Map in equatorial coordinates of the DCEPs in M31 (left panel) and M33 (right panel). The symbols are colour-coded based on the apparent DCEPs' $G$ magnitude. The two black crosses identify two RVTAU stars in M31.   
              }
         \label{fig:aitoffMapM31M33}
   \end{figure}

Figure~\ref{fig:aitoffMapM31M33} shows the 
spatial distribution of the Cepheids in M31 and M33. In this case we have basically only DCEPs, except for two RVTAU detected in M31. The spatial distribution of the M31 DCEPs 
follows neatly 
the galaxy spiral arms, where young stars are expected, while the DCEP distribution in M33 is less ordered, due to the different morphology of the galaxy and the different viewing angle from the Sun.

\subsection{Fourier parameters}

\begin{figure*}
   \centering
   \vbox{
   \includegraphics[width=\hsize]{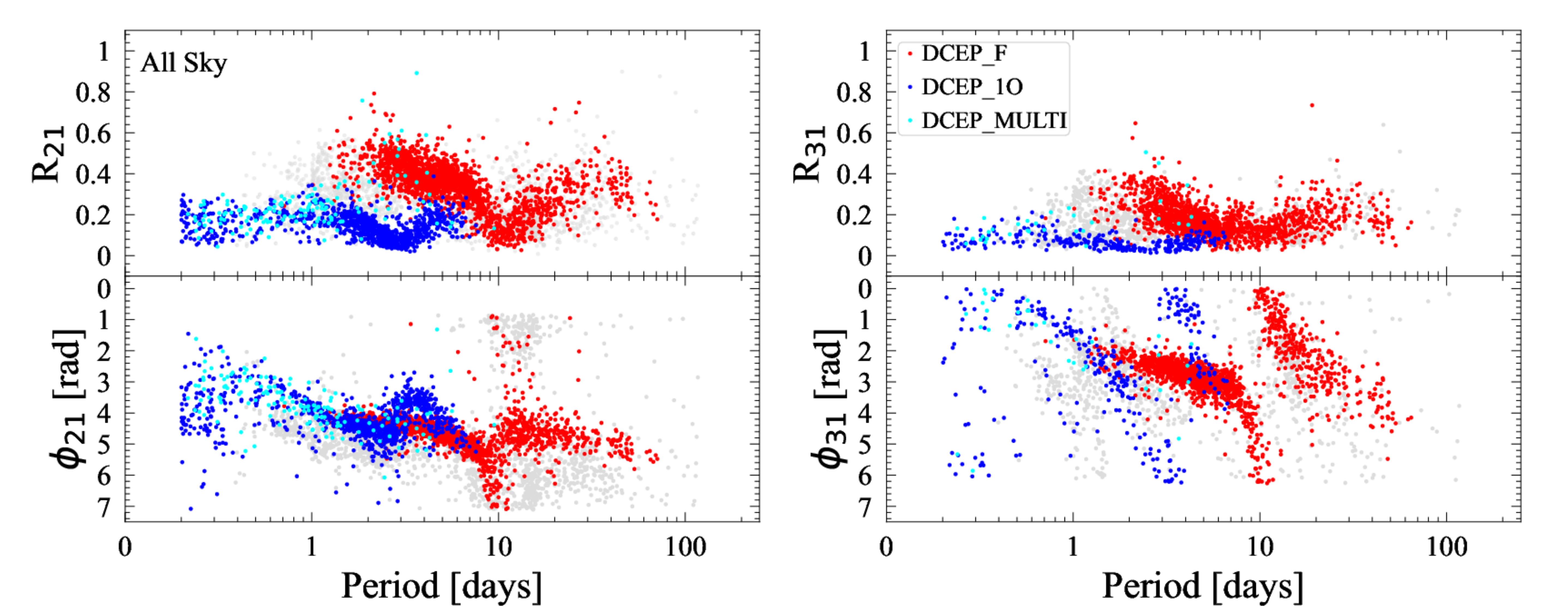}
   \includegraphics[width=\hsize]{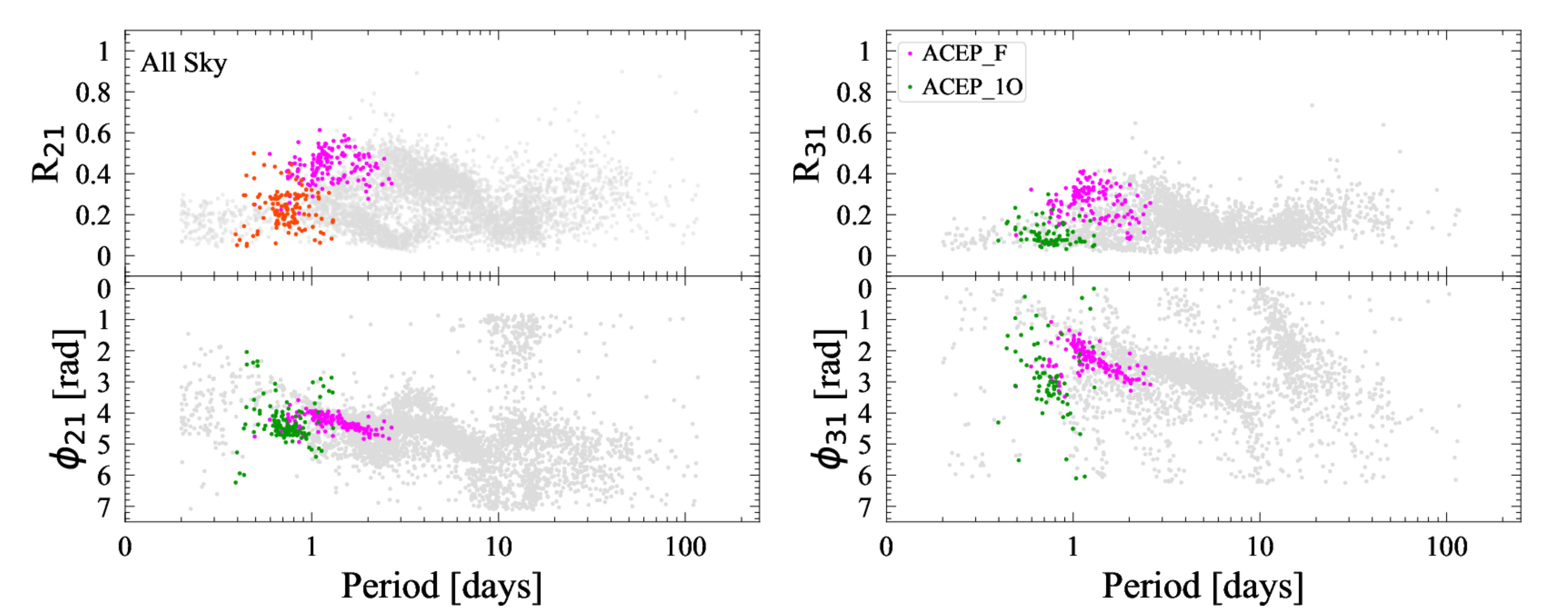}
   \includegraphics[width=\hsize]{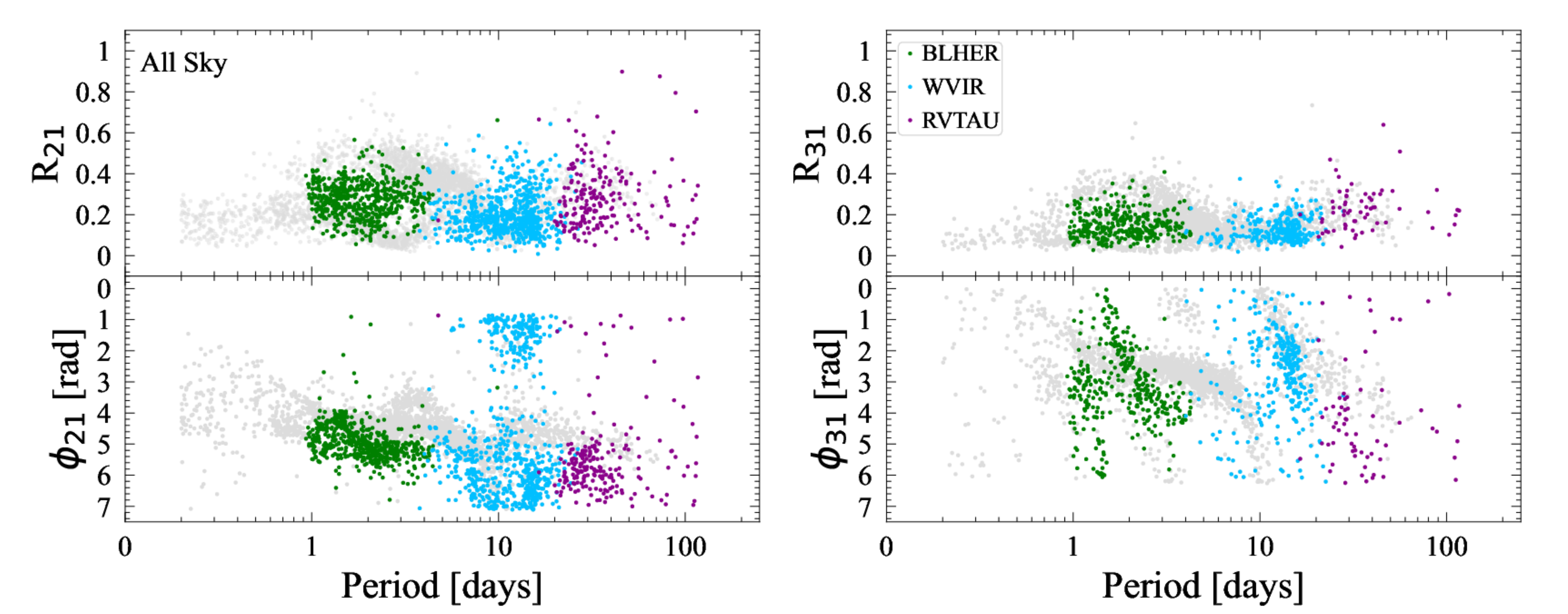}
   }
      \caption{Fourier parameters for the All-Sky sample. From top to bottom the different panels show the results for DCEPs, ACEPs and T2CEPs, respectively.
              }
         \label{fig:fourierMW}
   \end{figure*}

\begin{figure*}
   \centering
   \vbox{
   \includegraphics[width=\hsize]{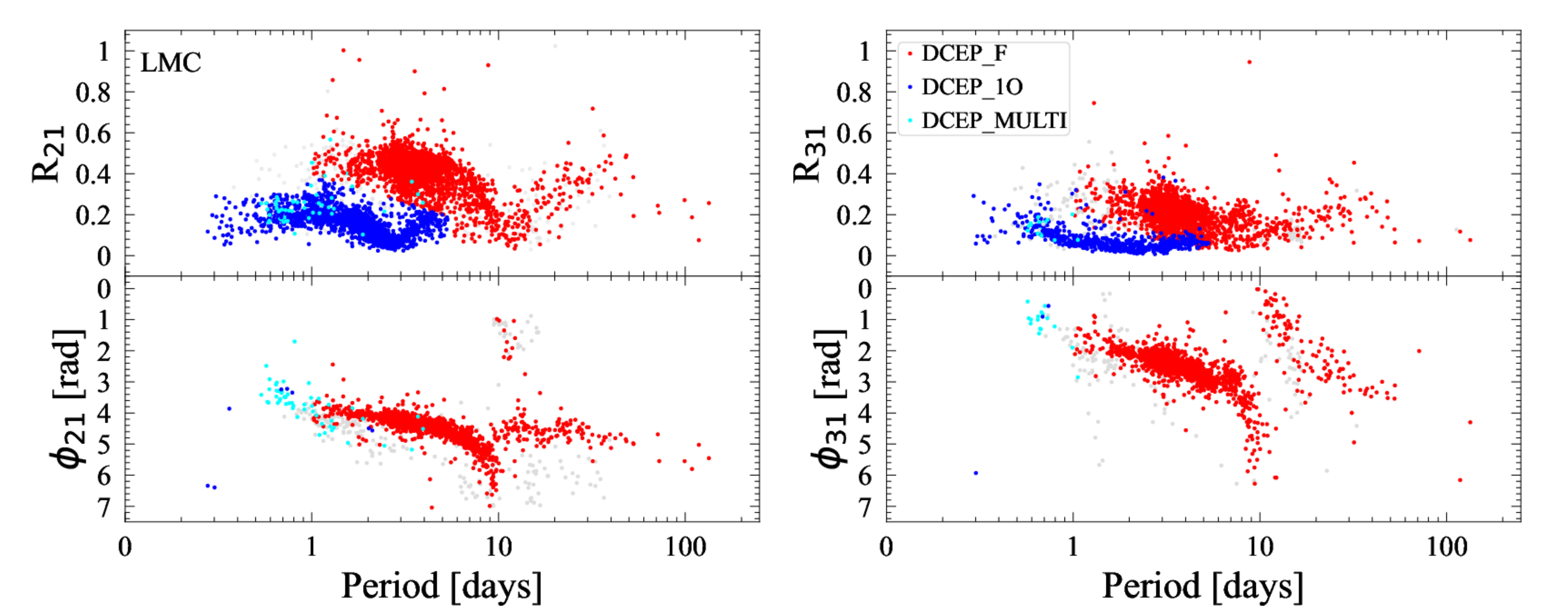}
   \includegraphics[width=\hsize]{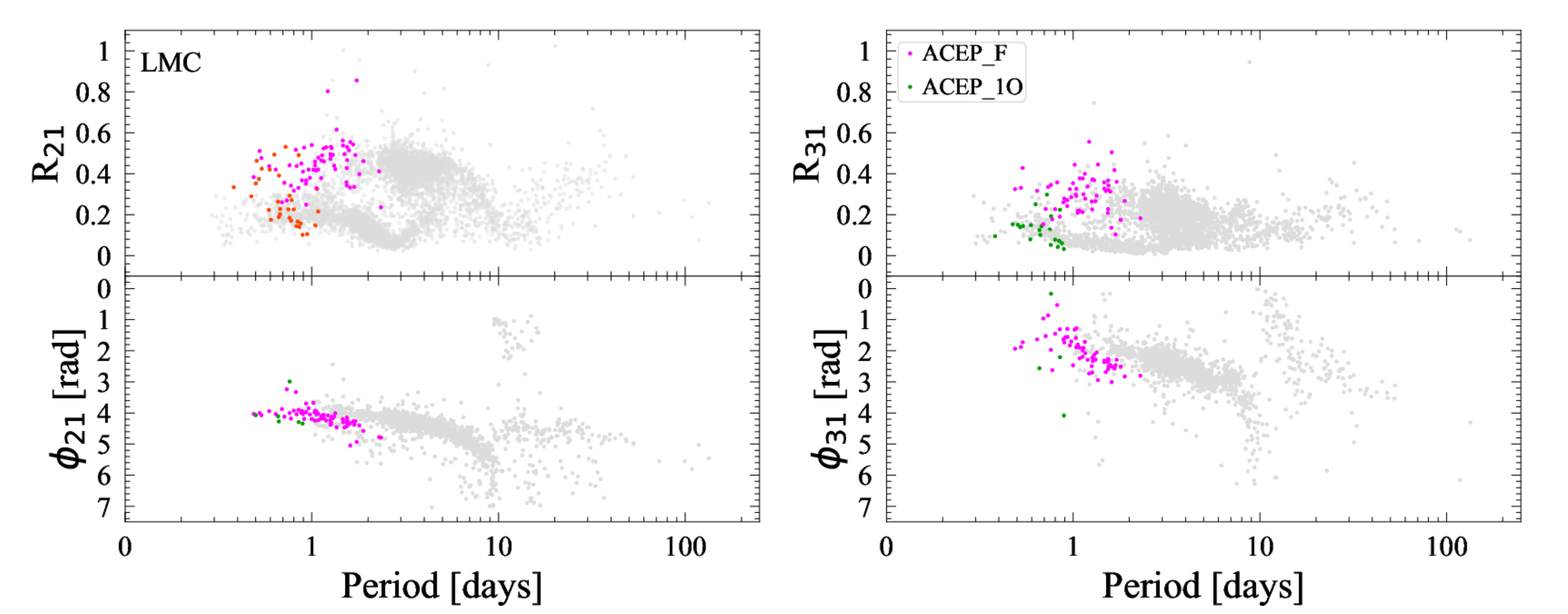}
   \includegraphics[width=\hsize]{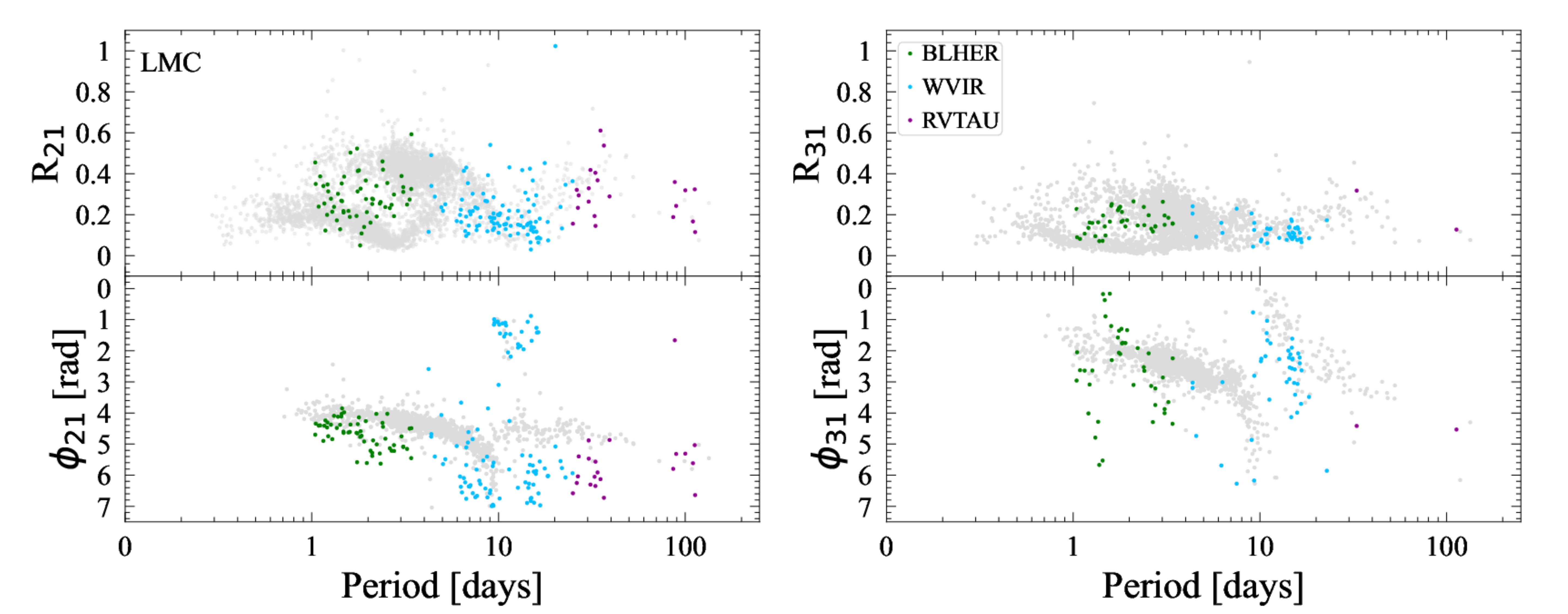}
}
      \caption{Same as in Fig.~\ref{fig:fourierMW} but for the LMC.}
         \label{fig:fourierLMC}
   \end{figure*}

\begin{figure*}
   \centering
   \vbox{
   \includegraphics[width=\hsize]{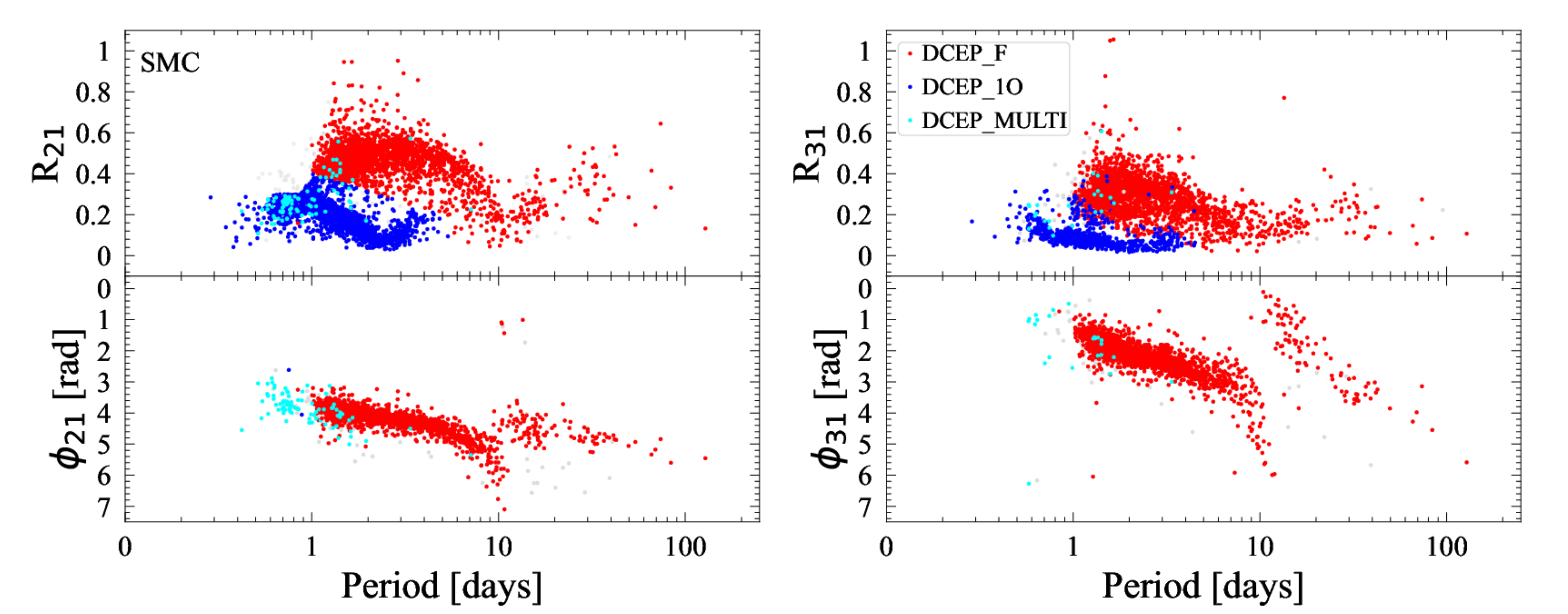}
   \includegraphics[width=\hsize]{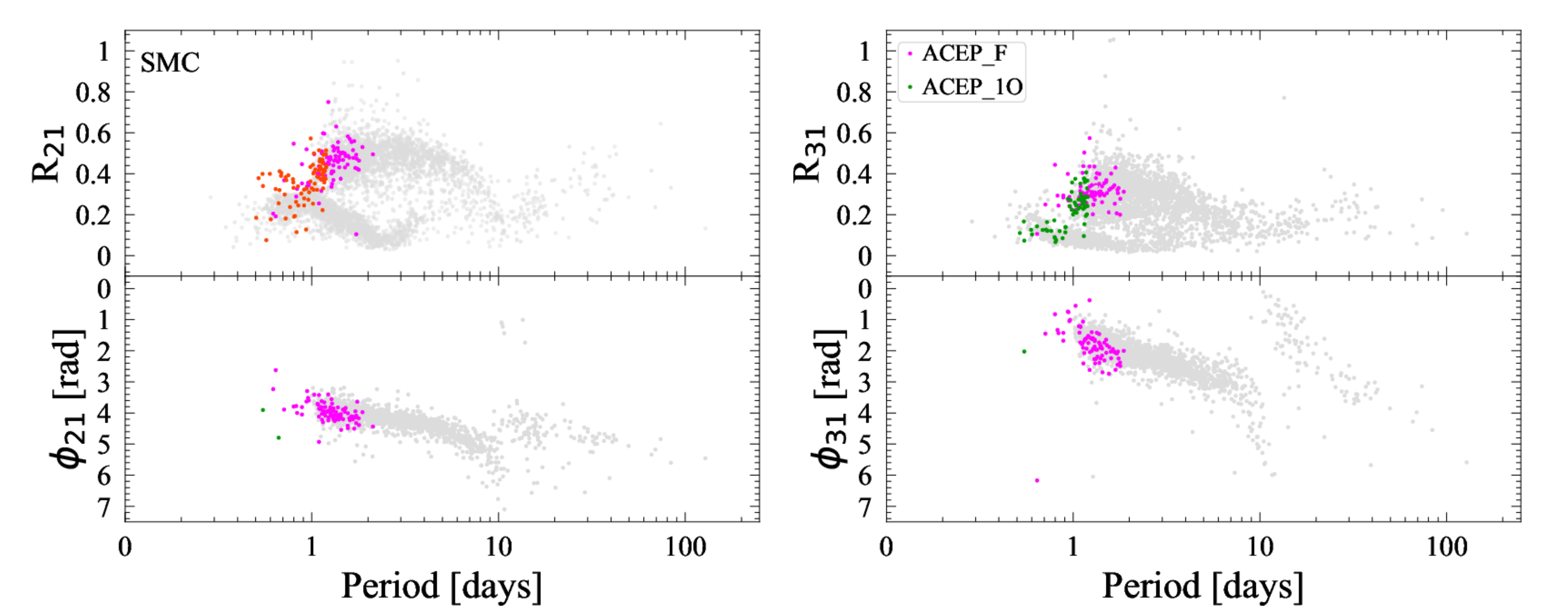}
   \includegraphics[width=\hsize]{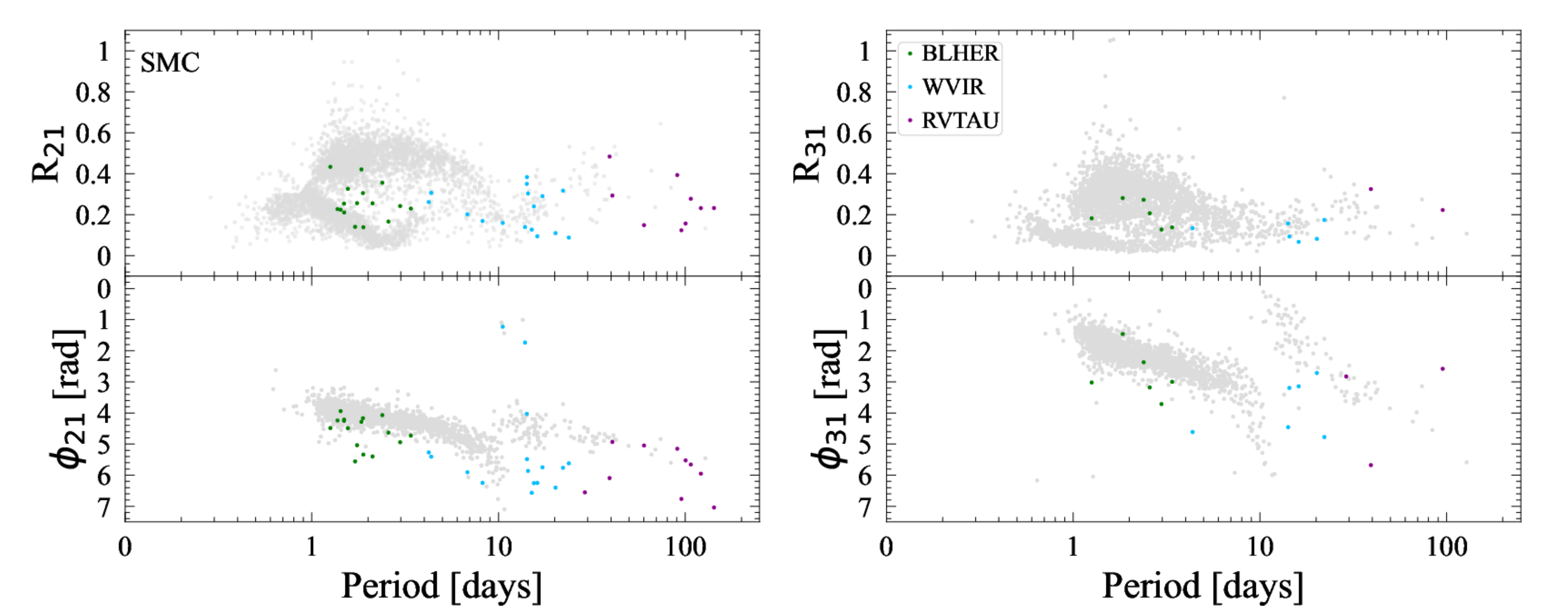}
      }
      \caption{Same as in Fig.~\ref{fig:fourierMW} but for the SMC.
              }
         \label{fig:fourierSMC}
   \end{figure*}

\begin{figure*}
   \centering
   \includegraphics[width=\hsize]{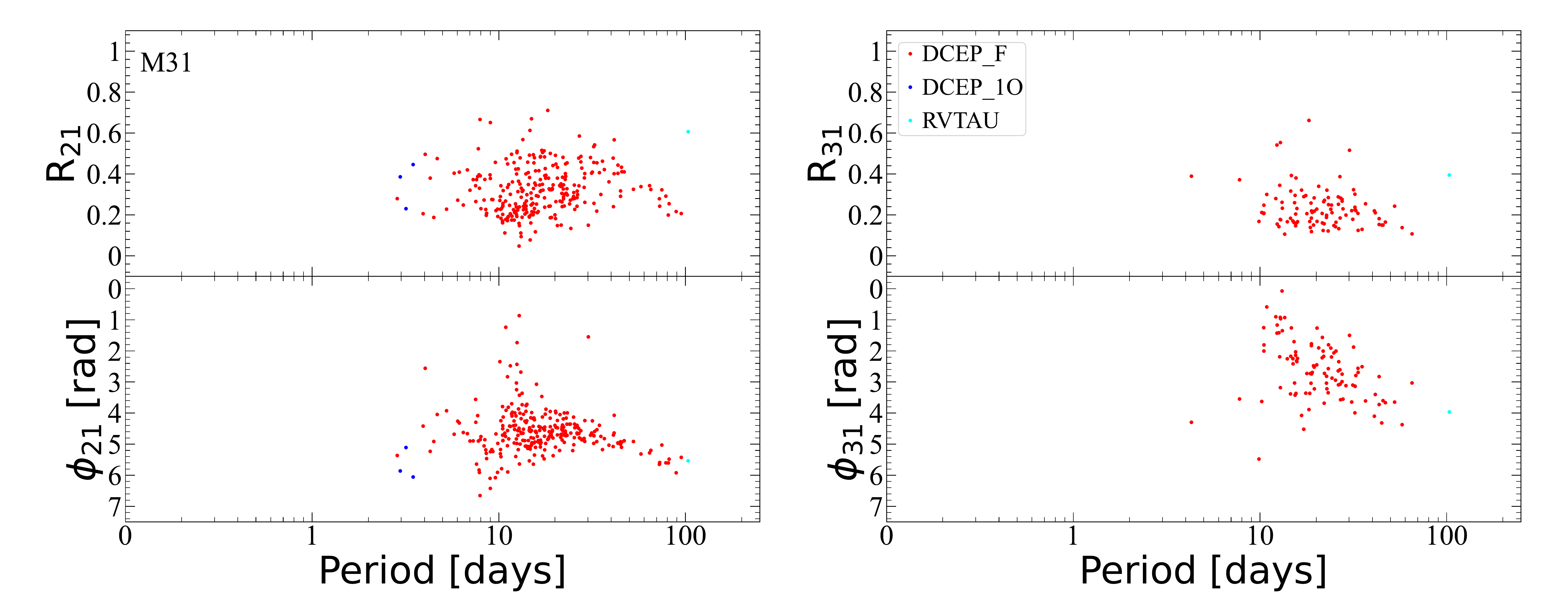}
      \caption{Fourier parameters for the M31 DCEPs.
              }
         \label{fig:fourierM31}
   \end{figure*}

\begin{figure*}
   \centering
   \includegraphics[width=\hsize]{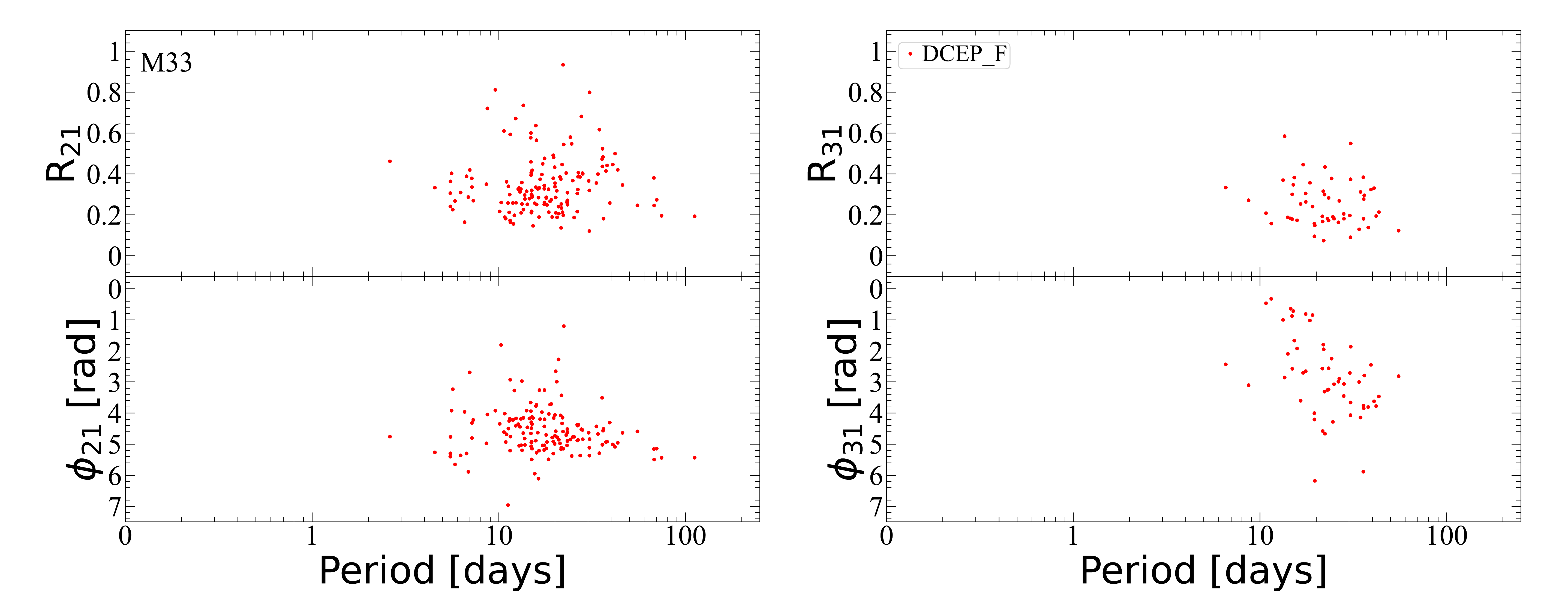}
      \caption{   Fourier parameters for the M33 DCEPs.
              }
         \label{fig:fourierM33}
   \end{figure*}

One of the outputs of the \sos\ pipeline are the Fourier parameters $R_{21}$, $R_{31}$, $\phi_{21}$ and $\phi_{31}$ which represent an important tool to distinguish the different types of variables. The Fourier parameters for the  Cepheids in the All-Sky sample  are shown in Fig.~\ref{fig:fourierMW},  separated in different panels for DCEPs, ACEPs and T2CEPs, for clarity reasons. The different distributions occupy the expected location for each variable type, confirming the goodness of our classification. The same kind of considerations attains for the LMC and SMC as shown in Figs.~\ref{fig:fourierLMC} and ~\ref{fig:fourierSMC}.
In the cases of M31 and M33 (Fig.~\ref{fig:fourierM31} and Fig.~\ref{fig:fourierM33}) the Fourier parameters show a less clear morphology, 
because light curves are in most part noisy, as we are analysing objects with magnitudes at the limits of \gaia\ capabilities. 
Nevertheless, it is remarkable that 
especially 
for M31, the morphology of the $P$-$R_{21}$ and $P$-$\phi_{21}$ relations is  similar to that displayed by the much closer All Sky, LMC and SMC samples.



\subsection{$PL$ and $PW$ diagrams}

\begin{figure*}
   \centering
   \includegraphics[width=\hsize]{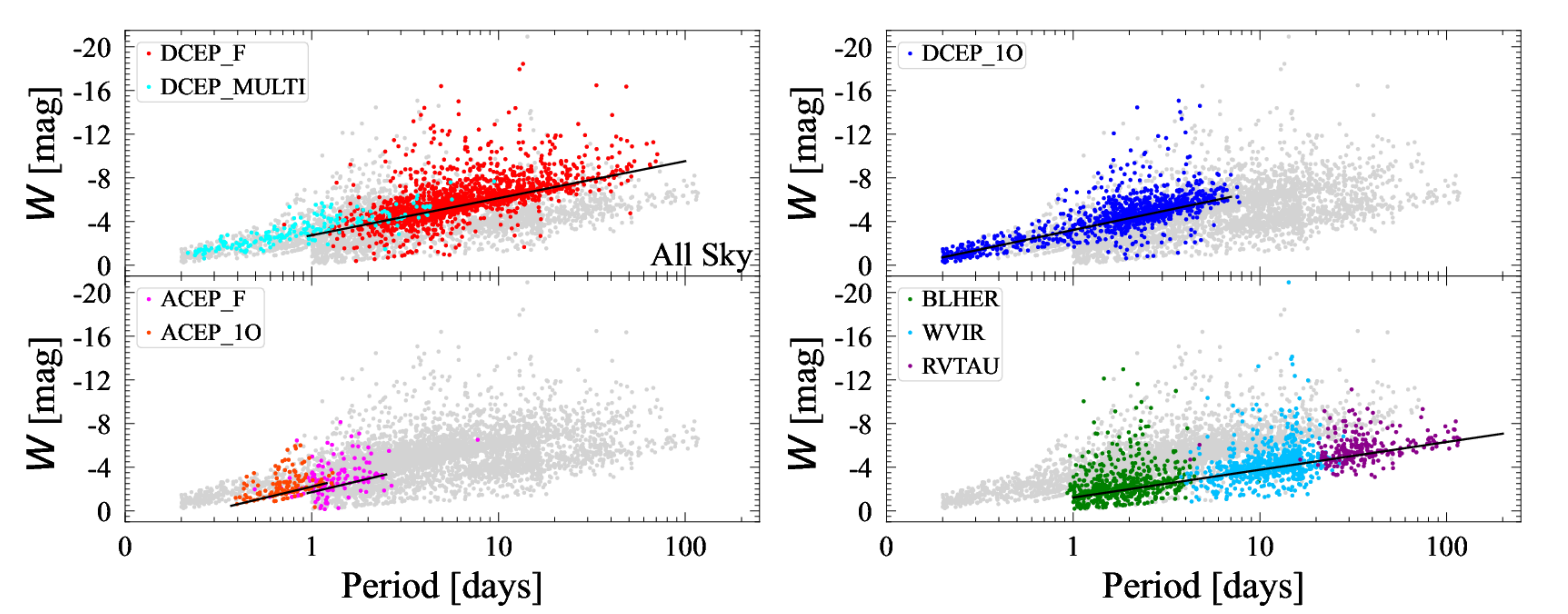}
      \caption{$PW$ relation for the All-Sky sample. Different types/modes of Cepheids displayed in the figures are labelled in each panel.   
              }
         \label{fig:PWmw}
   \end{figure*}

\begin{figure*}
   \centering
   \includegraphics[width=\hsize]{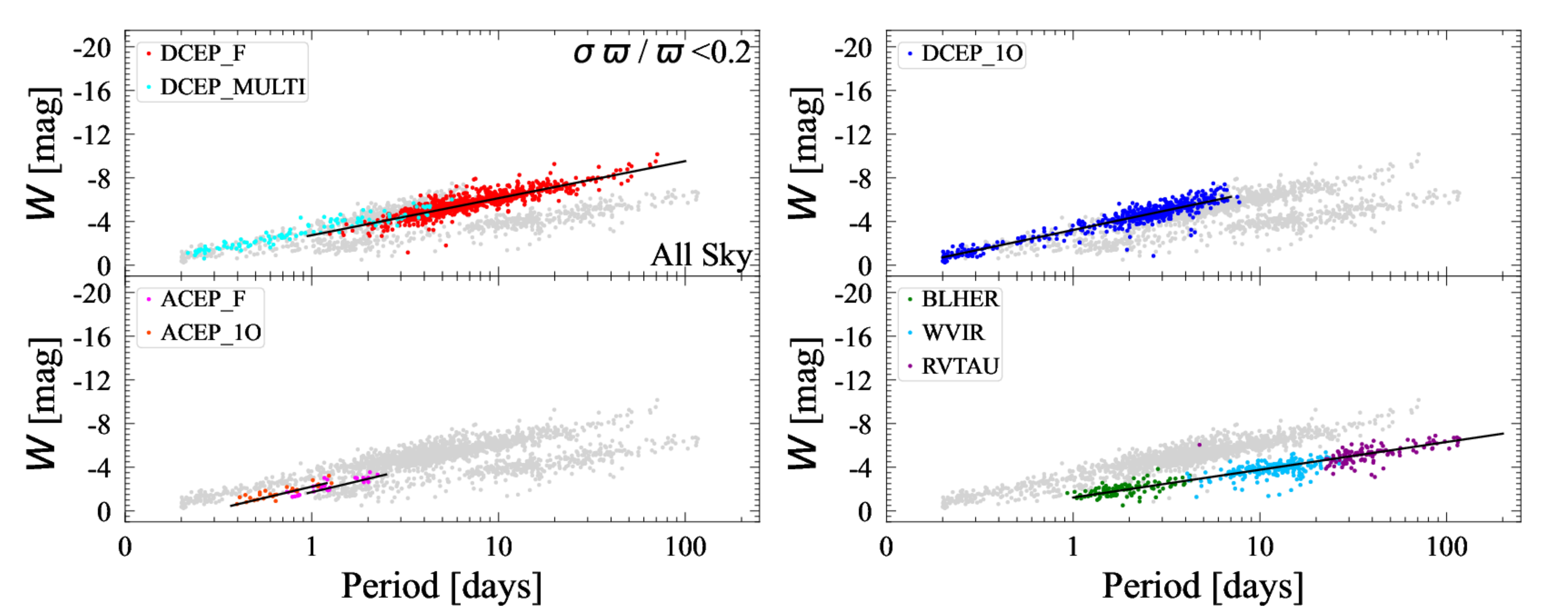}
      \caption{Same as in Fig.~\ref{fig:PWmw} but restricting the sample to objects with $\sigma \varpi / \varpi < 0.2$ 
              }
         \label{fig:PW_figa}
   \end{figure*}

Figure~\ref{fig:PWmw} shows the $PW$ relations for the All-Sky sample, separately for different Cepheid types/modes. These 
relationships were adopted by the \sos\ pipeline to select and classify the different types of Cepheids, as discussed in Sect.~\ref{sect:newPL}\footnote{We recall that these $PW$ relations are used with the ABL formulation in the \sos\ pipeline (see Sect.~\ref{sect:newPL}).}. There is a large scatter in Figure~\ref{fig:PWmw} 
as we plotted also objects with 
very large parallax errors 
(obviously pulsators with negative parallaxes cannot be shown in the figure). Much better defined $PW$ relationships can be obtained by plotting only objects with relative error in parallax better than 20\%, as shown in Fig.~\ref{fig:PW_figa}.

\begin{figure*}
   \centering
   \vbox{
   \includegraphics[width=\hsize]{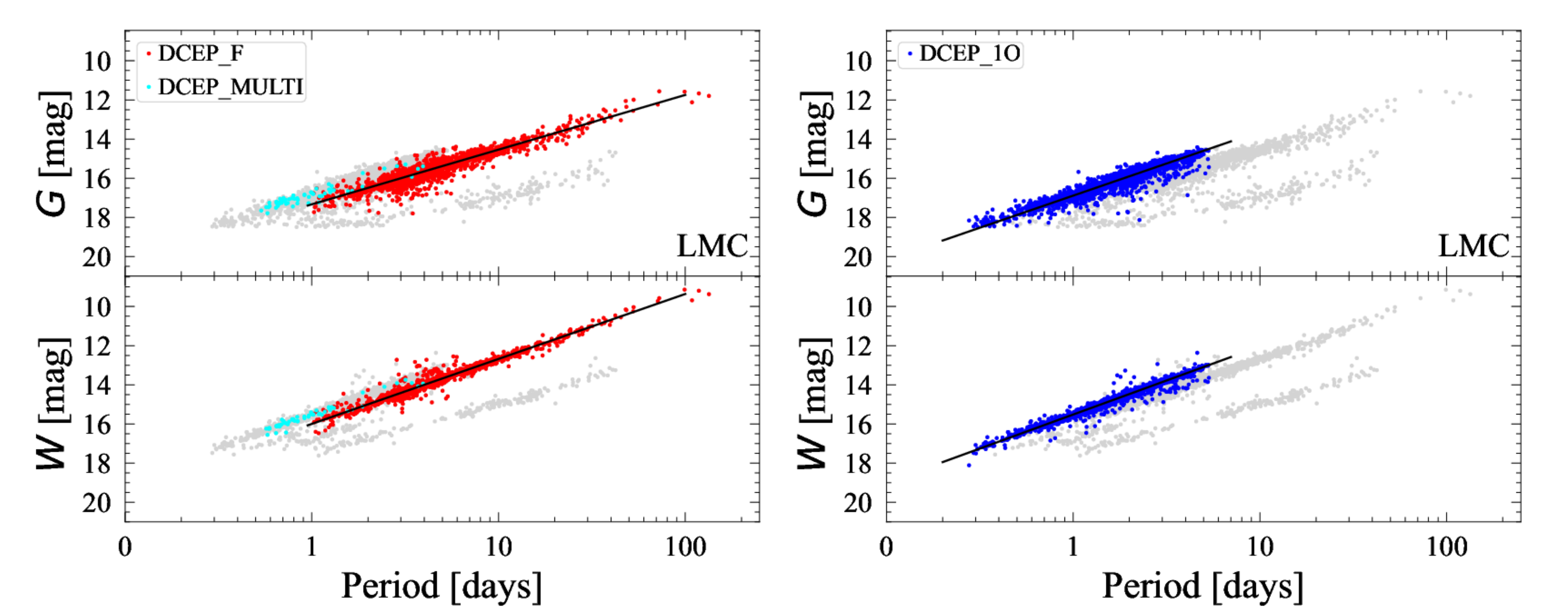}
   \includegraphics[width=\hsize]{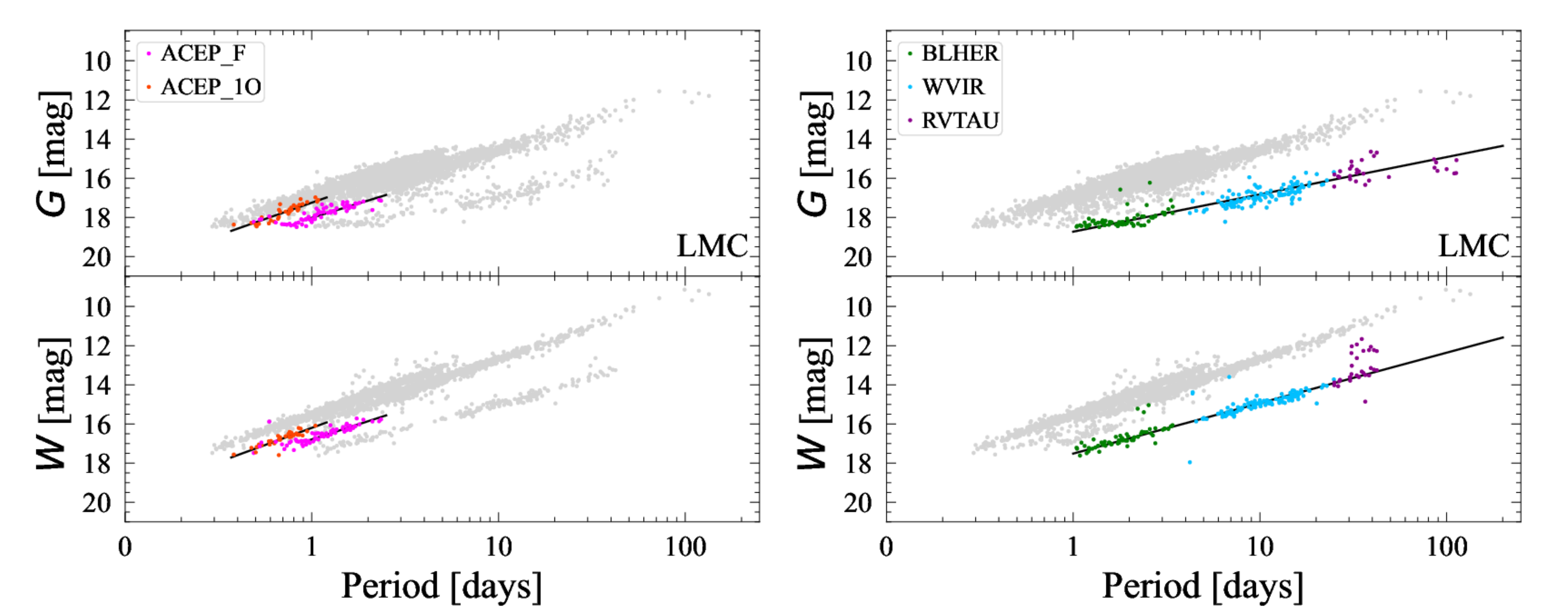}
      }
      \caption{$PL$ in the $G$-band and $PW$ relations for the LMC Cepheids. The top panels show results for the DCEPs, while the bottom panels display ACEPs and T2CEPs.  
              }
         \label{fig:PLlmc}
   \end{figure*}

\begin{figure*}
   \centering
   \includegraphics[width=\hsize]{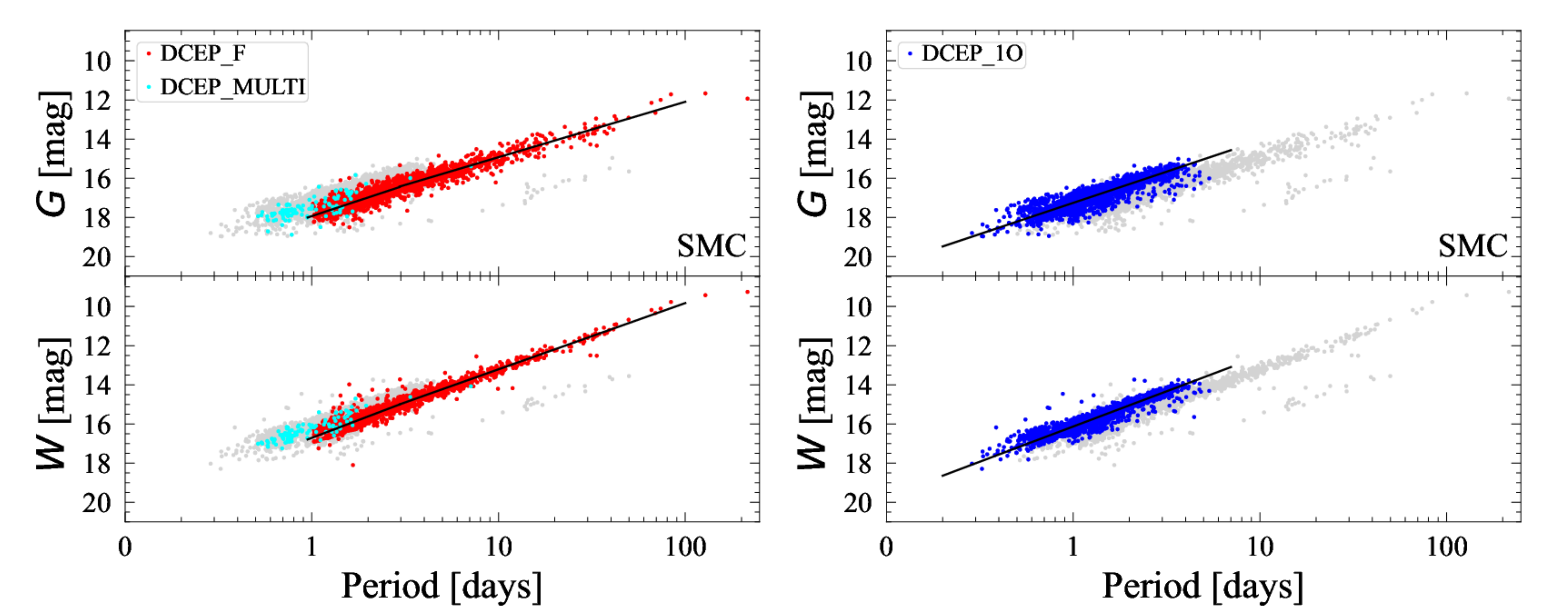}
   \includegraphics[width=\hsize]{figures/r_PL_LMC_ACEP_T2CEP.pdf}
      \caption{Same as in Fig.~\ref{fig:PLlmc} but for the SMC.  
              }
         \label{fig:PLsmc}
   \end{figure*}

Contrarily to the All-Sky sample, for the LMC and SMC, we can also use the $PL$ relations in the $G$ band, in addition to the $PW$ relations, as the reddening in these galaxies is in general rather low and approximately constant over each galaxy. The $PL$ diagrams are shown in Fig.~\ref{fig:PLlmc} and Fig.~\ref{fig:PLsmc} for the LMC and SMC, respectively. Both the $PL$ and the $PW$ diagrams are well defined, especially in the LMC, while the large depth 
along the line of sight increases significantly the dispersion in the SMC \citep[see][and references therein]{Ripepi2017}. 

For M31 and M33 the $PL$ relations are more accurate than the $PW$ relations
because 
the magnitudes in the \gbp\ and \grp\ bands, if any,  are less accurate that that in the $G$ band, thus causing 
a much larger dispersion in the $PW$ relations.
The $PL$ relations for both M31 and M33 show a remarkable linearity up to about $G\sim21$ mag.     

   \begin{figure*}
   \centering
   \includegraphics[width=\hsize]{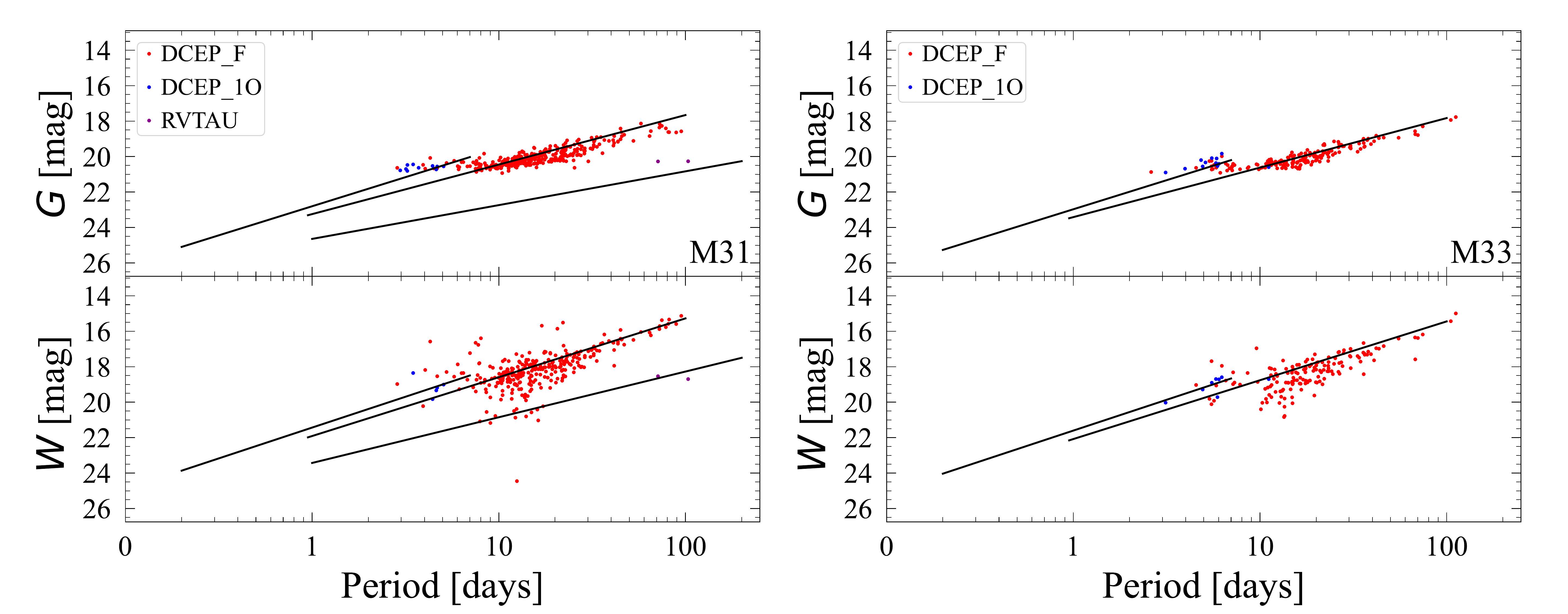}
      \caption{$PL$ in the $G$-band and $PW$ relations for the Cepheids in M31 (left panel) and M33 (right panel).  
              }
         \label{fig:PLm31m33}
   \end{figure*}    

   \begin{figure*}
   \centering
   \includegraphics[width=\hsize]{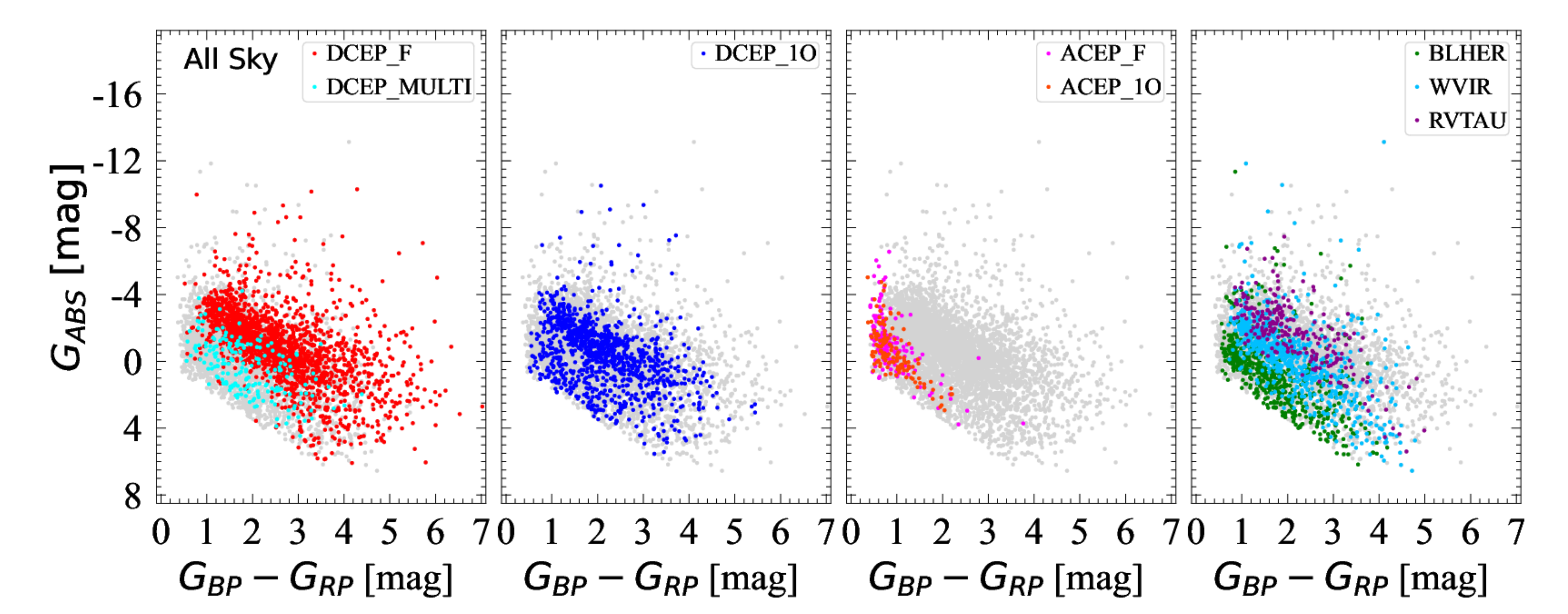}
      \caption{CMD of the All-Sky Cepheid sample 
              }
         \label{fig:cmdMW}
   \end{figure*}

   \begin{figure*}
   \centering
   \includegraphics[width=\hsize]{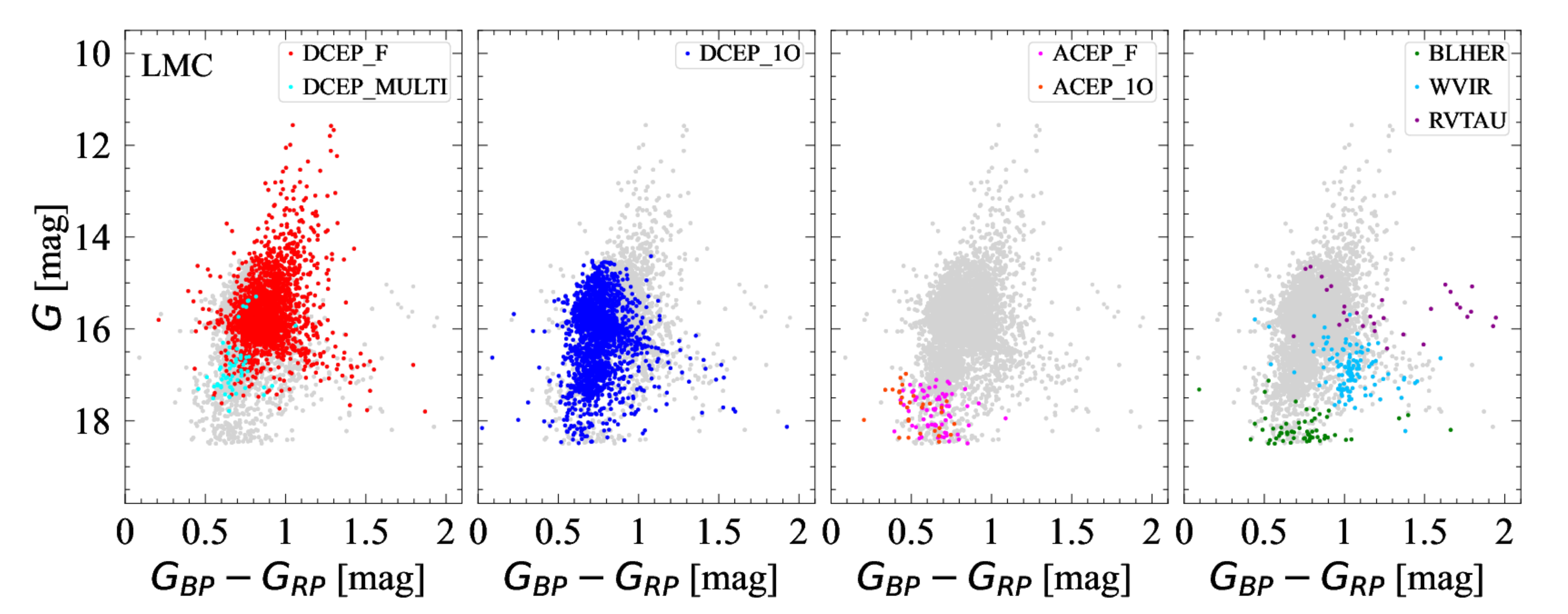}
      \caption{CMD in apparent $G$ magnitude of the LMC Cepheid sample.               }        \label{fig:cmdLMC}
   \end{figure*}   
   \begin{figure*}
   \centering
   \includegraphics[width=\hsize]{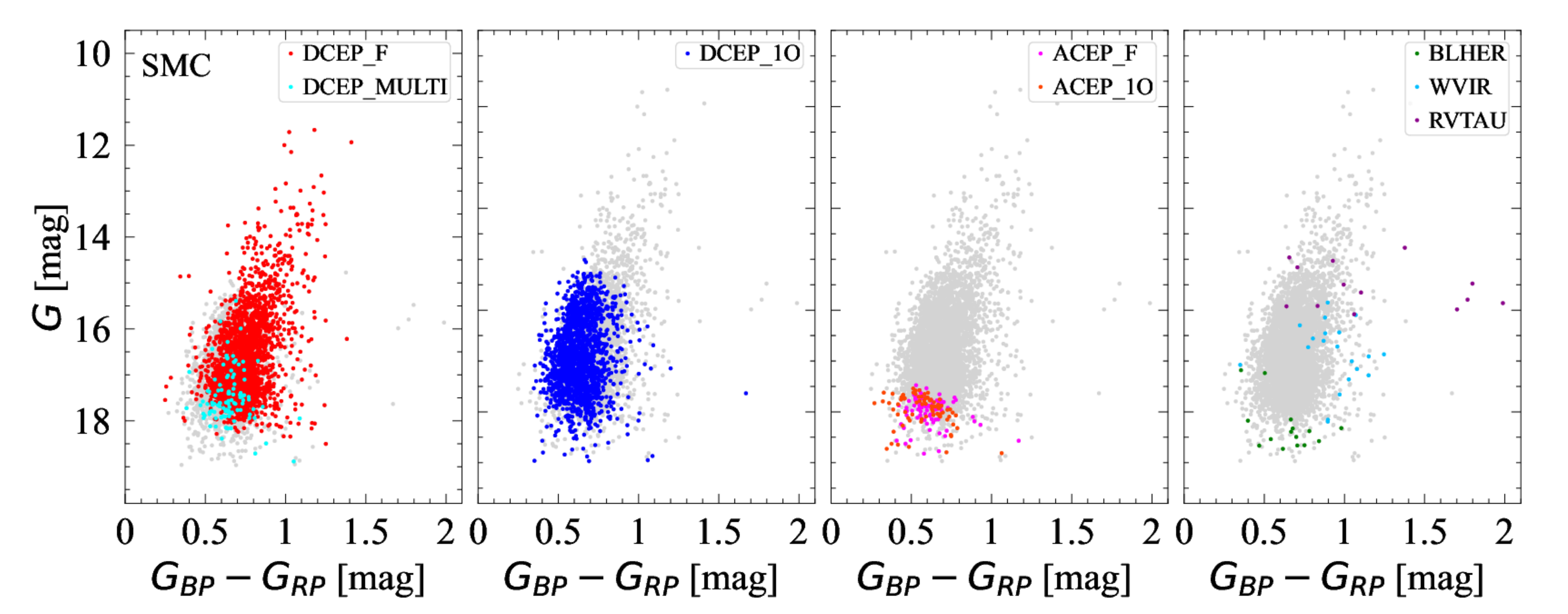}
      \caption{CMD in apparent $G$ magnitude  of the SMC Cepheid sample.              }
         \label{fig:cmdSMC}
   \end{figure*}  
   \begin{figure}
   \centering
   \includegraphics[width=\hsize]{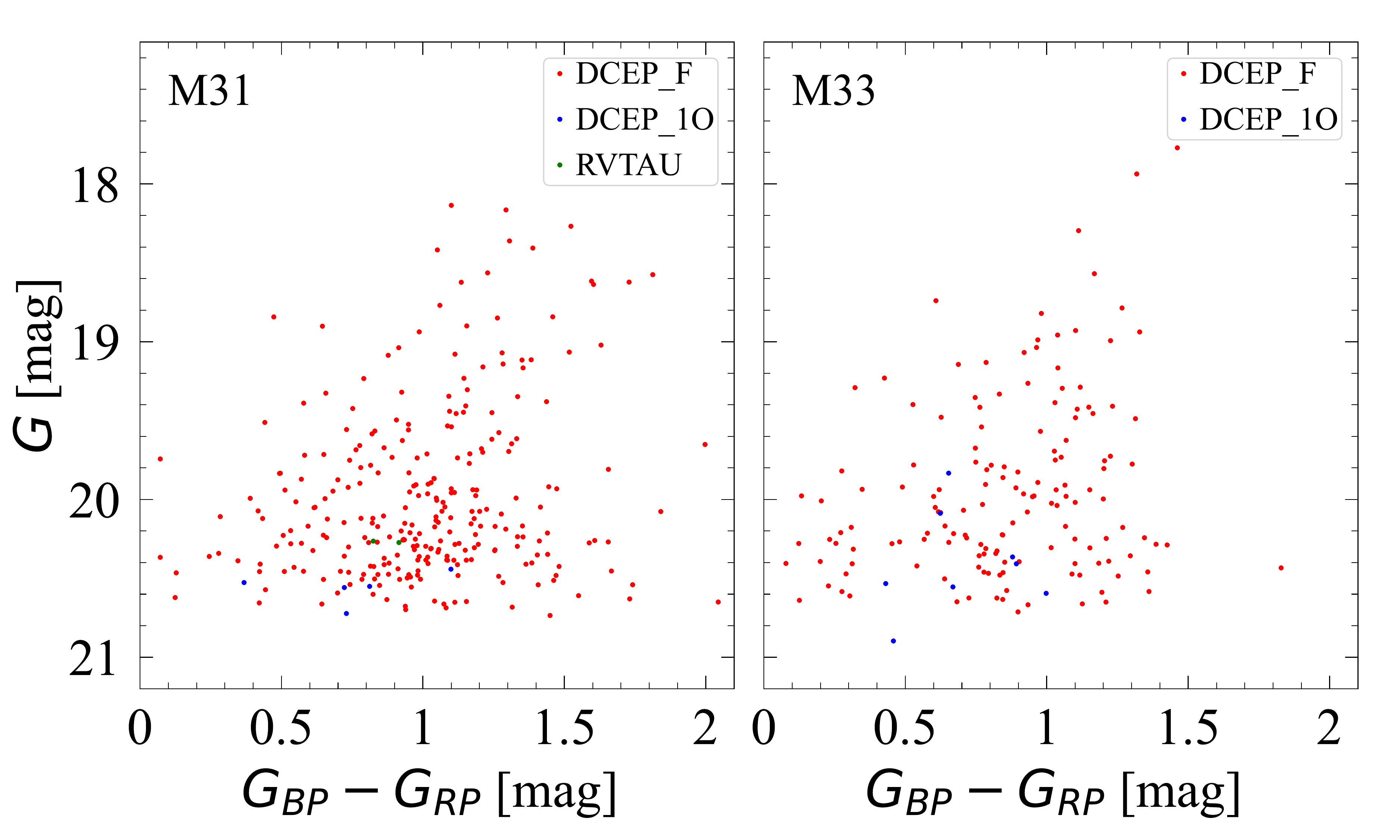}
      \caption{CMD in apparent $G$ magnitude  of the M31 (left pane)l and M33 (right panel) Cepheid samples.  
              }
         \label{fig:cmdM31M33}
   \end{figure}   
   \begin{figure*}
   \centering
   \includegraphics[width=\hsize]{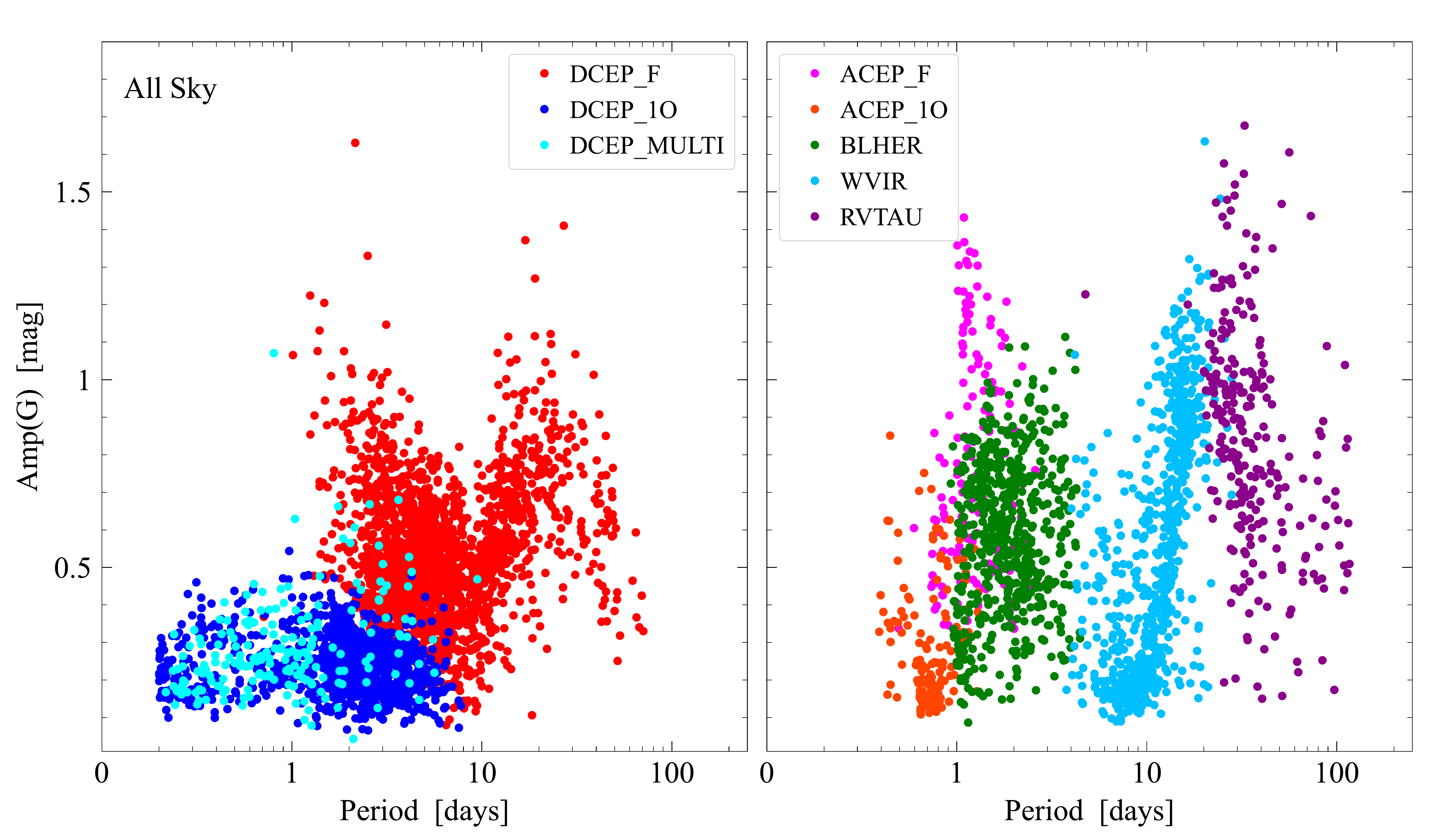}
      \caption{Period-Amplitude(G) diagram for the All-Sky sample.  
              }
         \label{fig:AmpPeriodMW}
   \end{figure*}  
   \begin{figure*}
   \centering
   \includegraphics[width=\hsize]{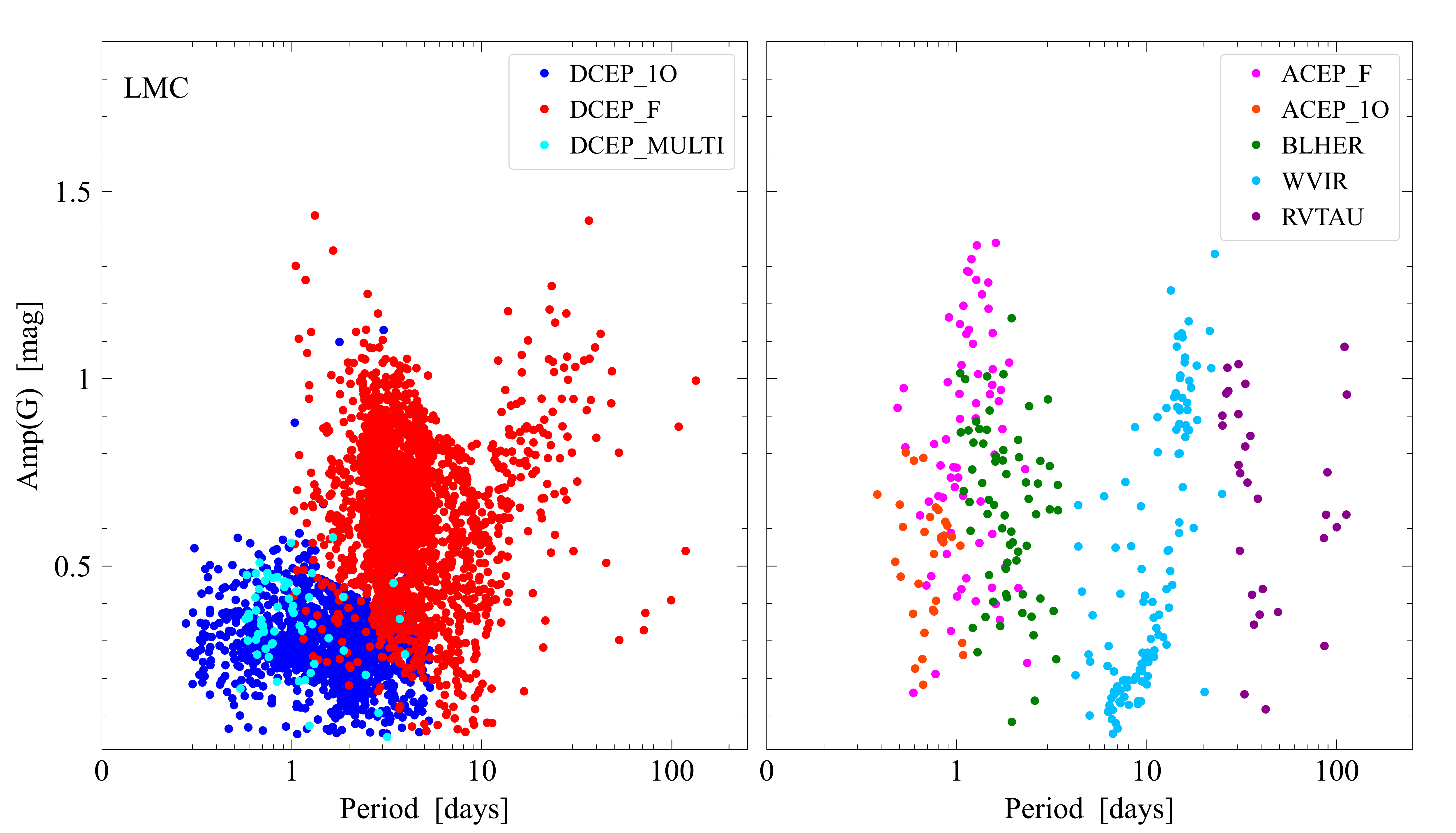}
      \caption{Period-Amplitude(G) diagram for the LMC sample.  
              }
         \label{fig:AmpPeriodLMC}
   \end{figure*}  
   \begin{figure*}
   \centering
   \includegraphics[width=\hsize]{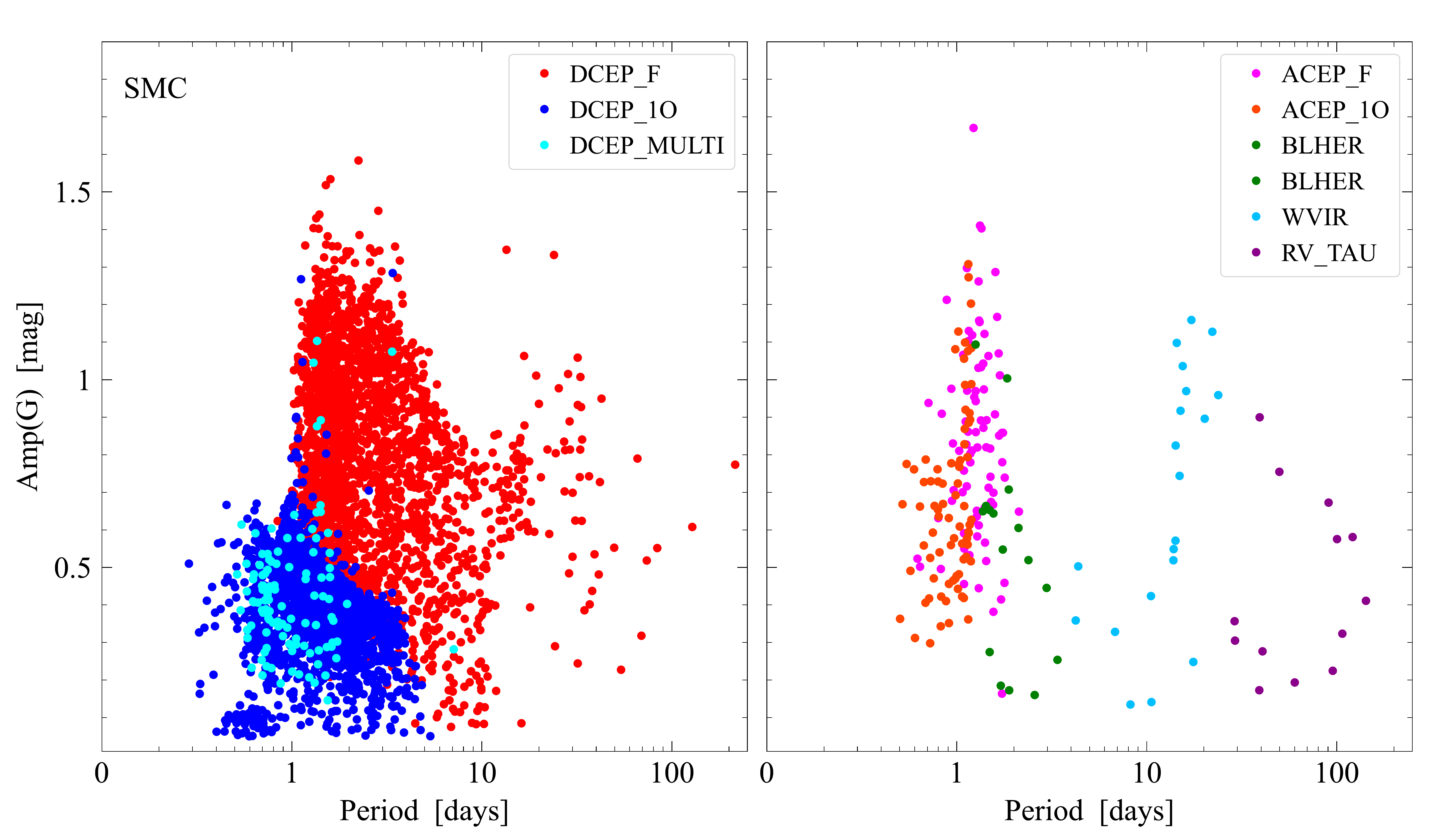}
      \caption{Period-Amplitude(G) diagram for the SMC sample.  
              }
         \label{fig:AmpPeriodSMC}
   \end{figure*} 
   \begin{figure*}
   \centering
   \includegraphics[width=\hsize]{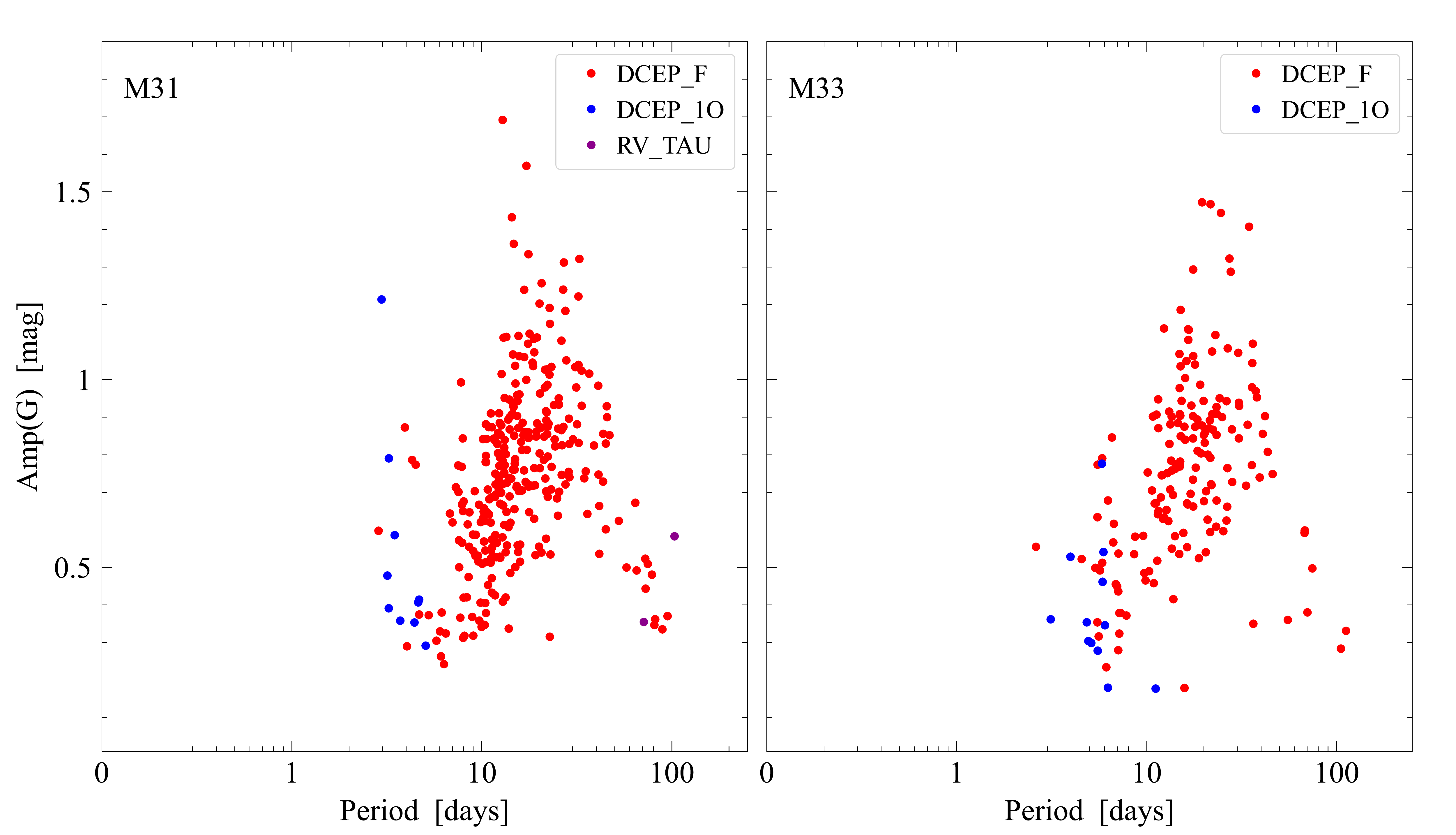}
      \caption{Period-Amplitude(G) diagram for the M31 (left panel) and M33 (right panel) samples, respectively.  
              }
         \label{fig:AmpPeriodM31M33}
   \end{figure*}

\subsection{Colour-magnitude diagrams}

Colour-magnitude diagrams (CMDs) for the Cepheids in all the sub-regions are shown in Figs.~\ref{fig:cmdMW} to ~\ref{fig:cmdM31M33}. The CMDs for the All-Sky sample show very large dispersions, as the reddening along the disc and the bulge, where most of the DCEPs and T2CEPs resides, can be of several magnitudes. Not surprisingly, the dispersion of ACEPs is smaller, as the majority of these objects is placed in the halo, where reddening is on average rather low. 

The MCs have approximately constant and low reddening, so that the CMDs of the Cepheids in these galaxies are more meaningful, with the DCEP\_1O clearly bluer than the DCEP\_F, as expected. The "spur" of LMC DCEPs of both modes extending up to \gbp\ - \grp\ $\sim$1.5 mag remind us that in the LMC there are regions with high reddening values. The range in colours spanned by ACEPs and different types of T2CEPs reflects their locations in the instability strip. 

The CMDs of M31 and M33 DCEPs are shown only for completeness, as the colours are in most cases totally unreliable. 

\subsection{Period-amplitude diagrams}

Figures from ~\ref{fig:AmpPeriodMW} to ~\ref{fig:AmpPeriodM31M33} display the 
period vs amplitude in the $G$  band  (P-Amp($G$))  relations 
for the different sub-regions and Cepheid types. The morphology of these plots for DCEPs in the All-Sky and MCs samples is as expected from the literature \citep[see e.g.][for the SMC and LMC, respectively]{Ripepi2017,Ripepi2022}. The DCEP\_1O, as well as most of the DCEP\_MULTI have Amp($G$)$<$0.5 mag, while the DCEP\_F show the characteristic double peak at periods of 2--3 days and 11-12 days in the All-Sky and LMC samples. The P-Amp($G$) distribution in the SMC is instead significantly different: the DCEP\_1Os show larger amplitudes and the first peak of the DCEP\_F pulsators occurs at shorter periods and larger amplitudes than in the All-Sky and LMC samples, while the second peak is only barely visible with much smaller amplitudes than the first one, again 
in contrast with the All-Sky and LMC samples. All these differences are most likely due to the much lower 
metallicity of the SMC DCEPs with respect to the MW and LMC samples \citep[see e.g.][]{DeSomma2022}.
The P-Amp($G$) diagrams for the DCEPs in the M31 and M33 galaxies appear rather different from 
the other samples. This is mainly because 
only a handful of stars with period shorter than 10 days were detected in these galaxies, so that the first amplitude peak 
for DCEP\_F is completely missed. Instead we observe the second peak, at least in M31, but shifted at about $P\sim$30 days. However, 
this feature needs 
confirmation, as in M31 \gaia\ is operating at the extreme limits of its capabilities.     

The P-Amp($G$) distributions of ACEPs and T2CEPs are also very interesting: i) as expected, ACEP\_1O have smaller amplitudes than ACEP\_F; ii) at periods in the range 1--2 days ACEP\_F can reach significantly higher amplitudes than both DCEP\_F and BLHER, providing us with an additional tool to distinguish them from the different Cepheid types; iii) at the period separation between different T2CEP types, also corresponds a difference in amplitudes, so that the WVIRs have a minimum and a maximum at the extreme periods characterising this class. These considerations are clearly visible in the All-Sky survey, due to the more numerous sample, but are also clearly discernible in the LMC, while in the SMC the paucity of T2CEPs prevents any conclusions.      

\subsection{Radial Velocities}

\begin{figure}
   \centering
   \includegraphics[width=\hsize]
    {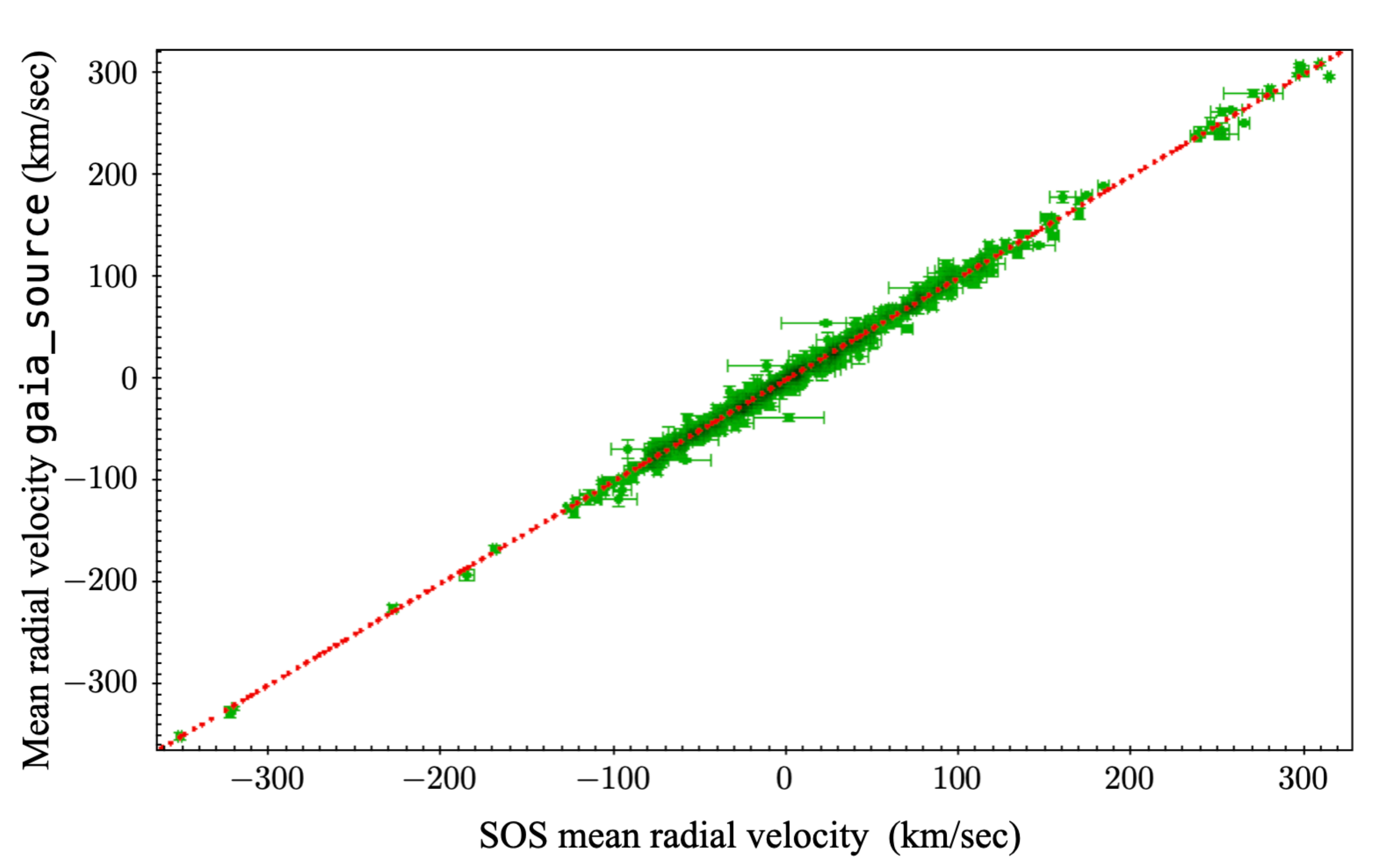}
\caption{Comparison between the average RV calculated by the \sos\ pipeline from fitting the RV curves and the mean values published in the {\tt gaia\_source} table \citep[see][for details]{Sartoretti}.}
              \label{Fig:RVcompCEP}%
    \end{figure}

One of the new products of \gaia\ DR3 is the publication of time series RV data. The final catalogue of Cepheids of all types includes 799 objects for which RV time-series are released. Among these, only for 786 objects the \sos\ pipeline 
obtained average RV and peak-to-peak amplitude values, as for 13 objects the number of epochs is smaller than seven, which is the minimum required for the RV curve fitting. In total the time-series are released for 582 DCEP\_F, 133 DCEP\_1O, 14 DCEP\_MULTI, 12 BLHER, 35 WVIR, 17 RVTAU, 3 ACEP\_F and 2 ACEP\_1O pulsators. Among the DCEP\_Fs, 15 and 9 objects belong to the LMC and SMC, respectively. 
In addition to the time-series, median RV values calculated 
by the general RV data processing in \gaia\ \citep{Sartoretti} are published for 3\,190 Cepheids of all types in the {\tt gaia\_source} table. 
As shown in Fig.~\ref{Fig:RVcompCEP}, there is an excellent agreement between the two estimates for the 736 stars in common between the two samples \citep[see][for further details]{DR3-DPACP-168}. Indeed, median and mean difference between the 
two average values are of 0.43 and 0.33 km/s, respectively, with  6.40 km/s  standard deviation.

\begin{figure}
\centering
\includegraphics[width=\hsize]{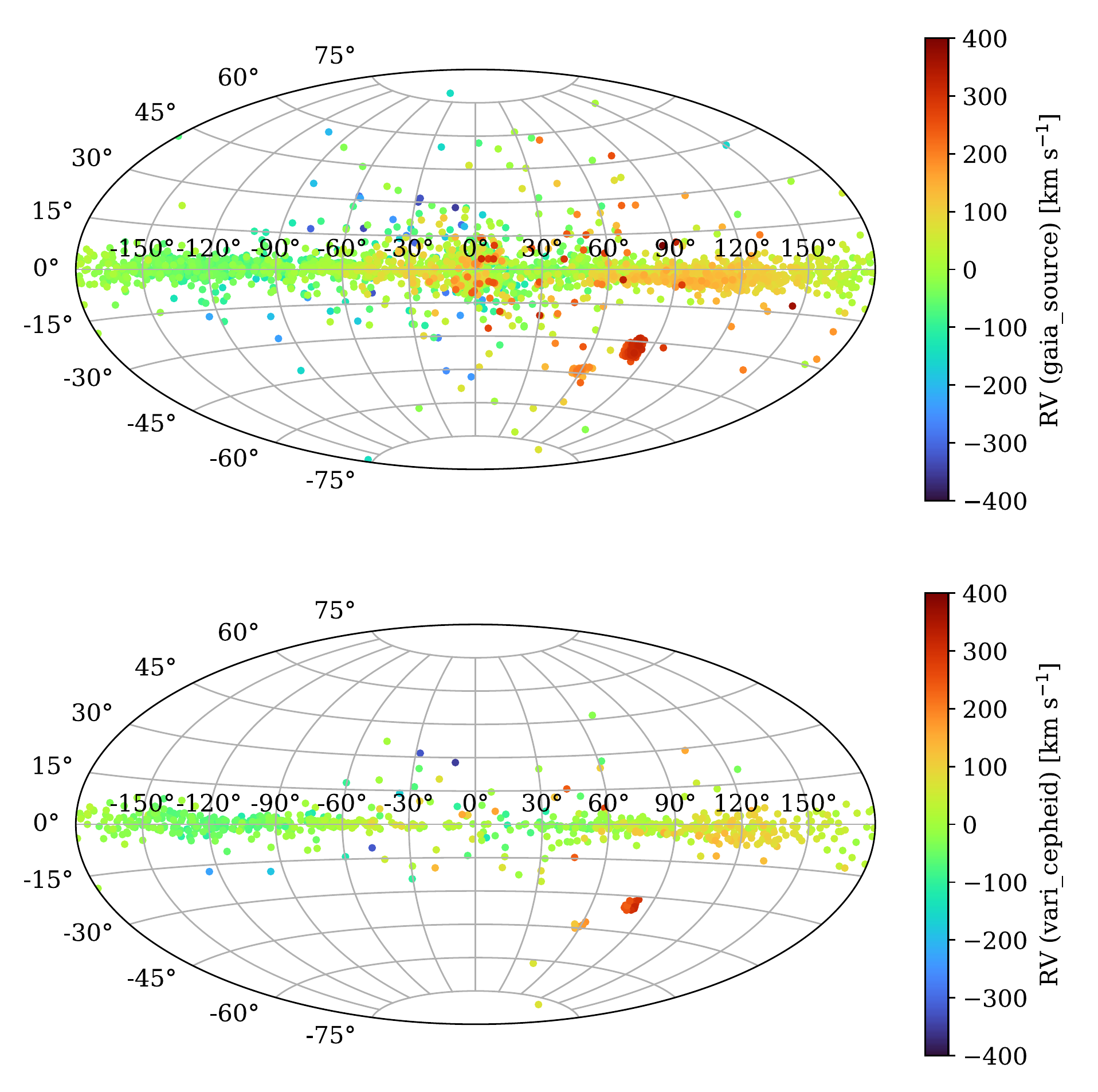}
\caption{RV maps defined by the 3\,190 Cepheids in the DR3 {\tt gaia\_source} table (top panel) and 786/799 Cepheids in the DR3 {\tt vari\_cepheid} table (bottom panel).} 
\label{Fig:RVmapCEP}%
\end{figure}

The spatial distributions of Cepheids with average RV values from both the general and the \sos\ pipelines are shown in Fig.~\ref{Fig:RVmapCEP}, colour coded according to the RV values. As expected, the objects lying in the disc  \citep[mainly DCEPs, see][for the an example of exploitation of these data]{DR3-DPACP-075} show low values of RV, while the halo Cepheids show both highly positive and negative RV values. The LMC and SMC are clearly identified by the RV values shared by all stars belonging to the two galaxies.   

\begin{figure}
   \centering
   \includegraphics[width=\hsize]{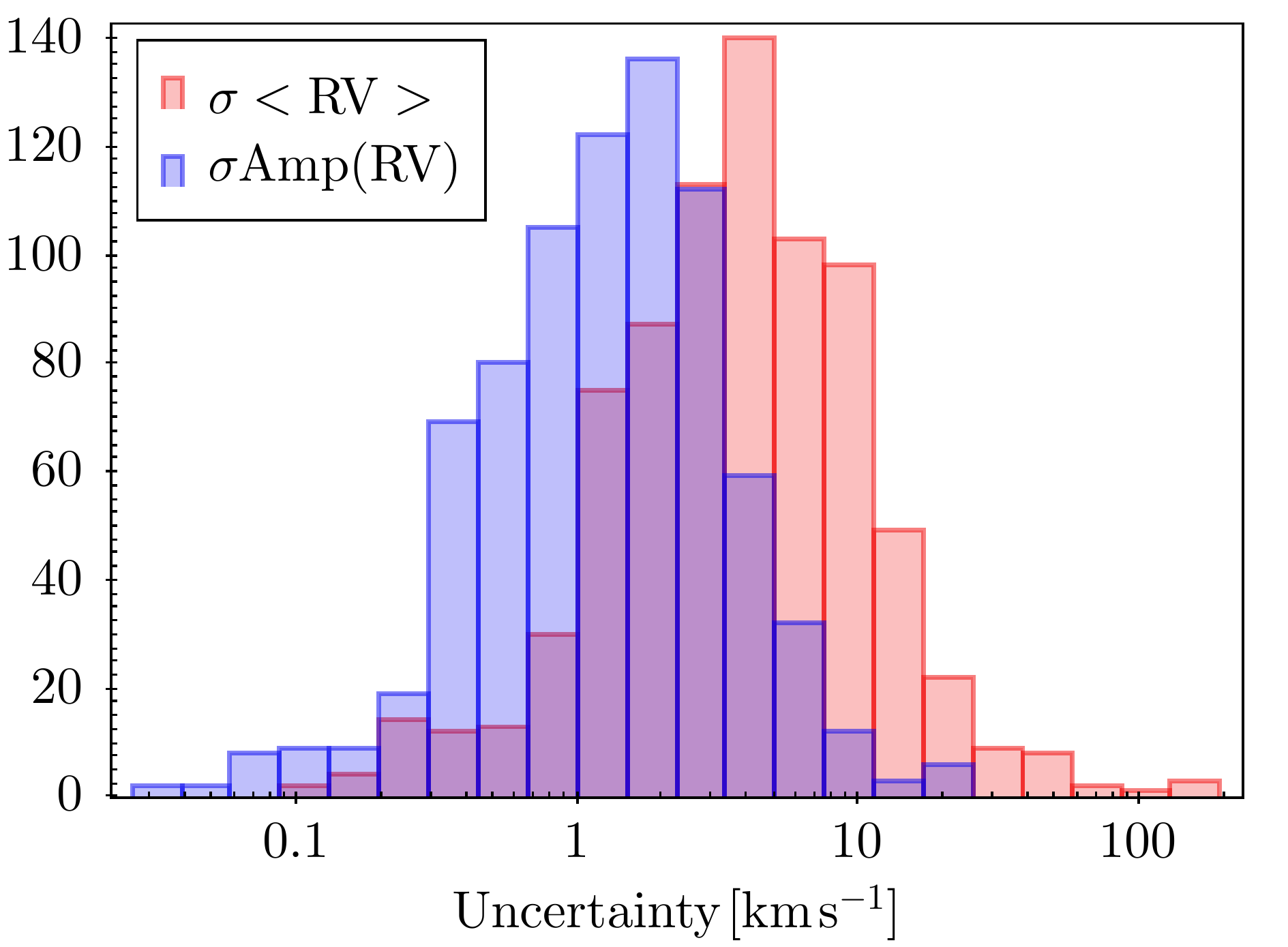}
      \caption{Uncertainties on the average and peak-to-peak RV values measured by the \sos\ pipeline for a sample of 786 Cepheids.              }
         \label{fig:uncertaintyRV}
   \end{figure}

\begin{figure}
   \centering
   \includegraphics[width=\hsize]{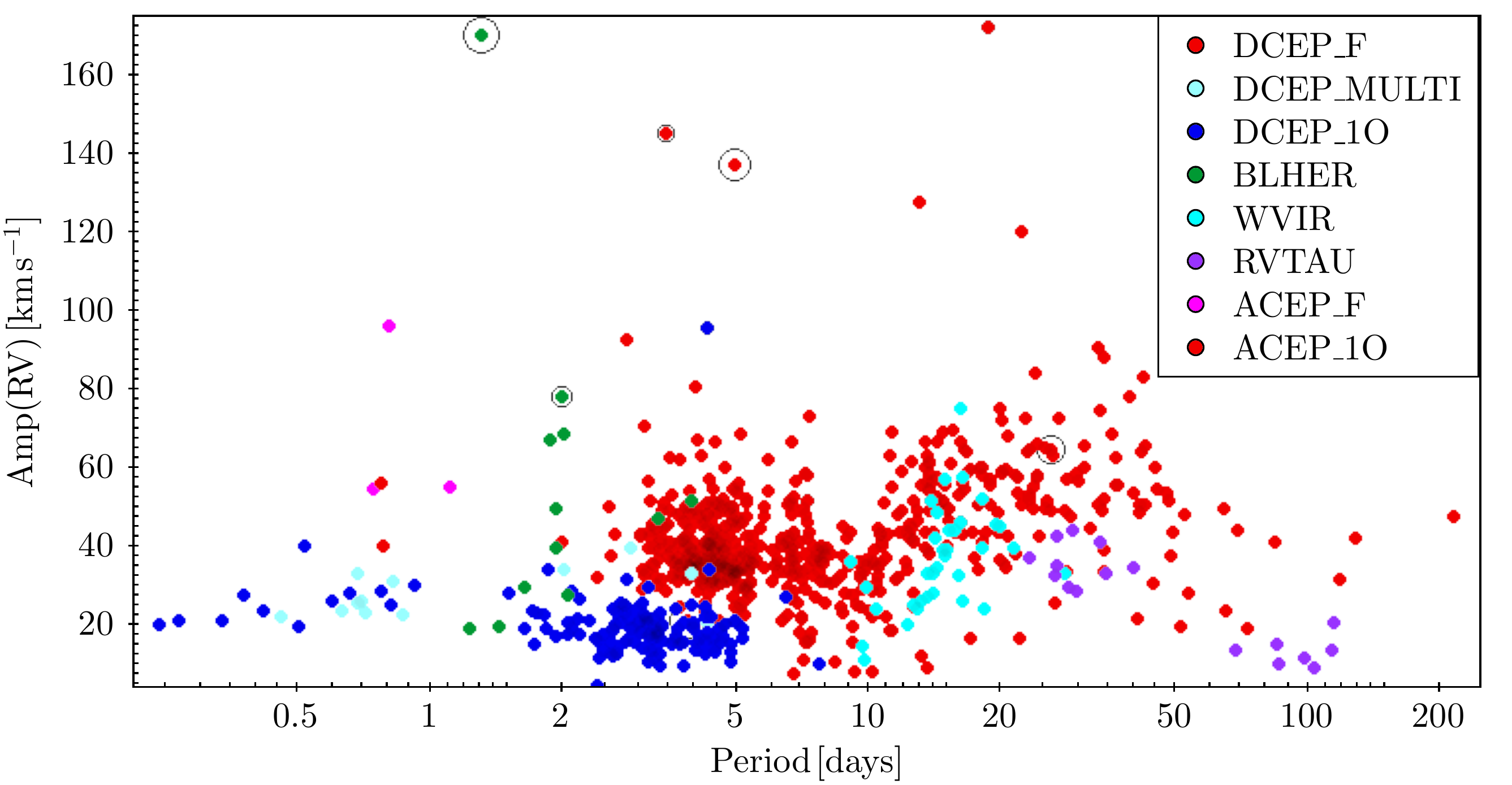}
      \caption{Period-amplitude (RV) for the 786 Cepheids whose RV curves were analysed by the \sos\ pipeline. The different Cepheid types are labelled. The size of the circles surrounding the symbols is proportional to the uncertainty in Amp(RV) (see also  Fig~\ref{fig:uncertaintyRV}.}
         \label{fig:periodAmp_RV}
   \end{figure}

The uncertainties measured by the \sos\ pipeline on the average RV ($\langle$RV$\rangle$) and on the RV peak-to-peak amplitude  (Amp(RV)) are shown in Fig.~\ref{fig:uncertaintyRV}. The typical uncertainties on $\langle$RV$\rangle$ are on the order of 1--1.5 km s$^{-1}$, as expected \citep[see][]{DR3-DPACP-168}. However there are a few objects showing large errors as measured by the bootstrap procedure. These cases are often correlated with the low number of RV epochs that are available for 
these Cepheids (see Fig.~\ref{fig:nEpochsRV}). Similarly, for the Amp(RV) the typical uncertainty is $\sim$3-4 km s$^{-1}$, but there are a few objects with uncertainties larger than 30-40 km s$^{-1}$ which can be 
indication of unreliable Amp(RV) values. 
This 
is verified in Fig.~\ref{fig:periodAmp_RV}, where, in analogy with the photometry, we show the relation between amplitude in RV and period. The general trend follows strictly that shown from photometry, with the DCEP\_F having larger amplitudes than DCEP\_1O or DCEP\_MULTI and showing the typical bell shape starting from a minimum amplitude at period $\sim$9 days and a maximum at $\sim$20 days. The figure shows that albeit the large uncertainties in the Amp(RV) of some objects, only a few Cepheids appear out of their expected position in this plot. We conclude that the RV amplitudes calculated by the \sos\ pipeline are generally reliable.  

\subsection{Metallicities}

\begin{figure}
   \centering
   \includegraphics[width=\hsize]{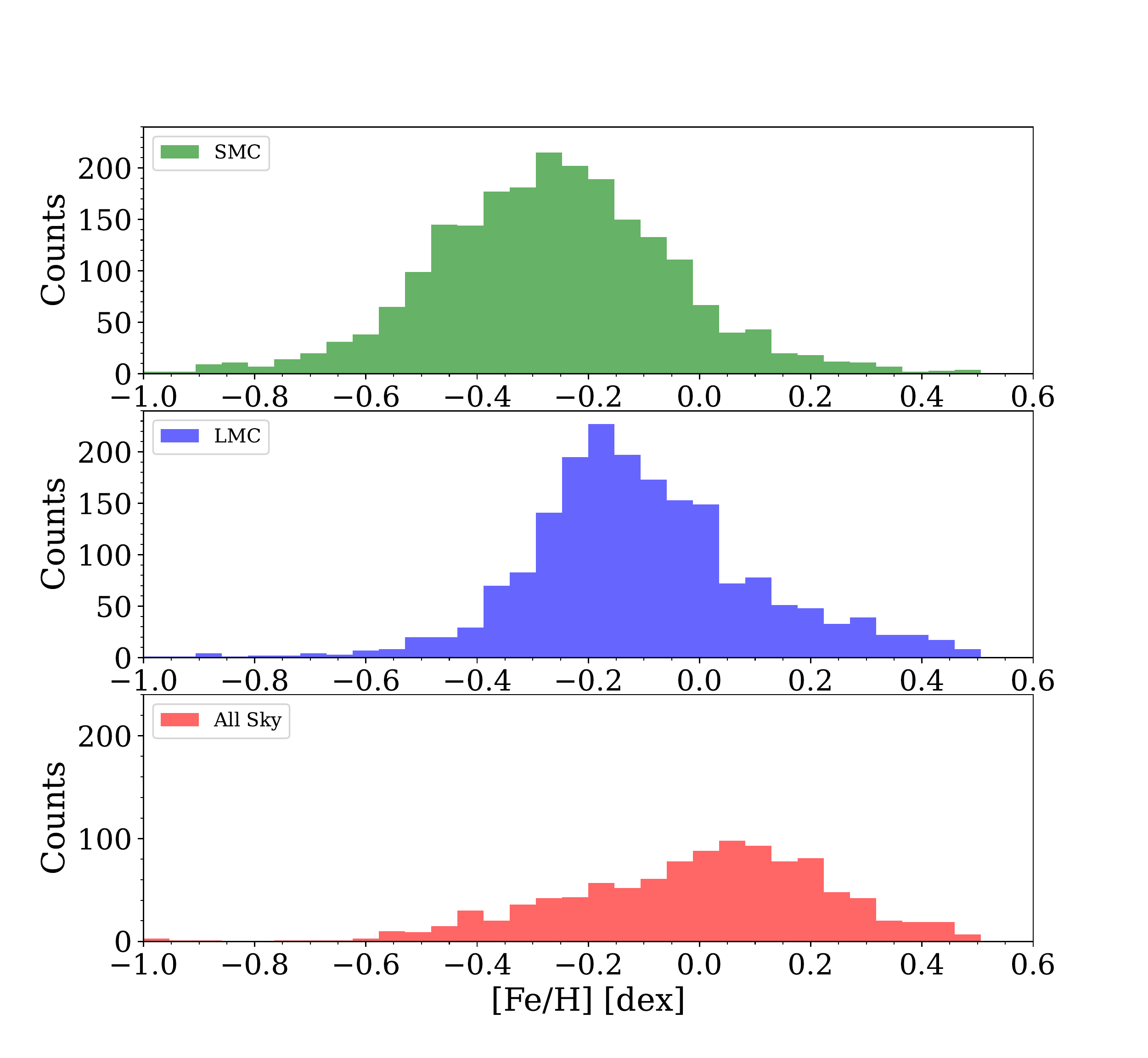}
      \caption{Photometric metallicities in the LMC, SMC and All-Sky samples. 
              }
         \label{fig:histoMetallicity}
   \end{figure}

An additional product of the \sos\ pipeline are photometric iron abundances inferred from the Fourier parameters $R_{21}$ and $R_{31}$ according to the calibration by \citet{Klagyivik2013}, which is valid for DCEP\_Fs 
with periods shorter than 6.3 days and for an interval of metallicity reaching the average [Fe/H] values of the LMC and SMC DCEPs \citep[see][for details]{Clementini2019}. As the metallicity estimates rely on  the $R_{21}$ and $R_{31}$ Fourier parameters, which sometimes have large errors calculated with the bootstrap technique, we suggest to use the [Fe/H] values 
with uncertainties larger than $\sim$0.5 dex with care. The catalogue includes a total of 5\,265 DCEP\_Fs with [Fe/H] estimates. However, since for the 142 objects reported in 
Table~\ref{tab:reclassification} we have changed the classification (see Sect.~\ref{sect:visualInspection}),   
some of these objects are no longer DCEP\_F, hence their metallicity estimates are incorrect and should not be used. 
The DCEP\_Fs with an [Fe/H] estimate are 1\,053, 1\,882, 2\,174, 7 and 7 in the All-Sky, LMC, SMC, M31 and M33 samples, respectively. The distribution of the metallicities in the SMC, LMC and All Sky samples is shown in Fig.~\ref{fig:histoMetallicity}. The figure shows that, as expected, on average the DCEPs in the All-Sky sample (exclusively MW objects) are more metal rich than the LMC ones, which, in turn are more metal abundant than those in the SMC. From the quantitative point of view, we can see that the peak of the All-Sky distribution is [Fe/H]$\sim$+0.05 dex, which is in general agreement with the literature \citep[see e.g.][]{Ripepi2019}. On the contrary, for the LMC and SMC, we have peaks of approximately $-$0.2 dex and $-$0.3 dex  for the LMC and SMC, respectively. These values are significantly larger than those found in the literature, namely, [Fe/H]$_{\rm LMC}=-0.41$ dex \citep[$\sigma$=0.08 dex][]{Romaniello2022} and [Fe/H]$_{\rm SMC}=-0.75$ dex \cite[$\sigma$=0.08 dex][]{Romaniello2008}. Therefore 
the photometric metallicities are not particularly reliable for values 
more metal poor than [Fe/H]$\sim-0.3$ dex, which is not a unespected, 
since the work by \citet{Klagyivik2013} relies on very few calibrators in this metallicity range. 

\begin{figure}
   \centering
   \includegraphics[width=\hsize]{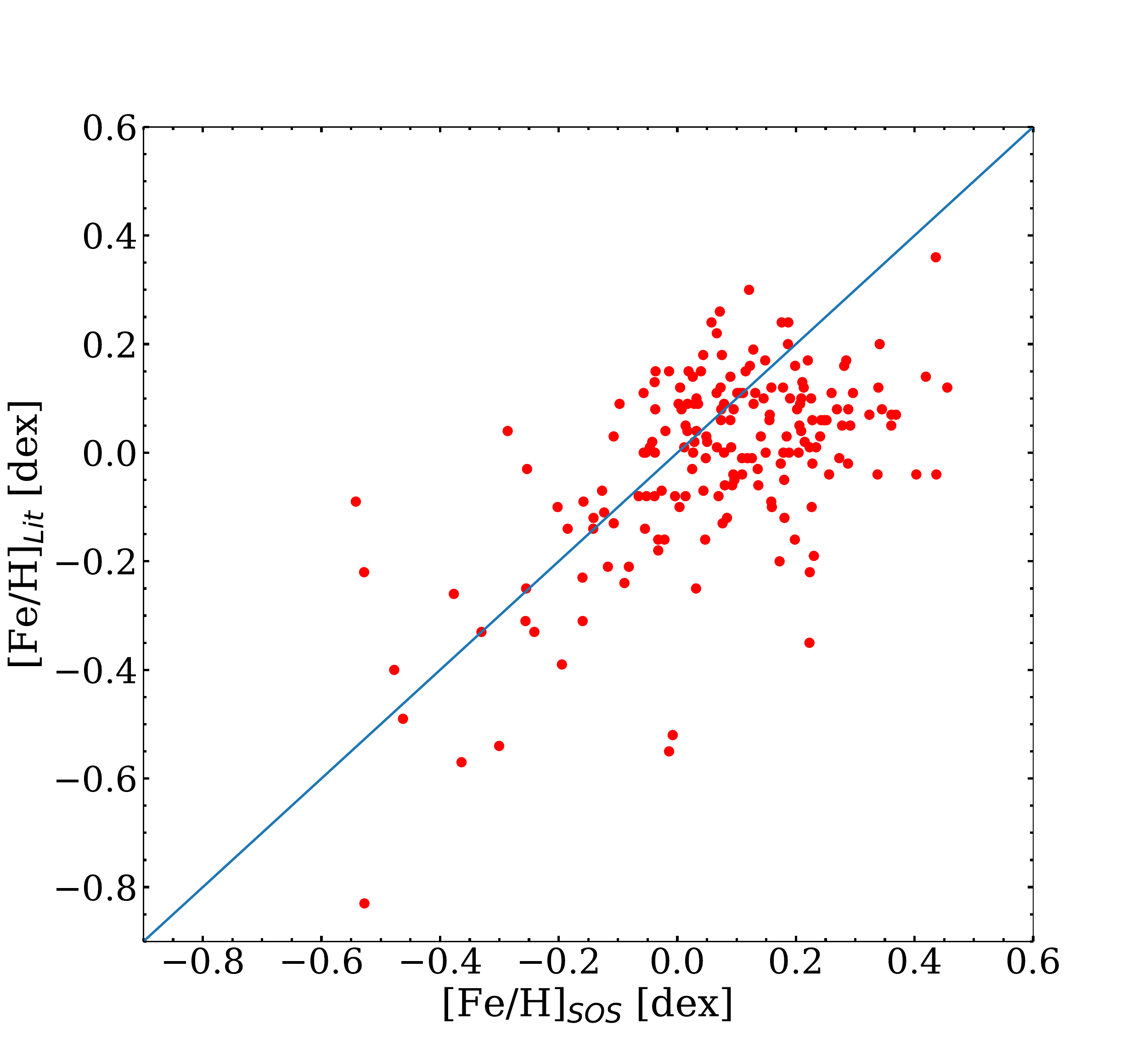}
      \caption{Comparison between photometric metallicities computed by the \sos\ pipeline ([Fe/H]$_{SOS}$) and metal abundances from high-resolution spectroscopy available in the literature ([Fe/H]$_{Lit}$). 
              }
         \label{fig:metallicityComparison}
   \end{figure}

We can perform a more detailed comparison between the photometric metallicities from the \sos\ pipeline and the literature by cross-matching the All-Sky sample with the list of DCEPs having metallicities measure from high-resolution spectroscopy recently published by \citet{Ripepi2022a}\footnote{In this and many other phases of this work, we made use of the {\tt TOPCAT} package \citep[Tool for OPerations on Catalogues And Tables][]{Taylor2005}}. The metallicity estimates for the 185 DCEPs in common between the two samples are displayed in Fig.~\ref{fig:metallicityComparison}. The photometric [Fe/H] values appears to be systematically more metal rich than the spectroscopic abundances. The average difference is [Fe/H]$_{\rm Lit}$-[Fe/H]$_{\rm SOS}$=$-$0.08 dex, with $\sigma$=0.16 dex and no apparent trend with the [Fe/H]$_{\rm Lit}$ value. The mean shift and relative dispersion are modest, so that, as far as the All-Sky sample is concerned, or at least in the metallicity range $-0.3< $[Fe/H]$<+0.4$ dex, the photometric metallicites can be used. We speculate that for more metal poor values, the metallicity sensitivity of the $R_{21}$ and $R_{31}$ parameters may vanish. This could explain the poor performance of the method for the LMC and SMC DCEP samples.

\begin{table}
\caption{Gaia source\_id of sources for which the \sos\ pipeline provides a metallicity estimate which should not be used as these stars are not DCEP\_F pulsators. Only the first ten lines are shown to guide the reader to the content of the table. The entire version of the table will be published at CDS (Centre de Donn\'ees astronomiques de Strasbourg, https://cds.u-strasbg.fr/).}             
\label{tab:noUseTheseMetallicities}      
\centering                          
\begin{tabular}{c}        
\hline\hline                 
Gaia source\_id  \\    
\hline                        
375318264077848448 \\
375435873166554752 \\
431184518613946112 \\
513074186146353536 \\
543759459725179136 \\
1208200864738741376 \\
1248397910338129664 \\
1374376207437762688 \\
1400474455952839168 \\
1682922431734385152  \\      
\hline                                   
\end{tabular}
\end{table}

\subsection{Cepheids hosted by stellar clusters and  satellite dwarf galaxies of the MW}

\label{sect:associationWithStellarSystems}

We searched for any association of Cepheids in the All-Sky sample with stellar clusters hosted by the MW or with dwarf galaxies orbiting our Galaxy. 
For the open clusters (OCs), we adopted the list of likely 
member stars by \citet{CantatGaudin2020} supplemented with new data by \citet{CastroGinard2021} and \citet{Tarricq2022}; for the globular clusters (GCs) we used the list by \citet{Clement2001} (continuously updated); for the dwarf galaxies we used a variety of literature sources including 
\citep{Sos2017,Sos2018}. 
Results are shown in Table~\ref{tab:association}. Additional 35 objects from the All-Sky sample can be associated with the MCs, 45 with Galactic GCs, 24 with OCs and one with the Draco dwarf spheroidal galaxy \citep[variable data for Draco by][]{Kinemuchi2008}.

\begin{table*}
\caption{Association of Cepheids in the All-Sky sample with open/globular clusters and with dwarf galaxies which are satellites of the MW.  
Only the first ten lines are shown
to guide the reader to the content of the table. The 
table is published in its entirety at the CDS (Centre de Donn\'ees astronomiques de
Strasbourg, https://cds.u-strasbg.fr/).}             
\label{tab:association}     
\begin{tabular}{rrrlll}
\hline
  \multicolumn{1}{c}{source\_id} &
  \multicolumn{1}{c}{RA (J2000)} &
  \multicolumn{1}{c}{Dec (J2000)} &
  \multicolumn{1}{c}{Mode} &
  \multicolumn{1}{c}{System} &
  \multicolumn{1}{c}{Other ID} \\
  \multicolumn{1}{c}{} &
  \multicolumn{1}{c}{(deg)} &
  \multicolumn{1}{c}{(deg)} &
  \multicolumn{1}{c}{} &
  \multicolumn{1}{c}{} &
  \multicolumn{1}{c}{} \\
\hline
  429385923752386944 & 0.246798 & 60.959002 & DCEP\_F & UBC\,406 & CG\,Cas\\
  4707044742055169152 & 9.219742 & -66.593232 & BLHER & SMC & OGLE-SMC-CEP-4693\\
  4684386345732125696 & 11.086946 & -76.195508 & DCEP\_F & SMC & OGLE-SMC-CEP-4710\\
  4702506576531479424 & 12.344029 & -69.50825 & DCEP\_F & SMC & OGLE-SMC-CEP-4723\\
  4635176637678468096 & 14.144298 & -77.920315 & DCEP\_F & SMC & OGLE-SMC-CEP-4967\\
  4691023998645738368 & 17.496078 & -69.937937 & DCEP\_F & SMC & OGLE-SMC-CEP-4816\\
  4691296265212110720 & 22.594176 & -69.427211 & RVTAU & SMC & --\\
  4636490008613940992 & 23.448788 & -77.656158 & ACEP\_F & SMC & OGLE-SMC-ACEP-091\\
  4691302823627312640 & 23.541042 & -69.256368 & DCEP\_F & SMC & OGLE-SMC-ACEP-092\\
  4698416118397803776 & 25.204992 & -67.494995 & ACEP\_F & SMC & --\\
\hline\end{tabular}
\end{table*}

\section{Validation} 

\subsection{Literature adopted for the validation}
\label{sect:validationLiterature}

To validate the results of the \sos\ pipeline classification we adopted different literature sources according to the different sub-regions of reference. Starting with the All-Sky, for the DCEPs we adopted the recent compilation by \citet{Piet2021} (hereafter, P21), including 3\,352 reliable bona fide DCEPs, mainly based on results from  the OGLE survey \citep[][]{Udalski2018,Sos2020}. For ACEPs and T2CEPs, we adopted the results of the OGLE survey \citep[][and references therein]{Sos2020} complemented by entries in \citet{Chen2020}, which is based on the ZTF (Zwicky Transient Factory) survey, and by \citet{Drake2014,Torrealba2015}, which are based on the Catalina sky survey (CSS). As the classification of the latter papers does not distinguish the mode/type of pulsation, we assigned the fundamental mode to the ACEP detected by CSS\footnote{For analogy with their studies on RR Lyrae stars, for which 
they only consider fundamental mode pulsators.} and separated BLHER from WVIR and WVIR from RVTAU using period thresholds of 4 and 24 days, respectively (in analogy with the \sos\ pipeline). The total sample 
having a positive cross-match with the \gaia\ DR3 catalogue includes 3\,917 Cepheids. Note that we have intentionally not included results from \gaia\ DR2 
re-classifications by \citet{Ripepi2019} to preserve the independence of the counterpart. We have also not included 
Cepheids by ASAS-SN \citep[All-Sky Automated Survey for Supernovae][]{Shappee2014,Jayasinghe2019} and ATLAS \citep[Asteroid Terrestrial-impact Last Alert System][]{Heinze2018} who adopt automatic classification procedures, not based on a careful visual inspection of the light curves. However, many stars originally detected by these surveys have been analysed by \citet{Piet2021} and are included in their catalogue. 

As for the MCs, we adopted the OGLE catalogue 
by \citet{Sos2019_MCs}, including 9\,650 DCEPs, 343 T2CEPs, and 278 ACEPs. A cross-match with \gaia\ DR3 results provides 4\,638 and 4\,608 matches for the LMC and SMC, respectively. 

For M31 we used the work by \citet{Kodric2018} who provided the classification for 2\,247 Cepheids, including DCEP\_F, DCEP\_1O and RVTAU. We have 262 stars in common with this work. As for M33, 112 over 185 objects classified as Cepheid from the \sos\ pipeline are present in the work by \citet{Pellerin2011}. However, 
these authors do not provide a classification in 
DCEPs or T2CEPs, hence, we 
refrained from 
any comparison.

\subsection{Accuracy of the classification, completeness and contamination}

On the basis of the literature data  discussed in the previous section, we produced confusion matrices for the LMC, SMC and the All-Sky samples. 

\begin{figure}
   \centering
   \includegraphics[width=\hsize]{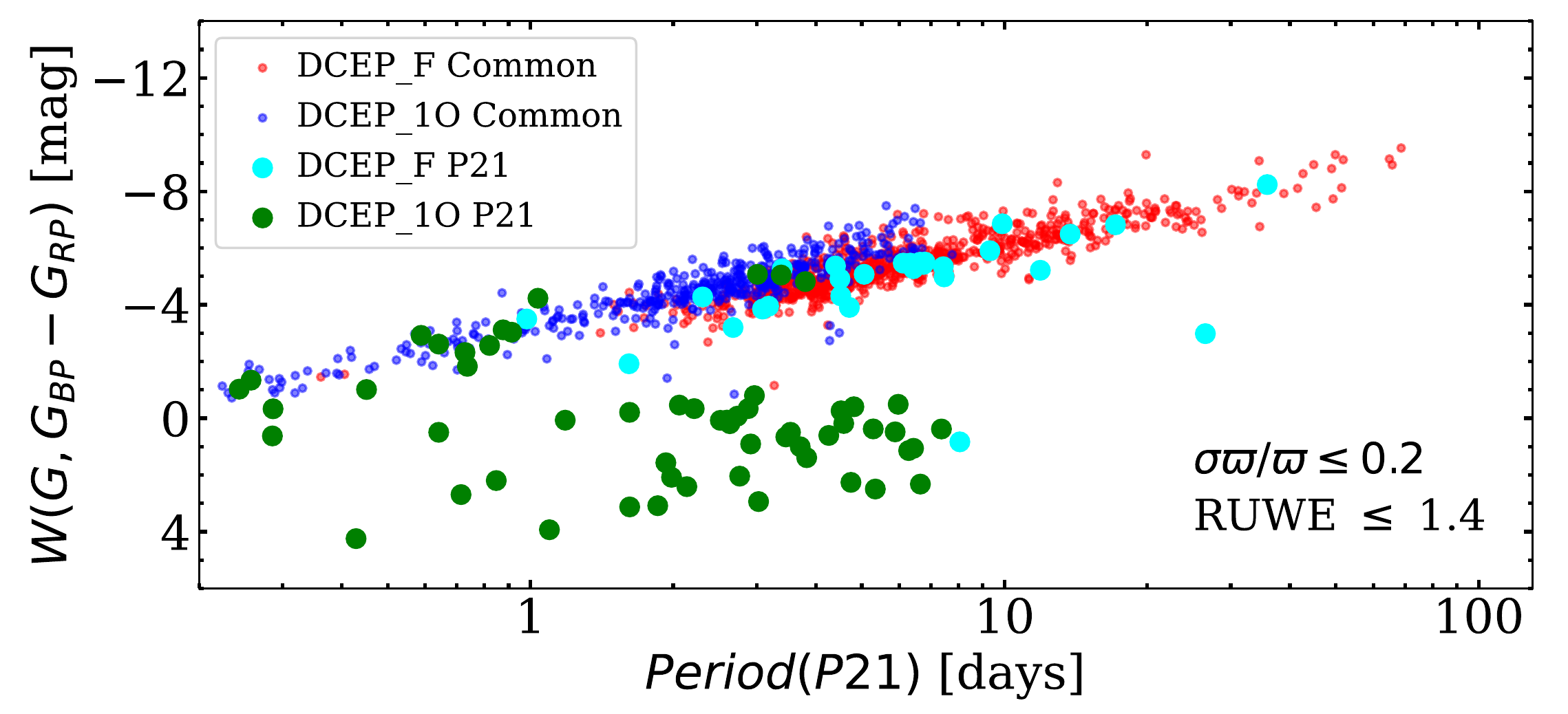}
      \caption{PW relation for a selected sample of DCEPs. Red and blue small filled circles show the DCEP\_Fs and DCEP\_1Os in common between \gaia\ DR3 and \citet[][abbreviated as P21 in the labels]{Piet2021}, respectively. Cyan and green large filled circles show DCEP\_Fs and DCEP\_1Os present in the P21 catalogue only. For all objects we applied a selection in parallax, requiring that the  relative precision is better than 20\%. We also required the RUWE parameter to be lower than 1.4, as to ensure a good astrometric solution (see text).   
              }
         \label{fig:sfriculiamentoOgle}
   \end{figure}

There are 2\,739 stars in common with P21, corresponding to 82\% of the sample. Further 130 objects are published in the general classification \citep[][]{DR3-DPACP-165} as the \sos\ pipeline found a wrong period for these objects. Therefore counting also the latter the completeness of the \gaia\ catalogue for the All Sky DCEP sample is of 85.6\% at least. However, the catalogue by \citet{Piet2021} is not exempted by contamination, especially for the DCEP\_1Os, that  can be easily confused with binaries if the distance is not used in the classification. This is shown in Fig.~\ref{fig:sfriculiamentoOgle} which shows the $PW$ relation for a selected sample of DCEPs with parallax relative errors better than 20\% and good astrometric solution (RUWE$\leq$1.4). The vast majority of the objects shown in the figure are in common between the \gaia\ DR3 catalogue and \citet{Piet2021}, and nicely depicts the expected linear relations for both DCEP\_F and DCEP\_1O pulsators. The second sample includes objects present only in the \citet{Piet2021} list. Most of the DCEP\_1O are clearly too faint to be DCEPs or any other type of Cepheids, and likely are binaries contaminating the DCEPs sample. Although the numbers of objects with good parallaxes is too small to obtain a sensible statistics, it is rather plausible that the completeness of the \gaia\ DR3 catalogue for DCEPs is larger than 85.6\% once the purity of the comparison samples is taken into account.

The completeness for ACEPs and T2CEPs is more difficult to establish as there are no homogeneous catalogues for these Cepheid types, except for regions of the sky covered by the OGLE survey. Therefore, we restricted our  estimates 
only to the Bulge and a portion of the disc \citep[see e.g.][]{Sos2020}, and calculated the ratio of the number of ACEPs and T2CEPs in DR3 and the OGLE catalogues. Given the small numbers involved compared with DCEPs, we summed up ACEPs and T2CEPs, obtaining an overall completeness of about 25\%. Such a lower completeness, compared to 
the 
DCEPs is because 
the large majority of the OGLE ACEPs and T2CEPs are  
in the Bulge, a region where \gaia\ has still a low number of epochs on average. In addition, the Bulge is also almost devoid of DCEPs, so that the \gaia\ low detection efficiency in this region does not impact the DCEP completeness.

The confusion matrix of the All-Sky sample is shown in Fig.~\ref{fig:confMatrixMW}. The apparent accuracy of our DCEPs classification ("Recall" column) is satisfactorily high, being 96\%, 92\% and 95\% for DCEP\_F, DCEP\_1O and DCEP\_MULTI, respectively. A similar result is obtained for T2CEP variables, namely $>$94\% for all Cepheid types. The percentages are less good for the ACEPs which are much more difficult to classify, given the similarities in light curve shape with DCEP and BLHER variables. Therefore we tend to classify more ACEPs than the literature, where usually the classification is only based on the light curve shape. As for the precision, it is again very high for T2CEPs and DCEPs with the exception of DCEP\_MULTI, of which we apparently missed about 30\%. This is not surprising, as for many pulsators we just do not have enough epochs to resolve more than one pulsating mode. For ACEPs the precision is about 70\%, which means that we are able to detect a large fraction of the literature ACEPs.    

The same kind of comparison is shown in Fig.~\ref{fig:confMatrixLMC} and ~\ref{fig:confMatrixSMC} for the LMC and SMC, respectively. The results are very good in the LMC for both accuracy and precision for all types, with the exception of the DCEP\_MULTI, which we massively missed and classified as DCEP\_1O because the low number of epochs prevented the detection of the second (or third) periodicity. The results are slightly worse in the SMC, where the elongation along the line of sight produces much less separated $PL$ and $PW$ relations. This impacts especially the ACEPs which were confused with DCEPs introducing a 2\% contamination among the latter. 
In the SMC we missed in percentage less DCEP\_MULTI. 

Concerning the overall completeness (e.g. ignoring the sub-classification in types or modes), in both LMC and SMC the \gaia\ DR3 catalogue includes 90\% of the known Cepheids of all types.

As for M31, we do not show the confusion matrix as the agreement between our classification and the literature is 100\%. Obviously the completeness is much less, because due to \gaia\ limiting magnitude we were able to detect reasonable light curves 
only for the brightest Cepheids in M31. This corresponded to only 12.1\% of the known Cepheids of all types. 
We do not have an accurate literature control sample for the M33 Cepheids, therefore we only mention that we detected about 23\% of the known Cepheids in this galaxy.

\subsection{Contamination by variables other than Cepheids}

In the previous section we established the reliability
of the Cepheid classification in the \gaia\ DR3 catalogue by comparison with high quality Cepheid catalogues in the literature. 
Then for the All-Sky sample, we use the same literature catalogues,  
namely, OGLE \citep[][]{Sos2019}, ZTF \citep{Chen2020} and CSS \citep{Drake2014}, that list also variability types other than Cepheids, to assess the possible contamination by non-Cepheids of the \gaia\ DR3 Catalogue.   
As a result we found 93 objects which are listed in Table~\ref{tab:contamination}. The main source of possible contamination is from RR Lyrae stars, eclipsing binaries and eruptive variables. Even if we restricted our comparison only to the aforementioned surveys, we can anyway conclude that contamination of the \gaia\ DR3 Cepheid catalogue is on the order of 1--2\%.

\subsection{The case of ARRD stars}
\label{sect:arrd}

Anomalous double-mode RR Lyrae stars  (ARRDs) differ from normal RRDs because of the smaller ratio between the 1O and F pulsation modes \citep[see][]{Sos2016,Sos2016arrd}. The ARRDs were originally discovered in the LMC, but \citet{Sos2019} reported the presence of many ARRDs also among the OGLE Bulge and Disc collection of RR Lyrae stars. Six of these ARRDs are in the All-Sky sample with classification as DCEP\_MULTI. The position of these stars in the Petersen diagram is highlighted in Fig.~\ref{fig:multi}. Five objects lie in the region where DCEPs pulsates in the F/1O multi-mode, while one (Gaia EDR3 4091104989668551936) is placed in the locus of 2O/3O pulsators. However, the two periods of the latter differ from those found by OGLE and could be wrong, as we have only 23 epochs in {\it Gaia}. Adoption of the OGLE periods would place also this sixth source close to the F/1O DCEP multi-mode pulsators. 

\begin{figure}
   \centering
   \includegraphics[width=\hsize]{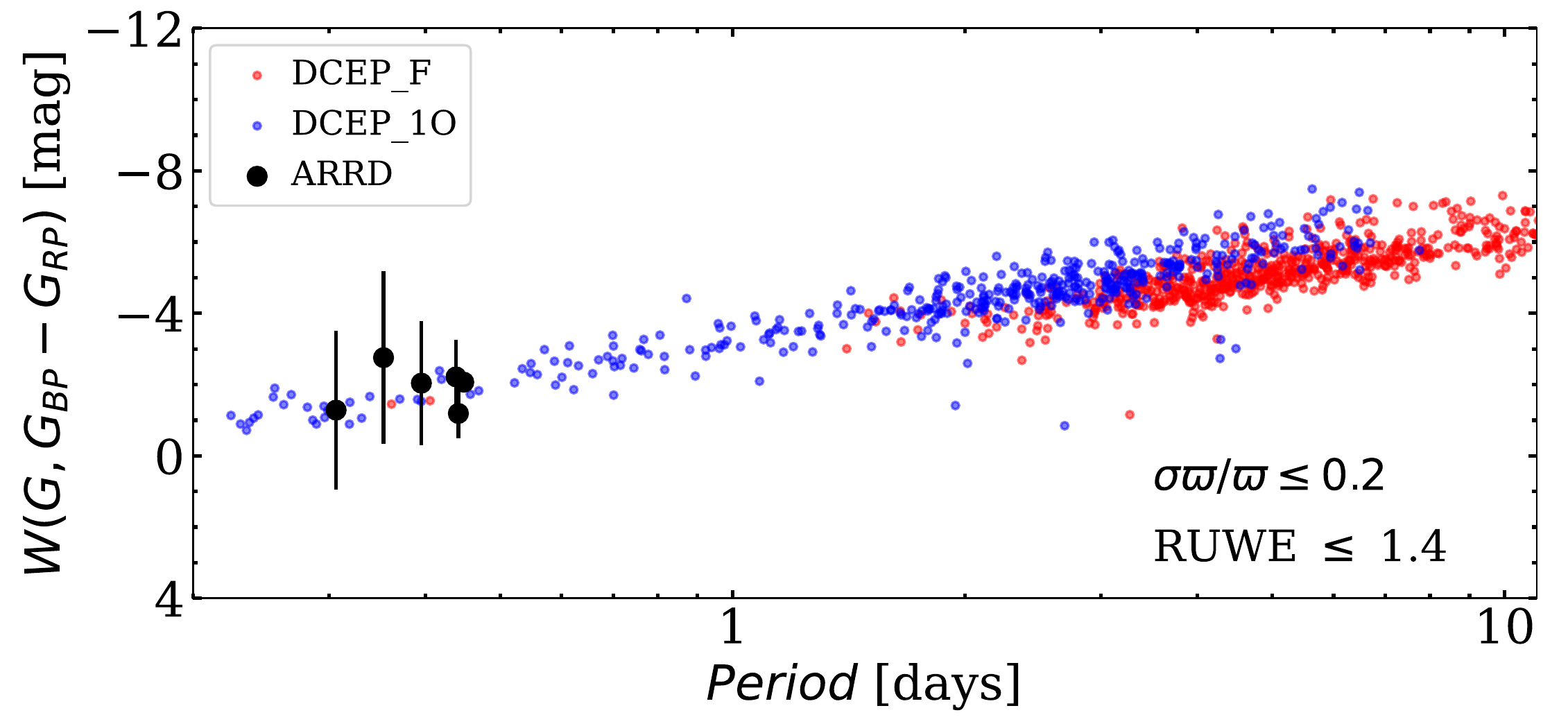}
      \caption{Red and blue dots are as in Fig.~\ref{fig:sfriculiamentoOgle}, black filled circles represent sources that are known in the literature as ARRD stars whereas we  classification them as DCEP\_MULTI.    
              }
         \label{fig:arrd}
   \end{figure}

\begin{figure}
   \centering
   \includegraphics[width=\hsize]{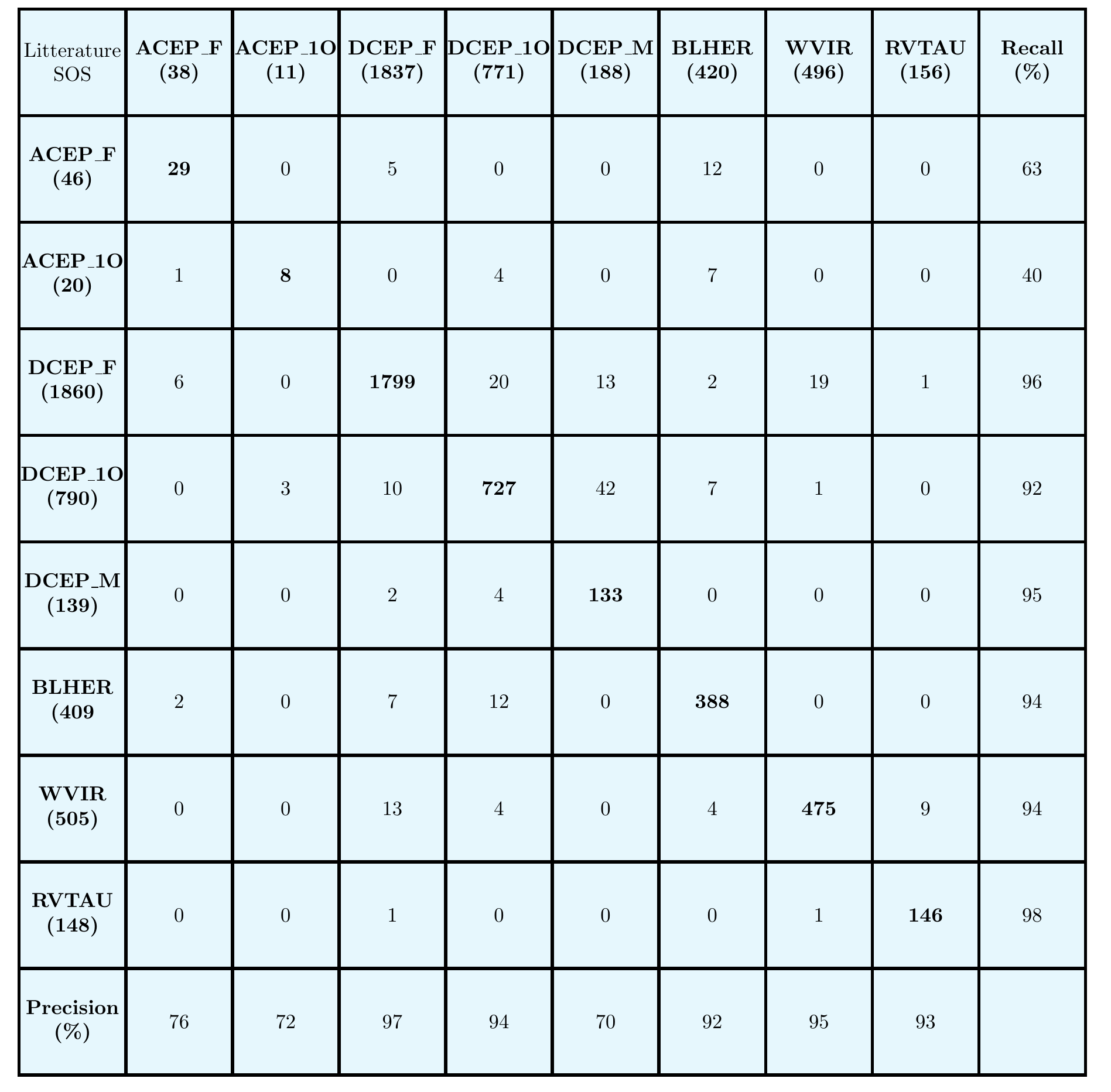}
      \caption{Confusion matrix for the All Sky sample.
              }
         \label{fig:confMatrixMW}
   \end{figure}

\begin{figure}
   \centering
   \includegraphics[width=\hsize]{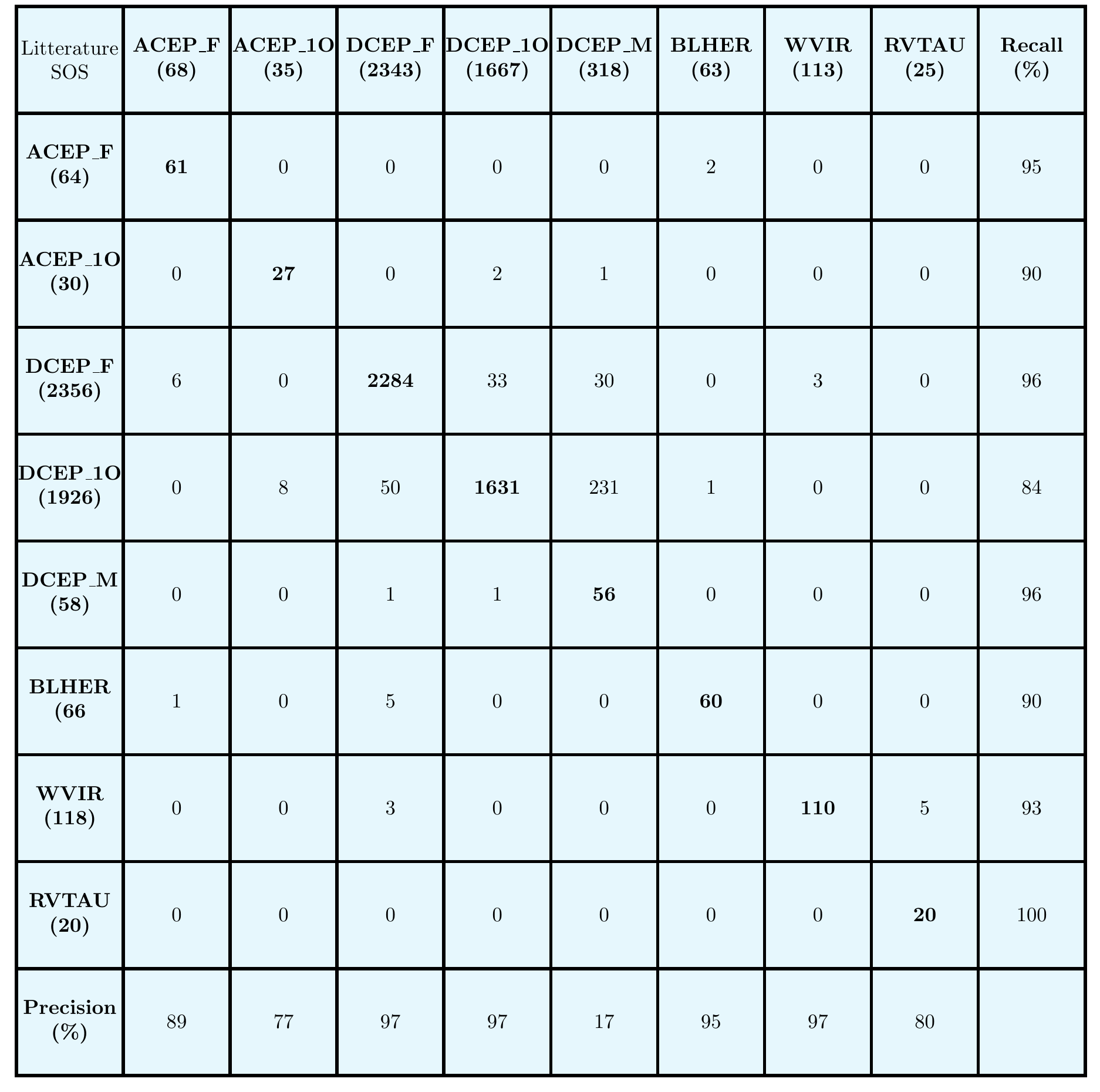}
      \caption{Confusion matrix for the LMC sample.
              }
         \label{fig:confMatrixLMC}
   \end{figure}

\begin{figure}
   \centering
   \includegraphics[width=\hsize]{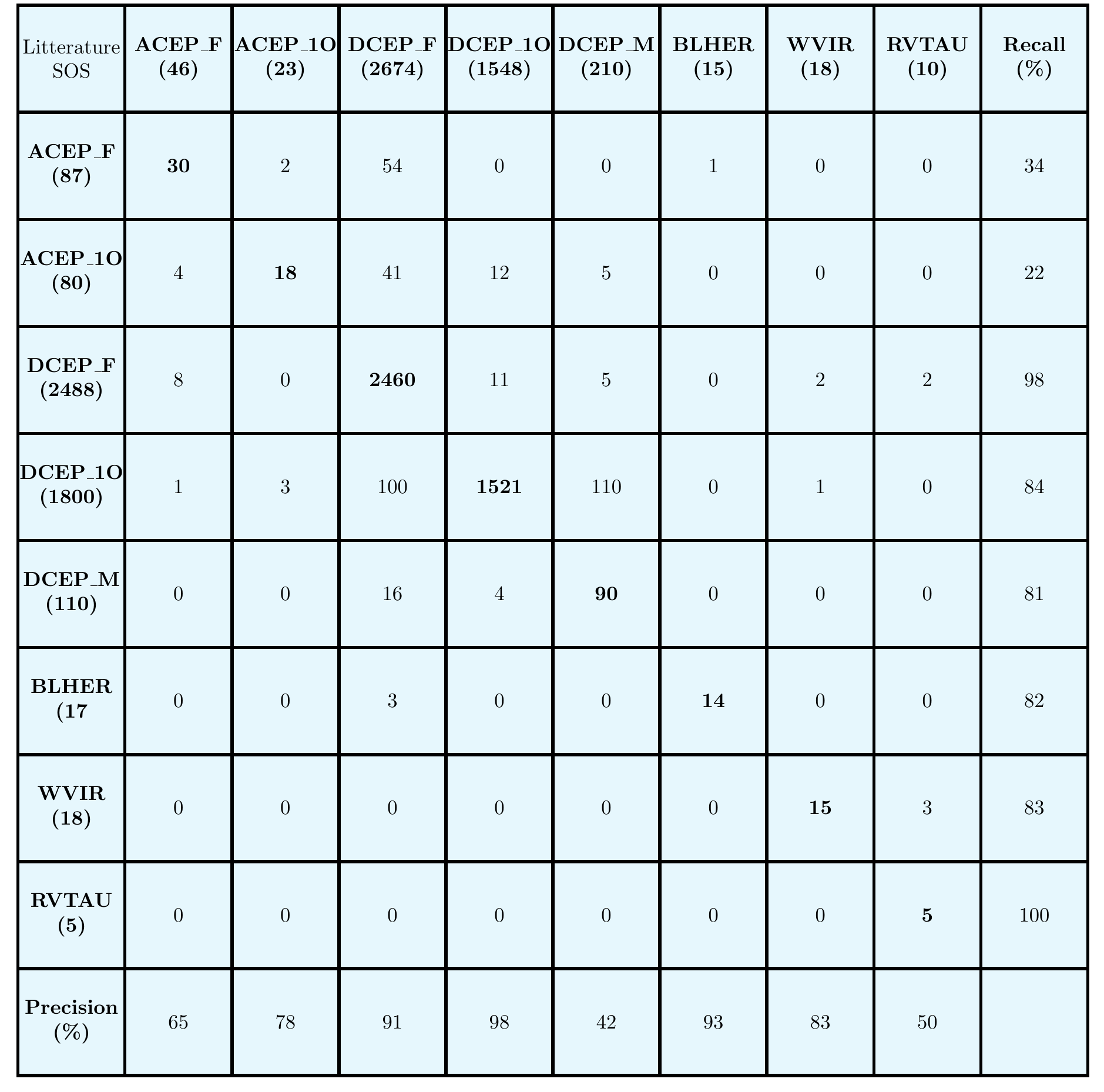}
      \caption{Confusion matrix for the SMC sample.
              }
         \label{fig:confMatrixSMC}
   \end{figure}

The location of the six objects in the $PW$ plane is shown  in Fig.~\ref{fig:arrd}. The uncertainty of the $W$ values for three objects are rather high due to the large uncertainty in their parallaxes, still, the location on the $PW$ relation of all six objects seems compatible with them being 
short-period DCEPs. We conclude that at least some of the objects classified as ARRDs in the MW are actually DCEPs and not RR Lyrae variables. This is due to the difficulty in determining the distances in the MW compared with the LMC, where all the objects are approximately at the same distance from us. 

\begin{table*}
\caption{Potential contaminants 
of type other than Cepheids. Only the first ten lines are shown to guide the reader to the content of the table. }             
\label{tab:contamination}      
\centering                          
\begin{tabular}{ccccccc}        
\hline\hline                 
        source\_id     & RA(J2000)  &  Dec(J2000)    &  P\_SOS  &  Class\_DR3 & Class\_Lit  & Lit\_source   \\
                      &   (deg)    &     (deg)      &  (days) &             &             &             \\
\hline            
  526109377526041344  &  12.70783  &   66.30146 &   0.34443   &   DCEP\_1O   &       RRC   &        ZTF  \\
  523287961970713728  &  16.24640  &   63.19386 &   0.51796   &   DCEP\_1O   &      RRAB   &        ZTF  \\
  525971972931888256  &  16.61012  &   66.18293 &   0.33474   &   DCEP\_1O   &       RRC   &        ZTF  \\
 2454747674236106240  &  18.90559  &  $-$16.24635 &   7.54712   &      WVIR   &        EW   &   CATALINA  \\
  514299866732307584  &  37.15235  &   63.32395 &   0.25799   &   DCEP\_1O   &        EW   &        ZTF  \\
  249876593076412032  &  55.50083  &   50.70071 &   0.65550   &   DCEP\_1O   &      RRAB   &        ZTF  \\
 4858120560289808256  &  60.61599  &  $-$35.48554 &   0.95495   &     BLHER   &      RRAB   &   CATALINA  \\
 3286936002024112896  &  66.40184  &    7.48227 &   0.73534   &    ACEP\_F   &      RRAB   &   CATALINA  \\
  258545623786523904  &  69.12216  &   49.27242 &  26.17633   &     RVTAU   &        SR   &        ZTF  \\
 3239597250445418624  &  73.44271  &    4.91058 &   0.71103   &   ACEP\_1O   &      RRAB   &   CATALINA  \\
 \hline            
\end{tabular}
\tablefoot{The meaning of the variability types listed in column "Class\_Lit" can be found at the following address: http://cdsarc.u-strasbg.fr/viz-bin/getCatFile\_Redirect/?-plus=-\%2b\&B/gcvs/./vartype.txt. \\
The "Lit\_source" acronyms are: CATALINA=\citet{Drake2014,Drake2017,Torrealba2015}; OGLE=\citet{Sos2019}; ZTF=\citet{Chen2020}.\\
The entire version of the table will be published at CDS (Centre de Donn\'ees astronomiques de Strasbourg, https://cds.u-strasbg.fr/).}
\end{table*}

\subsection{Validation with TESS photometry}

For validation we used photometric data collected by the Transiting Exoplanet Survey Satellite \citep[TESS,][]{TESS-2015JATIS...1a4003R}. TESS is collecting continuous photometry over a large, $24\degr\times96\degr$ area with four cameras with adjacent fields-of-view over 27 day long segments called sectors. In mission years 1, 2 and 3, the field of view was rotated around the centre of camera 4, positioned towards the southern, then the northern and then again the southern ecliptic pole, while avoiding a 12-degree band along the Ecliptic. In year 4 five sectors were rotated so that all cameras were pointing towards the Ecliptic and observations cover a roughly 230\degr segment of it. We searched the full-frame image data up to Sector 43, which was the fourth sector in year 4 and the second along Ecliptic. Sampling cadence of the full-frame images was initially 30 min in years 1--2, which was lowered to 10 min in the first extended mission (years 3--4). 

The spatial resolution of TESS is limited to 21"/px. Therefore, although it is capable to reach the brighter Cepheids in the LMC and SMC, the images suffer from severe crowding and blending \citep{Plachy-2021ApJS..253...11P}. To avoid that, we only looked at Galactic Cepheids in this study. We cross-matched the \gaia\ coordinates with the sector coverage using the Web TESS Viewing Tool\footnote{\url{https://heasarc.gsfc.nasa.gov/cgi-bin/tess/webtess/wtv.py}} and then queried the TESS Quick Look Pipeline (QLP) database for light curves \citep{QLP-0-2021RNAAS...5..234K,QLP-I-2020RNAAS...4..204H,QLP-II-2020RNAAS...4..206H}. The pre-processed QLP light curves have a faint limit of $T = 15$~mag, equivalent to the same $G_{\rm RP}$ magnitude, and are produced primarily for exoplanet transit search. As a consequence, not all Cepheid candidates have good QLP light curves. Therefore we also extracted photometry from the full-frame images with the \texttt{eleanor} software, which is capable of both pixel aperture and PSF photometry and post-processing of the light curves via regression against a systematics model or via principal-component analysis \citep{eleanor-2019PASP..131i4502F}. We then selected the best light curves from the QLP and the four \texttt{eleanor} results (raw, corrected, PCA-corrected and PSF photometry), and applied further corrections: sigma-clipping to remove outliers and detrending, removing and residual slow variations. For the trend removal we used the method described by \citet{Bodi-2022PASP..134a4503B}. In brief the algorithm searches for the dominant periodicity in the light curve, computes the phase dispersion of the folded data and then fits a polynomial to the data by minimising against the phase dispersion. This way even high-order polynomials can be fitted that still follow the changes in average brightness and are much less affected by the effects of incomplete pulsation cycles at the edges. 

We then calculated the pulsation periods and $A_{i1}$ and $\phi_{i0}$ relative Fourier coefficients of the first few harmonics from the processed light curves and compared them to that of the OGLE \textit{I}-band measurements \citep{Sos2015,Sos2015b,Sos2018,Sos2019,Sos2021}. This validation only focused on the periods, light curve shapes and Fourier coefficients and we did not use positions on the $PL$ or $PW$ relations for classification here. If the software failed to calculate the Fourier coefficients, we only classified the star if we deemed the light curve shape conclusive enough through visual inspection. For the DCEP\_MULTI candidates we fitted all possible pulsation frequencies and calculated the frequency ratios. We also checked if significant secondary periodicities are present in the single-mode stars and calculated period ratios for any potential DCEP\_MULTI stars. Since TESS sectors are 27 d long, we were effectively limited to $<20$~d periods. For some long-period stars we were able to stitch data from consecutive sectors but this was limited to high and low ecliptic latitudes and was prone to brightness differences and other systematics.

Overall we searched for light curves for 4\,690 stars and were able to classify 2\,378 (51\%) of those. The validation results show strong agreement between the \gaia\ and TESS classifications. The largest discrepancy occurs among the 1O/2O DCEP\_MULTI stars, where we identified a significant number of further stars classified as single-mode DCEP\_1O in DR3. We also identified six stars as 1O/2O/3O DCEP\_MULTI pulsators. This subclass is not included in DR3 but is known among the OGLE Cepheids.

\begin{figure}
   \centering
      \includegraphics[width=\hsize]{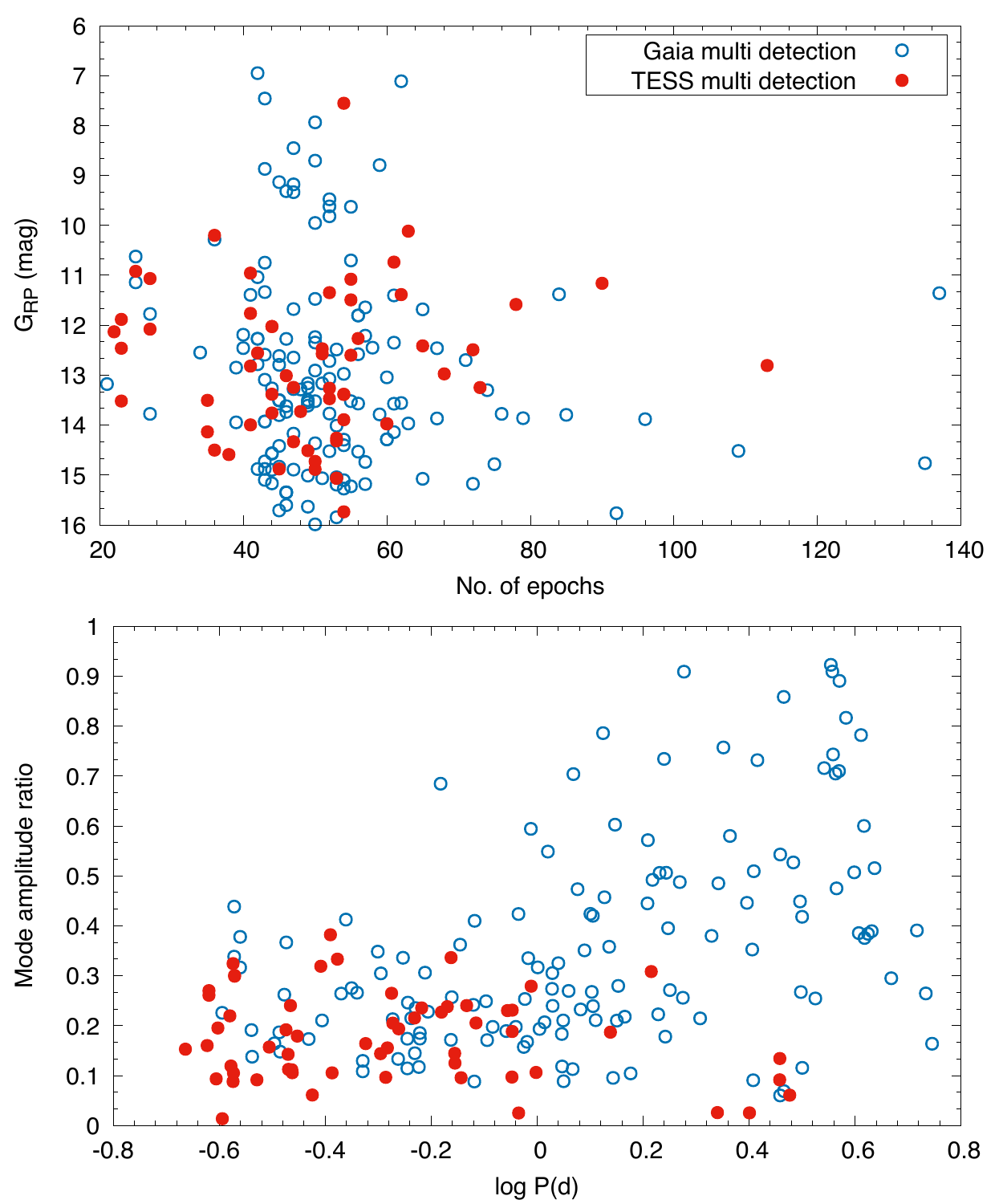}
            \caption{Comparison of the parameters of the multi-mode stars detected in \textit{Gaia} (blue circles) or from the TESS light curves (red dots). The upper plot compares the $G_\mathrm{RP}$ brightness (which is close to the TESS passband) and the number of photometric epochs available in DR3. The lower plot compares the amplitude ratio of the modes and the logarithm of the longer pulsation period.
            }
         \label{fig:gaia-tess-cep}
   \end{figure}

Finally, we investigated the possible reasons for missing a significant amount of DM Cepheids in the DR3 classification. Figure~\ref{fig:gaia-tess-cep} displays four diagnostic quantities: the upper panel shows the brightness of the stars (in $G_\mathrm{RP}$ band) against the number of epochs in the light curves; the lower panel shows the amplitude ratio of the modes (calculated from the Fourier amplitudes of the pulsation frequencies) against the logarithm of the periods. The plots indicate that the number of epochs and brightness had little effect on the detection with the brightest or most well sampled stars having the highest positive detection rate in \textit{Gaia}. The main driver for detection success appears to be the mode amplitude ratio, with all stars above 40\% identified from the \textit{Gaia} data. Longer-period DCEP\_MULTI stars also seem to be easier to discover from the sparse photometry collected by the mission. 

\subsection{Validation of RV data}

\begin{figure}
   \centering
   \includegraphics[width=\hsize]{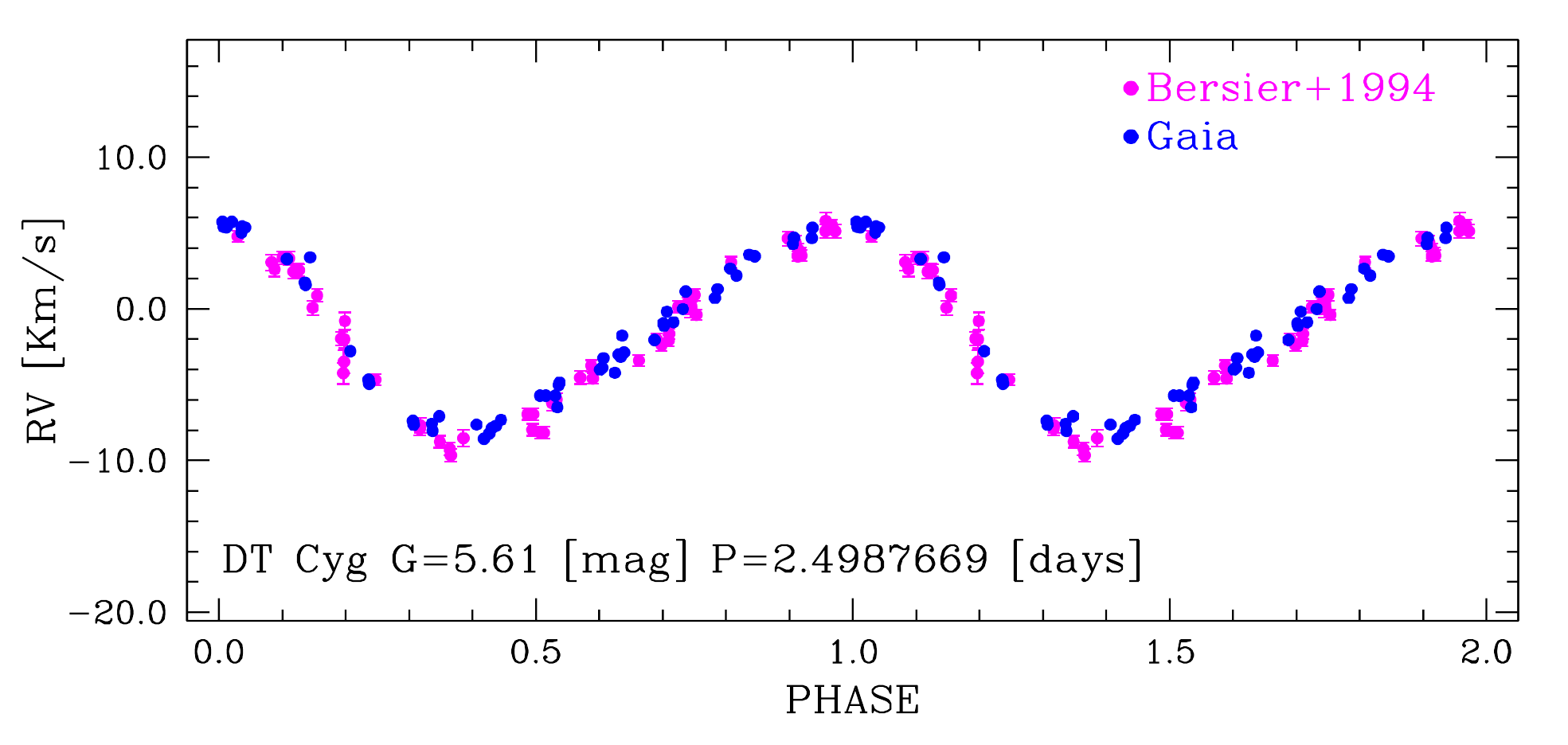}
   \includegraphics[width=\hsize]{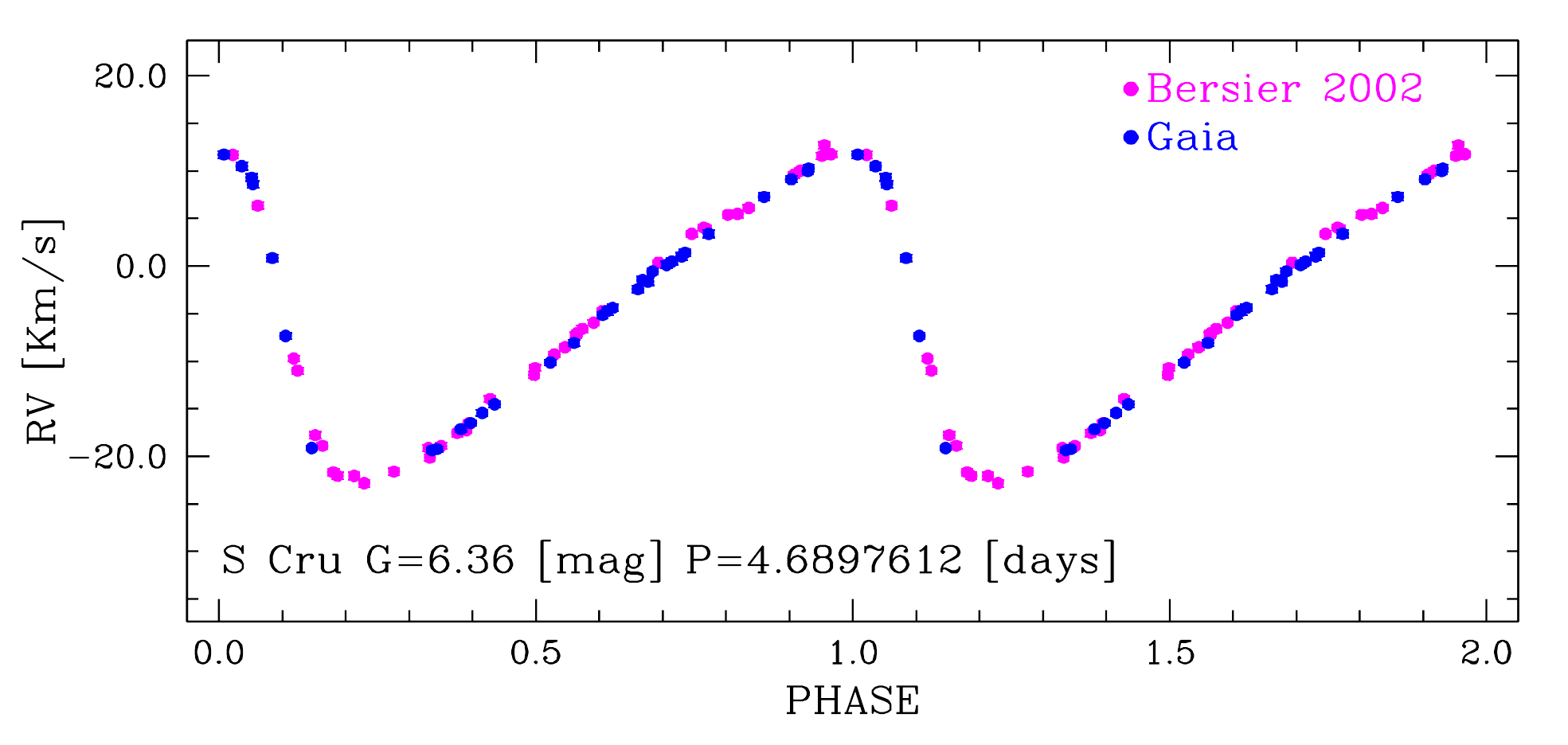}
      \caption{Examples of the comparison between the \gaia\ and the literature RV curves for a DCEP\_1O (DT\,Cyg) and a DCEP\_F (S\,Cru)}
         \label{fig:rv_comparison}
   \end{figure}

It is important to validate the RV curves of Cepheids published in \gaia\ DR3, as they have important applications, especially in the case of DCEPs. For example, the Baade-Wesselink method is widely used to estimate the radius and the distance of radial pulsators of different types and Cepheids in particular, from the combination of light and RV curves \citep[e.g.][and references therein]{Wesselink1946,Gautschy1987,Ripepi1997,Gieren2018}. We searched the literature for Cepheids with complete and reliable RV curves. As a result, we considered fourteen DCEPs whose complete properties are listed in Table~\ref{tab:rv_validation}. The comparison between the centre-of-mass velocity estimated from the \gaia\ RVs and those from the literature shows a good agreement within 1-2$\sigma$. The only discrepant object is V340\,Nor for which the \sos\ pipeline uncertainty is perhaps underestimated. To summarise, the Cepheid RV time-series released with \gaia\ DR3 are reliable and can be used to derive the star's intrinsic parameters.     
Examples of the comparison between \gaia\ and literature RVs are shown in Fig.~\ref{fig:rv_comparison}.

\begin{sidewaystable*}
  \begin{tiny}
    \setlength\tabcolsep{3pt}
\caption{Data for the validation of 14 DCEPs with high accuracy RV
  curves  available in the literature.}
\smallskip
\begin{tabular}{clccccccccccccccccc}
\hline\hline             
\noalign{\smallskip} 
      Source\_id  &   ~~Name & Mode & Period  & $\langle G \rangle$& N$_{\rm  RVS}$ & Amp(RV)& $\gamma_{\rm RVS}$&  $\gamma_{\rm Lit.}$ &
${\rm {[Fe/H]_{Lit.}}}$ & $\langle T_{\rm eff}\rangle_{\rm Lit.}$  & $\langle log g \rangle_{\rm Lit.} $  &  V$_t$ & Bin.  & P$_{\rm orbit}$  &Ref1&Ref2&Ref3& Ins. \\ 
 & & & (days) & (mag) & & (km $s^{-1}$) & (km $s^{-1}$) & (km $s^{-1}$) & (dex) & (K) & (km $s^{-1}$) & & & (days) & && & \\ 
(1) &~~~~(2) &(3) &~~~~~(4) &(5)~~&(6) &(7)~~~&(8)~ &(9)~ &(10)~~ &(11) &(12) &(13) &(14) &(15)&(16)&(17)&(18)&(19)\\
\noalign{\smallskip} \hline \noalign{\smallskip}
 5519380081746387328  &  AH Vel    &  1O & 4.227132   &   5.53   &   16  &   16.4$\pm$0.4   &     26.34$\pm$0.10  &      24.4$\pm$0.5 &     0.09  &   6\,040 &   2.2  &  4.3  &   B  &    $>$1\,000             &   1,2    &   8  &  12    &    a,b\\
5597379741549105280  &  AQ Pup    &  F  & 30.18194   &   8.12   &   24  &   56.7$\pm$3.6   &     65.45$\pm$1.47  &      61.0$\pm$0.8 &  $-$0.14  &   4\,940 &   0.6  &  5.6  &   B  &    $\sim$1\,400          &   16,17  &  11  &   -    &    b\\   
5848500161483878400  &  AV Cir    &  1O & 3.06526    &   7.10   &   31  &   14.3$\pm$0.2   &      5.11$\pm$0.04  &       4.8$\pm$0.4 &     0.14  &   6\,170 &   2.1  &  3.3  &   V  &      -                   &    2     &   8  &   -    &     a\\
5873984023533350400  &  AX Cir    &  F  & 5.27337    &   5.63   &   74  &   30.2$\pm$0.9   &  $-$15.78$\pm$0.28  &   $-$14.7$\pm$0.7 &  $-$0.04  &   5\,760 &   1.8  &  3.7  &   OV &    6\,532$\pm$35         &    3     &   8  &  13    &    b\\
6058439910929477120  &  BG Cru    &  1O & 3.3425241  &   3.34   &   23  &    9.9$\pm$0.1   &  $-$19.42$\pm$0.07  &   $-$19.7$\pm$0.5 &  $-$0.11  &   6\,250 &   2.1  &  4.3  &   B  &    4\,050,4\,950,6\,650  &   1,2    &   9  &  14    &    a,b\\
5877460679352962048  &  BP Cir    &  1O & 2.398106   &   7.29   &   38  &   16.7$\pm$0.2   &  $-$18.18$\pm$0.05  &   $-$18.0$\pm$0.5 &  $-$0.01  &   6\,530 &   2.4  &  3.7  &   B  &     -                    &   1,3    &   8  &   -    &    b \\
2011892703004353792  &  CF Cas    &  F  & 4.875122   &  10.73   &   12  &   36.8$\pm$10.6  &  $-$71.87$\pm$3.64  &   $-$78.4$\pm$1.1 &  $-$0.01  &   5\,510 &   1.7  &  4.0  &   -  &     -                    &     4    &  10  &   -    &   a\\
1853025642297186688  &  DT Cyg    &  1O & 2.498763   &   5.61   &   57  &   13.8$\pm$0.1   &   $-$1.04$\pm$0.03  &    $-$1.9$\pm$0.6 &     0.16  &   6\,270 &   2.4  &  3.6  &   -  &     -                    &     4    &  10  &   -    &   a\\
5546476927338700416  &  RS Pup    &  F  & 41.49233   &   6.46   &   25  &   50.8$\pm$8.3   &     25.80$\pm$1.58  &      25.0$\pm$0.8 &     0.21  &   5\,070 &   1.0  &  5.0  &   -  &     -                    &   5,6    &  11  &   -    &   b\\
3409635486731094400  &  SZ Tau    &  1O & 3.148786   &   6.23   &   20  &   20.7$\pm$0.4   &      0.27$\pm$0.12  &   $-$0.61$\pm$0.5 &     0.15  &   5\,990 &   2.2  &  3.6  &   B  &     1\,244               &     4    &  10  &   15   &   a\\
6060173364074372352  &  S Cru     &  F  & 4.689765   &   6.36   &   31  &   44.9$\pm$3.0   &   $-$6.58$\pm$0.40  &    $-$5.1$\pm$0.5 &     0.08  &   6\,460 &   2.1  &  4.1  &   -  &      -                   &     7    &  11  &    -   &    a\\
6054829806275577216  &  T Cru     &  F  & 6.73324    &   6.38   &   23  &   28.8$\pm$0.8   &   $-$5.94$\pm$0.44  &    $-$9.6$\pm$0.5 &     0.11  &   5\,590 &   1.7  &  4.3  &   B  &      -                   &     7    &   9  &    -   &    a\\
5932569709575669504  &  V340 Nor  &  F  & 11.28895   &   7.98   &   37  &   18.8$\pm$0.3   &  $-$38.89$\pm$0.07  &   $-$30.3$\pm$0.6 &     0.16  &   5\,730 &   2.0  &  5.3  &   -  &      -                   &     4    &   8  &    -   &   a\\
2027263738133623168  &  X Vul     &  F  & 6.3196418  &   8.23   &   35  &   42.3$\pm$5.2   &  $-$14.94$\pm$0.94  &   $-$15.7$\pm$0.6 &     0.13  &   5\,650 &   1.7  &  3.9  &   -  &     -                    &     4    &  10  &    -   &   a\\
\noalign{\smallskip} \hline \noalign{\smallskip}
\end{tabular}
\tablefoot{Meaning  of the different columns is as follows: (1) Gaia DR3 source id;
  (2) Literature Name; (3) Mode of pulsation (for brevity F=Fundamental; 1O= First
Overtone); (4) Pulsation period (P), as re-evaluated in the present
analysis; (5) Intensity-averaged $G$-band mean magnitude, as derived
from the \sos\ pipeline; (6)
number of valid RVS RV measurements;  (7) Peak-to-peak amplitude of
the RVS RV curve and relative uncertainty; (8) Center of mass RV
($\gamma$) as estimated by \sos\ pipeline and uncertainty; (9) as for
the previous column but for the literature; (10) Iron abundance; (11) Mean effective temperature; (12) Mean
gravity; (13) Microturbulent velocity; (14) Binary 
type; (15) Orbital period (P$_{\rm orbit}$); (16) References for the
literature RVs; (17) References for the stellar parameters; (18)  References for  P$_{\rm orbit}$; (19) Instrument type: a=CORAVEL; b=Other spectrograph. 
Metallicities in col. (10) are taken from \citet{Genovali2014}.   
The meaning of the numbers in columns (16), (17) and (18)  is as follows: 
1=\citet{Gallenne2019};
2=\citet{Kienzle1999};
3=\citet{Petterson2005};
4=\citet{Bersier1994};
5=\citet{Anderson2014};
6=\citet{Storm2004};
7=\citet{Bersier2002};
8=\citet{Luck2011};
9=\citet{Usenko2014};
10=\citet{Andrievsky2002};
11=\citet{Andrievsky2013};
12=\citet{Gieren1977};
13=\citet{Petterson2004};
14=\citet{Szabados1989};
15=\citet{Gorynya1996};
16=\citet{Anderson2016};
17=\citet{Storm2011}.
The binary types listed in col. (14) are taken from the web site
https://konkoly.hu/CEP/nagytab3.html maintained by L. Szabados. The
different symbols mean: B=spectroscopic binary; O=spectroscopic binary with known orbit;
V=visual binary.}
\label{tab:rv_validation}
\end{tiny}
\end{sidewaystable*}

\section{Conclusions}

In this paper we have presented the \gaia\ DR3 catalogue of Cepheids of all types. We have discussed the changes in the \sos\ pipeline with respect to DR2, including the derivation of a full set of $PL$ and $PW$ relations adopted in the pipeline. The major novelties in DR3 compared to the previous release is the analysis of DCEPs in the distant galaxies M31 and M33, and the analysis of the RV data for a sub-sample of 799 Cepheids of all types, including 24 objects belonging to LMC and SMC.  

We have described the techniques adopted to carry out a first gross cleaning of the sample from the large amount of spurious objects retrieved from the general classification catalogue. In this process we made also use of machine learning techniques which helped significantly to single out the most promising candidates for further analysis. 

To obtain the maximum purity in the sample, we analysed visually almost all the candidates and corrected the classification provided from the \gaia\ \sos\ pipeline when it was wrong. In this context, the $G$ time-series of a number of suspect multi-mode pulsators were re-processed to determine correct pulsation periods. 

In total, the \gaia\ DR3 catalogue counts 
15,006 Cepheids of all types, among which 327 objects were known variable stars with a different classification in the literature, while, to our knowledge, 474 stars either had not been reported previously or had non-Cepheid type classifciation in the literature, and therefore they are likely new Cepheid discoveries by {\it Gaia}. 

The validation of the DR3 catalogue was carried out via  comparison with literature results and through analysis of a consistent sample of light curves from TESS. The overall purity of the sample is very high and certainly larger than 90-95\%. The completeness varies significantly as a function of region in the sky and Cepheid type. It is overall larger than 90\% in LMC and SMC, while is on the order of 10-20\% in M31 and M33. Concerning the All-Sky sample, largely dominated by MW objects, the completeness for DCEPs is likely between 85\% and 90\%, with a few \% contamination. The completeness is lower for ACEPs and especially T2CEPs, which are located in large numbers in the MW Bulge, a region for which \gaia\ has not collected a sufficient number of epoch data yet. 

The validation of the  RV curves with literature data showed that the \gaia\ RV curves for Cepheids are generally accurate and usable for astrophysical purposes.

Compared to DR2, the Cepheids in DR3 represent a huge improvement both quantitatively, given the addition of $\sim$5\,000 Cepheids of all types, and qualitatively, as the DR3 Cepheid catalogue has a much improved purity, especially for the All-Sky sample. In addition a significant benefit of DR3 is represented by the release of RV time-series for 799 Cepheids of all types.

The following release (DR4) will present further improvements compared to DR3, mainly due to the additional 24 months of data leading, in turn, to more accurate period determinations. For the next release we plan to use thoroughly the machine learning technique that were implemented to clean the DR3 sample. In this respect, the present \gaia\ DR3 Cepheid sample, with its high purity will represent an excellent training set.

\begin{acknowledgements}

This work has made use of data from the European Space Agency (ESA) mission
{\it Gaia} (\url{https://www.cosmos.esa.int/gaia}), processed by the {\it Gaia}
Data Processing and Analysis Consortium (DPAC,
\url{https://www.cosmos.esa.int/web/gaia/dpac/consortium}).
Funding for the DPAC has been provided by national institutions, in particular the institutions participating in the {\it Gaia} Multilateral Agreement.
The Italian participation in DPAC has been supported by Istituto Nazionale di
Astrofisica (INAF) and the Agenzia Spaziale Italiana (ASI) through grants
I/037/08/0,  I/058/10/0,  2014-025-R.0, 2014-025-R.1.2015 and  2018-24-HH.0 to INAF (PI M.G. Lattanzi). \\

The Swiss participation by the Swiss State Secretariat for Education, Research
and Innovation  through the ``Activit\'{e}s Nationales Compl\'{e}mentaires''.\\

UK community participation in this work has been supported by funding
from the UK Space Agency, and from the UK Science and Technology
Research Council.\\

This work was supported in part by the French Centre National de la Recherche Scientifique (CNRS), the Centre National d'Etudes Spatiales (CNES), the Institut des Sciences de l' Univers (INSU) through the Service National d'Observation (SNO) Gaia. \\

This research was supported by the `SeismoLab' KKP-137523 \'Elvonal grant of the Hungarian Research, Development and Innovation Office (NKFIH), and by the LP2018-7 Lend\"ulet grant of the Hungarian Academy of Sciences.\\

This research has made use of the SIMBAD database, operated at CDS, Strasbourg, France.\\

It is a pleasure to thank M.B. Taylor for developing the {\tt TOPCAT} software, which was very useful in carrying out this work. \\

\end{acknowledgements}

%
%

\bibliographystyle{aa} 
\bibliography{myBib} 

\begin{thebibliography}{103}
\expandafter\ifx\csname natexlab\endcsname\relax\def\natexlab#1{#1}\fi

\bibitem[{{Anderson}(2014)}]{Anderson2014}
{Anderson}, R.~I. 2014, \aap, 566, L10

\bibitem[{{Anderson} {et~al.}(2016){Anderson}, {Casertano}, {Riess}, {Melis},
  {Holl}, {Semaan}, {Papics}, {Blanco-Cuaresma}, {Eyer}, {Mowlavi},
  {Palaversa}, \& {Roelens}}]{Anderson2016}
{Anderson}, R.~I., {Casertano}, S., {Riess}, A.~G., {et~al.} 2016, \apjs, 226,
  18

\bibitem[{{Andrievsky} {et~al.}(2002){Andrievsky}, {Kovtyukh}, {Luck},
  {L{\'e}pine}, {Maciel}, \& {Beletsky}}]{Andrievsky2002}
{Andrievsky}, S.~M., {Kovtyukh}, V.~V., {Luck}, R.~E., {et~al.} 2002, \aap,
  392, 491

\bibitem[{{Andrievsky} {et~al.}(2013){Andrievsky}, {L{\'e}pine}, {Korotin},
  {Luck}, {Kovtyukh}, \& {Maciel}}]{Andrievsky2013}
{Andrievsky}, S.~M., {L{\'e}pine}, J.~R.~D., {Korotin}, S.~A., {et~al.} 2013,
  \mnras, 428, 3252

\bibitem[{{Arenou} \& {Luri}(1999)}]{Arenou1999}
{Arenou}, F. \& {Luri}, X. 1999, in Astronomical Society of the Pacific
  Conference Series, Vol. 167, Harmonizing Cosmic Distance Scales in a
  Post-HIPPARCOS Era, ed. D.~{Egret} \& A.~{Heck}, 13--32

\bibitem[{{Bersier}(2002)}]{Bersier2002}
{Bersier}, D. 2002, \apjs, 140, 465

\bibitem[{{Bersier} {et~al.}(1994){Bersier}, {Burki}, {Mayor}, \&
  {Duquennoy}}]{Bersier1994}
{Bersier}, D., {Burki}, G., {Mayor}, M., \& {Duquennoy}, A. 1994, \aaps, 108,
  25

\bibitem[{{B{\'o}di} {et~al.}(2022){B{\'o}di}, {Szab{\'o}}, {Plachy},
  {Moln{\'a}r}, \& {Szab{\'o}}}]{Bodi-2022PASP..134a4503B}
{B{\'o}di}, A., {Szab{\'o}}, P., {Plachy}, E., {Moln{\'a}r}, L., \&
  {Szab{\'o}}, R. 2022, \pasp, 134, 014503

\bibitem[{{Cantat-Gaudin} {et~al.}(2020){Cantat-Gaudin}, {Anders},
  {Castro-Ginard}, {Jordi}, {Romero-G{\'o}mez}, {Soubiran}, {Casamiquela},
  {Tarricq}, {Moitinho}, {Vallenari}, {Bragaglia}, {Krone-Martins}, \&
  {Kounkel}}]{CantatGaudin2020}
{Cantat-Gaudin}, T., {Anders}, F., {Castro-Ginard}, A., {et~al.} 2020, \aap,
  640, A1

\bibitem[{{Cappellari} {et~al.}(2013){Cappellari}, {Scott}, {Alatalo}, {Blitz},
  {Bois}, {Bournaud}, {Bureau}, {Crocker}, {Davies}, {Davis}, {de Zeeuw},
  {Duc}, {Emsellem}, {Khochfar}, {Krajnovi{\'c}}, {Kuntschner}, {McDermid},
  {Morganti}, {Naab}, {Oosterloo}, {Sarzi}, {Serra}, {Weijmans}, \&
  {Young}}]{Cappellari2013}
{Cappellari}, M., {Scott}, N., {Alatalo}, K., {et~al.} 2013, \mnras, 432, 1709

\bibitem[{{Caputo}(1998)}]{Caputo1998}
{Caputo}, F. 1998, \aapr, 9, 33

\bibitem[{{Caputo} {et~al.}(2004){Caputo}, {Castellani}, {Degl'Innocenti},
  {Fiorentino}, \& {Marconi}}]{Caputo2004}
{Caputo}, F., {Castellani}, V., {Degl'Innocenti}, S., {Fiorentino}, G., \&
  {Marconi}, M. 2004, \aap, 424, 927

\bibitem[{{Caputo} {et~al.}(2000){Caputo}, {Marconi}, {Musella}, \&
  {Santolamazza}}]{Caputo2000}
{Caputo}, F., {Marconi}, M., {Musella}, I., \& {Santolamazza}, P. 2000, \aap,
  359, 1059

\bibitem[{{Castro-Ginard} {et~al.}(2022){Castro-Ginard}, {Jordi}, {Luri},
  {Cantat-Gaudin}, {Carrasco}, {Casamiquela}, {Anders},
  {Balaguer-N{\'u}{\~n}ez}, \& {Badia}}]{CastroGinard2021}
{Castro-Ginard}, A., {Jordi}, C., {Luri}, X., {et~al.} 2022, \aap, 661, A118

\bibitem[{{Chen} {et~al.}(2020){Chen}, {Wang}, {Deng}, {de Grijs}, {Yang}, \&
  {Tian}}]{Chen2020}
{Chen}, X., {Wang}, S., {Deng}, L., {et~al.} 2020, \apjs, 249, 18

\bibitem[{{Clement} {et~al.}(2001){Clement}, {Muzzin}, {Dufton}, {Ponnampalam},
  {Wang}, {Burford}, {Richardson}, {Rosebery}, {Rowe}, \& {Hogg}}]{Clement2001}
{Clement}, C.~M., {Muzzin}, A., {Dufton}, Q., {et~al.} 2001, \aj, 122, 2587

\bibitem[{{Clementini} {et~al.}(2016){Clementini}, {Ripepi}, {Leccia},
  {Mowlavi}, {Lecoeur-Taibi}, {Marconi}, {Szabados}, {Eyer}, {Guy},
  {Rimoldini}, {Jevardat de Fombelle}, {Holl}, {Busso}, {Charnas}, {Cuypers},
  {De Angeli}, {De Ridder}, {Debosscher}, {Evans}, {Klagyivik}, {Musella},
  {Nienartowicz}, {Ord{\'o}{\~n}ez}, {Regibo}, {Riello}, {Sarro}, \&
  {S{\"u}veges}}]{Clementini2016}
{Clementini}, G., {Ripepi}, V., {Leccia}, S., {et~al.} 2016, \aap, 595, A133

\bibitem[{{Clementini} {et~al.}(2019){Clementini}, {Ripepi}, {Molinaro},
  {Garofalo}, {Muraveva}, {Rimoldini}, {Guy}, {Jevardat de Fombelle},
  {Nienartowicz}, {Marchal}, {Audard}, {Holl}, {Leccia}, {Marconi}, {Musella},
  {Mowlavi}, {Lecoeur-Taibi}, {Eyer}, {De Ridder}, {Regibo}, {Sarro},
  {Szabados}, {Evans}, \& {Riello}}]{Clementini2019}
{Clementini}, G., {Ripepi}, V., {Molinaro}, R., {et~al.} 2019, \aap, 622, A60

\bibitem[{{Clementini et al.}(2022)}]{DR3-DPACP-168}
{Clementini et al.} 2022, \aap\ submitted

\bibitem[{{Conn} {et~al.}(2012){Conn}, {Ibata}, {Lewis}, {Parker}, {Zucker},
  {Martin}, {McConnachie}, {Irwin}, {Tanvir}, {Fardal}, {Ferguson}, {Chapman},
  \& {Valls-Gabaud}}]{Conn2012}
{Conn}, A.~R., {Ibata}, R.~A., {Lewis}, G.~F., {et~al.} 2012, \apj, 758, 11

\bibitem[{{De Somma et al.}(2022)}]{DeSomma2022}
{De Somma et al.} 2022, \apjs\ submitted

\bibitem[{{Drake} {et~al.}(2017){Drake}, {Djorgovski}, {Catelan}, {Graham},
  {Mahabal}, {Larson}, {Christensen}, {Torrealba}, {Beshore}, {McNaught},
  {Garradd}, {Belokurov}, \& {Koposov}}]{Drake2017}
{Drake}, A.~J., {Djorgovski}, S.~G., {Catelan}, M., {et~al.} 2017, \mnras, 469,
  3688

\bibitem[{{Drake} {et~al.}(2014){Drake}, {Graham}, {Djorgovski}, {Catelan},
  {Mahabal}, {Torrealba}, {Garc{\'\i}a-{\'A}lvarez}, {Donalek}, {Prieto},
  {Williams}, {Larson}, {Christen sen}, {Belokurov}, {Koposov}, {Beshore},
  {Boattini}, {Gibbs}, {Hill}, {Kowalski}, {Johnson}, \& {Shelly}}]{Drake2014}
{Drake}, A.~J., {Graham}, M.~J., {Djorgovski}, S.~G., {et~al.} 2014, \apjs,
  213, 9

\bibitem[{{Eyer et al.}(2022)}]{DR3-DPACP-162}
{Eyer et al.} 2022, \aap\ in prep.

\bibitem[{{Feast} \& {Catchpole}(1997)}]{Feast1997}
{Feast}, M.~W. \& {Catchpole}, R.~M. 1997, \mnras, 286, L1

\bibitem[{{Feast} {et~al.}(2008){Feast}, {Laney}, {Kinman}, {van Leeuwen}, \&
  {Whitelock}}]{Feast2008}
{Feast}, M.~W., {Laney}, C.~D., {Kinman}, T.~D., {van Leeuwen}, F., \&
  {Whitelock}, P.~A. 2008, \mnras, 386, 2115

\bibitem[{{Feinstein} {et~al.}(2019){Feinstein}, {Montet}, {Foreman-Mackey},
  {Bedell}, {Saunders}, {Bean}, {Christiansen}, {Hedges}, {Luger}, {Scolnic},
  \& {Cardoso}}]{eleanor-2019PASP..131i4502F}
{Feinstein}, A.~D., {Montet}, B.~T., {Foreman-Mackey}, D., {et~al.} 2019,
  \pasp, 131, 094502

\bibitem[{{Gaia Collaboration} {et~al.}(2018){Gaia Collaboration}, {Brown},
  {Vallenari}, {Prusti}, {de Bruijne}, {Babusiaux}, {Bailer-Jones}, {Biermann},
  {Evans}, {Eyer}, {Jansen}, {Jordi}, {Klioner}, {Lammers}, {Lindegren},
  {Luri}, {Mignard}, {Panem}, {Pourbaix}, {Randich}, {Sartoretti}, {Siddiqui},
  {Soubiran}, {van Leeuwen}, {Walton}, {Arenou}, {Bastian}, {Cropper},
  {Drimmel}, {Katz}, {Lattanzi}, {Bakker}, {Cacciari}, {Casta{\~n}eda},
  {Chaoul}, {Cheek}, {De Angeli}, {Fabricius}, {Guerra}, {Holl}, {Masana},
  {Messineo}, {Mowlavi}, {Nienartowicz}, {Panuzzo}, {Portell}, {Riello},
  {Seabroke}, {Tanga}, {Th{\'e}venin}, {Gracia-Abril}, {Comoretto},
  {Garcia-Reinaldos}, {Teyssier}, {Altmann}, {Andrae}, {Audard},
  {Bellas-Velidis}, {Benson}, {Berthier}, {Blomme}, {Burgess}, {Busso},
  {Carry}, {Cellino}, {Clementini}, {Clotet}, {Creevey}, {Davidson}, {De
  Ridder}, {Delchambre}, {Dell'Oro}, {Ducourant},
  {Fern{\'a}ndez-Hern{\'a}ndez}, {Fouesneau}, {Fr{\'e}mat}, {Galluccio},
  {Garc{\'\i}a-Torres}, {Gonz{\'a}lez-N{\'u}{\~n}ez}, {Gonz{\'a}lez-Vidal},
  {Gosset}, {Guy}, {Halbwachs}, {Hambly}, {Harrison}, {Hern{\'a}ndez},
  {Hestroffer}, {Hodgkin}, {Hutton}, {Jasniewicz}, {Jean-Antoine-Piccolo},
  {Jordan}, {Korn}, {Krone-Martins}, {Lanzafame}, {Lebzelter}, {L{\"o}ffler},
  {Manteiga}, {Marrese}, {Mart{\'\i}n-Fleitas}, {Moitinho}, {Mora}, {Muinonen},
  {Osinde}, {Pancino}, {Pauwels}, {Petit}, {Recio-Blanco}, {Richards},
  {Rimoldini}, {Robin}, {Sarro}, {Siopis}, {Smith}, {Sozzetti}, {S{\"u}veges},
  {Torra}, {van Reeven}, {Abbas}, {Abreu Aramburu}, {Accart}, {Aerts},
  {Altavilla}, {{\'A}lvarez}, {Alvarez}, {Alves}, {Anderson}, {Andrei},
  {Anglada Varela}, {Antiche}, {Antoja}, {Arcay}, {Astraatmadja}, {Bach},
  {Baker}, {Balaguer-N{\'u}{\~n}ez}, {Balm}, {Barache}, {Barata}, {Barbato},
  {Barblan}, {Barklem}, {Barrado}, {Barros}, {Barstow}, {Bartholom{\'e}
  Mu{\~n}oz}, {Bassilana}, {Becciani}, {Bellazzini}, {Berihuete}, {Bertone},
  {Bianchi}, {Bienaym{\'e}}, {Blanco-Cuaresma}, {Boch}, {Boeche}, {Bombrun},
  {Borrachero}, {Bossini}, {Bouquillon}, {Bourda}, {Bragaglia}, {Bramante},
  {Breddels}, {Bressan}, {Brouillet}, {Br{\"u}semeister}, {Brugaletta},
  {Bucciarelli}, {Burlacu}, {Busonero}, {Butkevich}, {Buzzi}, {Caffau},
  {Cancelliere}, {Cannizzaro}, {Cantat-Gaudin}, {Carballo}, {Carlucci},
  {Carrasco}, {Casamiquela}, {Castellani}, {Castro-Ginard}, {Charlot},
  {Chemin}, {Chiavassa}, {Cocozza}, {Costigan}, {Cowell}, {Crifo}, {Crosta},
  {Crowley}, {Cuypers}, {Dafonte}, {Damerdji}, {Dapergolas}, {David}, {David},
  {de Laverny}, {De Luise}, {De March}, {de Martino}, {de Souza}, {de Torres},
  {Debosscher}, {del Pozo}, {Delbo}, {Delgado}, {Delgado}, {Di Matteo},
  {Diakite}, {Diener}, {Distefano}, {Dolding}, {Drazinos}, {Dur{\'a}n},
  {Edvardsson}, {Enke}, {Eriksson}, {Esquej}, {Eynard Bontemps}, {Fabre},
  {Fabrizio}, {Faigler}, {Falc{\~a}o}, {Farr{\`a}s Casas}, {Federici},
  {Fedorets}, {Fernique}, {Figueras}, {Filippi}, {Findeisen}, {Fonti},
  {Fraile}, {Fraser}, {Fr{\'e}zouls}, {Gai}, {Galleti}, {Garabato},
  {Garc{\'\i}a-Sedano}, {Garofalo}, {Garralda}, {Gavel}, {Gavras}, {Gerssen},
  {Geyer}, {Giacobbe}, {Gilmore}, {Girona}, {Giuffrida}, {Glass}, {Gomes},
  {Granvik}, {Gueguen}, {Guerrier}, {Guiraud}, {Guti{\'e}rrez-S{\'a}nchez},
  {Haigron}, {Hatzidimitriou}, {Hauser}, {Haywood}, {Heiter}, {Helmi}, {Heu},
  {Hilger}, {Hobbs}, {Hofmann}, {Holland}, {Huckle}, {Hypki}, {Icardi},
  {Jan{\ss}en}, {Jevardat de Fombelle}, {Jonker}, {Juh{\'a}sz}, {Julbe},
  {Karampelas}, {Kewley}, {Klar}, {Kochoska}, {Kohley}, {Kolenberg},
  {Kontizas}, {Kontizas}, {Koposov}, {Kordopatis}, {Kostrzewa-Rutkowska},
  {Koubsky}, {Lambert}, {Lanza}, {Lasne}, {Lavigne}, {Le Fustec}, {Le
  Poncin-Lafitte}, {Lebreton}, {Leccia}, {Leclerc}, {Lecoeur-Taibi},
  {Lenhardt}, {Leroux}, {Liao}, {Licata}, {Lindstr{\o}m}, {Lister}, {Livanou},
  {Lobel}, {L{\'o}pez}, {Managau}, {Mann}, {Mantelet}, {Marchal}, {Marchant},
  {Marconi}, {Marinoni}, {Marschalk{\'o}}, {Marshall}, {Martino}, {Marton},
  {Mary}, {Massari}, {Matijevi{\v{c}}}, {Mazeh}, {McMillan}, {Messina},
  {Michalik}, {Millar}, {Molina}, {Molinaro}, {Moln{\'a}r}, {Montegriffo},
  {Mor}, {Morbidelli}, {Morel}, {Morris}, {Mulone}, {Muraveva}, {Musella},
  {Nelemans}, {Nicastro}, {Noval}, {O'Mullane}, {Ord{\'e}novic},
  {Ord{\'o}{\~n}ez-Blanco}, {Osborne}, {Pagani}, {Pagano}, {Pailler},
  {Palacin}, {Palaversa}, {Panahi}, {Pawlak}, {Piersimoni}, {Pineau}, {Plachy},
  {Plum}, {Poggio}, {Poujoulet}, {Pr{\v{s}}a}, {Pulone}, {Racero}, {Ragaini},
  {Rambaux}, {Ramos-Lerate}, {Regibo}, {Reyl{\'e}}, {Riclet}, {Ripepi}, {Riva},
  {Rivard}, {Rixon}, {Roegiers}, {Roelens}, {Romero-G{\'o}mez}, {Rowell},
  {Royer}, {Ruiz-Dern}, {Sadowski}, {Sagrist{\`a} Sell{\'e}s}, {Sahlmann},
  {Salgado}, {Salguero}, {Sanna}, {Santana-Ros}, {Sarasso}, {Savietto},
  {Schultheis}, {Sciacca}, {Segol}, {Segovia}, {S{\'e}gransan}, {Shih},
  {Siltala}, {Silva}, {Smart}, {Smith}, {Solano}, {Solitro}, {Sordo}, {Soria
  Nieto}, {Souchay}, {Spagna}, {Spoto}, {Stampa}, {Steele},
  {Steidelm{\"u}ller}, {Stephenson}, {Stoev}, {Suess}, {Surdej}, {Szabados},
  {Szegedi-Elek}, {Tapiador}, {Taris}, {Tauran}, {Taylor}, {Teixeira},
  {Terrett}, {Teyssandier}, {Thuillot}, {Titarenko}, {Torra Clotet}, {Turon},
  {Ulla}, {Utrilla}, {Uzzi}, {Vaillant}, {Valentini}, {Valette}, {van Elteren},
  {Van Hemelryck}, {van Leeuwen}, {Vaschetto}, {Vecchiato}, {Veljanoski},
  {Viala}, {Vicente}, {Vogt}, {von Essen}, {Voss}, {Votruba}, {Voutsinas},
  {Walmsley}, {Weiler}, {Wertz}, {Wevers}, {Wyrzykowski}, {Yoldas},
  {{\v{Z}}erjal}, {Ziaeepour}, {Zorec}, {Zschocke}, {Zucker}, {Zurbach}, \&
  {Zwitter}}]{Gaia2018}
{Gaia Collaboration}, {Brown}, A.~G.~A., {Vallenari}, A., {et~al.} 2018, \aap,
  616, A1

\bibitem[{{Gaia Collaboration} {et~al.}(2021{\natexlab{a}}){Gaia
  Collaboration}, {Brown}, {Vallenari}, {Prusti}, {de Bruijne}, {Babusiaux},
  {Biermann}, {Creevey}, {Evans}, {Eyer}, {Hutton}, {Jansen}, {Jordi},
  {Klioner}, {Lammers}, {Lindegren}, {Luri}, {Mignard}, {Panem}, {Pourbaix},
  {Randich}, {Sartoretti}, {Soubiran}, {Walton}, {Arenou}, {Bailer-Jones},
  {Bastian}, {Cropper}, {Drimmel}, {Katz}, {Lattanzi}, {van Leeuwen}, {Bakker},
  {Cacciari}, {Casta{\~n}eda}, {De Angeli}, {Ducourant}, {Fabricius},
  {Fouesneau}, {Fr{\'e}mat}, {Guerra}, {Guerrier}, {Guiraud}, {Jean-Antoine
  Piccolo}, {Masana}, {Messineo}, {Mowlavi}, {Nicolas}, {Nienartowicz},
  {Pailler}, {Panuzzo}, {Riclet}, {Roux}, {Seabroke}, {Sordo}, {Tanga},
  {Th{\'e}venin}, {Gracia-Abril}, {Portell}, {Teyssier}, {Altmann}, {Andrae},
  {Bellas-Velidis}, {Benson}, {Berthier}, {Blomme}, {Brugaletta}, {Burgess},
  {Busso}, {Carry}, {Cellino}, {Cheek}, {Clementini}, {Damerdji}, {Davidson},
  {Delchambre}, {Dell'Oro}, {Fern{\'a}ndez-Hern{\'a}ndez}, {Galluccio},
  {Garc{\'\i}a-Lario}, {Garcia-Reinaldos}, {Gonz{\'a}lez-N{\'u}{\~n}ez},
  {Gosset}, {Haigron}, {Halbwachs}, {Hambly}, {Harrison}, {Hatzidimitriou},
  {Heiter}, {Hern{\'a}ndez}, {Hestroffer}, {Hodgkin}, {Holl}, {Jan{\ss}en},
  {Jevardat de Fombelle}, {Jordan}, {Krone-Martins}, {Lanzafame},
  {L{\"o}ffler}, {Lorca}, {Manteiga}, {Marchal}, {Marrese}, {Moitinho}, {Mora},
  {Muinonen}, {Osborne}, {Pancino}, {Pauwels}, {Petit}, {Recio-Blanco},
  {Richards}, {Riello}, {Rimoldini}, {Robin}, {Roegiers}, {Rybizki}, {Sarro},
  {Siopis}, {Smith}, {Sozzetti}, {Ulla}, {Utrilla}, {van Leeuwen}, {van
  Reeven}, {Abbas}, {Abreu Aramburu}, {Accart}, {Aerts}, {Aguado}, {Ajaj},
  {Altavilla}, {{\'A}lvarez}, {{\'A}lvarez Cid-Fuentes}, {Alves}, {Anderson},
  {Anglada Varela}, {Antoja}, {Audard}, {Baines}, {Baker},
  {Balaguer-N{\'u}{\~n}ez}, {Balbinot}, {Balog}, {Barache}, {Barbato},
  {Barros}, {Barstow}, {Bartolom{\'e}}, {Bassilana}, {Bauchet},
  {Baudesson-Stella}, {Becciani}, {Bellazzini}, {Bernet}, {Bertone}, {Bianchi},
  {Blanco-Cuaresma}, {Boch}, {Bombrun}, {Bossini}, {Bouquillon}, {Bragaglia},
  {Bramante}, {Breedt}, {Bressan}, {Brouillet}, {Bucciarelli}, {Burlacu},
  {Busonero}, {Butkevich}, {Buzzi}, {Caffau}, {Cancelliere}, {C{\'a}novas},
  {Cantat-Gaudin}, {Carballo}, {Carlucci}, {Carnerero}, {Carrasco},
  {Casamiquela}, {Castellani}, {Castro-Ginard}, {Castro Sampol}, {Chaoul},
  {Charlot}, {Chemin}, {Chiavassa}, {Cioni}, {Comoretto}, {Cooper}, {Cornez},
  {Cowell}, {Crifo}, {Crosta}, {Crowley}, {Dafonte}, {Dapergolas}, {David},
  {David}, {de Laverny}, {De Luise}, {De March}, {De Ridder}, {de Souza}, {de
  Teodoro}, {de Torres}, {del Peloso}, {del Pozo}, {Delbo}, {Delgado},
  {Delgado}, {Delisle}, {Di Matteo}, {Diakite}, {Diener}, {Distefano},
  {Dolding}, {Eappachen}, {Edvardsson}, {Enke}, {Esquej}, {Fabre}, {Fabrizio},
  {Faigler}, {Fedorets}, {Fernique}, {Fienga}, {Figueras}, {Fouron},
  {Fragkoudi}, {Fraile}, {Franke}, {Gai}, {Garabato}, {Garcia-Gutierrez},
  {Garc{\'\i}a-Torres}, {Garofalo}, {Gavras}, {Gerlach}, {Geyer}, {Giacobbe},
  {Gilmore}, {Girona}, {Giuffrida}, {Gomel}, {Gomez}, {Gonzalez-Santamaria},
  {Gonz{\'a}lez-Vidal}, {Granvik}, {Guti{\'e}rrez-S{\'a}nchez}, {Guy},
  {Hauser}, {Haywood}, {Helmi}, {Hidalgo}, {Hilger}, {H{\l}adczuk}, {Hobbs},
  {Holland}, {Huckle}, {Jasniewicz}, {Jonker}, {Juaristi Campillo}, {Julbe},
  {Karbevska}, {Kervella}, {Khanna}, {Kochoska}, {Kontizas}, {Kordopatis},
  {Korn}, {Kostrzewa-Rutkowska}, {Kruszy{\'n}ska}, {Lambert}, {Lanza}, {Lasne},
  {Le Campion}, {Le Fustec}, {Lebreton}, {Lebzelter}, {Leccia}, {Leclerc},
  {Lecoeur-Taibi}, {Liao}, {Licata}, {Lindstr{\o}m}, {Lister}, {Livanou},
  {Lobel}, {Madrero Pardo}, {Managau}, {Mann}, {Marchant}, {Marconi}, {Marcos
  Santos}, {Marinoni}, {Marocco}, {Marshall}, {Martin Polo},
  {Mart{\'\i}n-Fleitas}, {Masip}, {Massari}, {Mastrobuono-Battisti}, {Mazeh},
  {McMillan}, {Messina}, {Michalik}, {Millar}, {Mints}, {Molina}, {Molinaro},
  {Moln{\'a}r}, {Montegriffo}, {Mor}, {Morbidelli}, {Morel}, {Morris},
  {Mulone}, {Munoz}, {Muraveva}, {Murphy}, {Musella}, {Noval}, {Ord{\'e}novic},
  {Orr{\`u}}, {Osinde}, {Pagani}, {Pagano}, {Palaversa}, {Palicio}, {Panahi},
  {Pawlak}, {Pe{\~n}alosa Esteller}, {Penttil{\"a}}, {Piersimoni}, {Pineau},
  {Plachy}, {Plum}, {Poggio}, {Poretti}, {Poujoulet}, {Pr{\v{s}}a}, {Pulone},
  {Racero}, {Ragaini}, {Rainer}, {Raiteri}, {Rambaux}, {Ramos}, {Ramos-Lerate},
  {Re Fiorentin}, {Regibo}, {Reyl{\'e}}, {Ripepi}, {Riva}, {Rixon}, {Robichon},
  {Robin}, {Roelens}, {Rohrbasser}, {Romero-G{\'o}mez}, {Rowell}, {Royer},
  {Rybicki}, {Sadowski}, {Sagrist{\`a} Sell{\'e}s}, {Sahlmann}, {Salgado},
  {Salguero}, {Samaras}, {Sanchez Gimenez}, {Sanna}, {Santove{\~n}a},
  {Sarasso}, {Schultheis}, {Sciacca}, {Segol}, {Segovia}, {S{\'e}gransan},
  {Semeux}, {Shahaf}, {Siddiqui}, {Siebert}, {Siltala}, {Slezak}, {Smart},
  {Solano}, {Solitro}, {Souami}, {Souchay}, {Spagna}, {Spoto}, {Steele},
  {Steidelm{\"u}ller}, {Stephenson}, {S{\"u}veges}, {Szabados}, {Szegedi-Elek},
  {Taris}, {Tauran}, {Taylor}, {Teixeira}, {Thuillot}, {Tonello}, {Torra},
  {Torra}, {Turon}, {Unger}, {Vaillant}, {van Dillen}, {Vanel}, {Vecchiato},
  {Viala}, {Vicente}, {Voutsinas}, {Weiler}, {Wevers}, {Wyrzykowski}, {Yoldas},
  {Yvard}, {Zhao}, {Zorec}, {Zucker}, {Zurbach}, \& {Zwitter}}]{Gaia2021}
{Gaia Collaboration}, {Brown}, A.~G.~A., {Vallenari}, A., {et~al.}
  2021{\natexlab{a}}, \aap, 649, A1

\bibitem[{{Gaia Collaboration} {et~al.}(2016{\natexlab{a}}){Gaia
  Collaboration}, {Brown}, {Vallenari}, {Prusti}, {de Bruijne}, {Mignard},
  {Drimmel}, {Babusiaux}, {Bailer-Jones}, {Bastian}, {Biermann}, {Evans},
  {Eyer}, {Jansen}, {Jordi}, {Katz}, {Klioner}, {Lammers}, {Lindegren}, {Luri},
  {O'Mullane}, {Panem}, {Pourbaix}, {Randich}, {Sartoretti}, {Siddiqui},
  {Soubiran}, {Valette}, {van Leeuwen}, {Walton}, {Aerts}, {Arenou}, {Cropper},
  {H{\o}g}, {Lattanzi}, {Grebel}, {Holland}, {Huc}, {Passot}, {Perryman},
  {Bramante}, {Cacciari}, {Casta{\~n}eda}, {Chaoul}, {Cheek}, {De Angeli},
  {Fabricius}, {Guerra}, {Hern{\'a}ndez}, {Jean-Antoine-Piccolo}, {Masana},
  {Messineo}, {Mowlavi}, {Nienartowicz}, {Ord{\'o}{\~n}ez-Blanco}, {Panuzzo},
  {Portell}, {Richards}, {Riello}, {Seabroke}, {Tanga}, {Th{\'e}venin},
  {Torra}, {Els}, {Gracia-Abril}, {Comoretto}, {Garcia-Reinaldos}, {Lock},
  {Mercier}, {Altmann}, {Andrae}, {Astraatmadja}, {Bellas-Velidis}, {Benson},
  {Berthier}, {Blomme}, {Busso}, {Carry}, {Cellino}, {Clementini}, {Cowell},
  {Creevey}, {Cuypers}, {Davidson}, {De Ridder}, {de Torres}, {Delchambre},
  {Dell'Oro}, {Ducourant}, {Fr{\'e}mat}, {Garc{\'\i}a-Torres}, {Gosset},
  {Halbwachs}, {Hambly}, {Harrison}, {Hauser}, {Hestroffer}, {Hodgkin},
  {Huckle}, {Hutton}, {Jasniewicz}, {Jordan}, {Kontizas}, {Korn}, {Lanzafame},
  {Manteiga}, {Moitinho}, {Muinonen}, {Osinde}, {Pancino}, {Pauwels}, {Petit},
  {Recio-Blanco}, {Robin}, {Sarro}, {Siopis}, {Smith}, {Smith}, {Sozzetti},
  {Thuillot}, {van Reeven}, {Viala}, {Abbas}, {Abreu Aramburu}, {Accart},
  {Aguado}, {Allan}, {Allasia}, {Altavilla}, {{\'A}lvarez}, {Alves},
  {Anderson}, {Andrei}, {Anglada Varela}, {Antiche}, {Antoja}, {Ant{\'o}n},
  {Arcay}, {Bach}, {Baker}, {Balaguer-N{\'u}{\~n}ez}, {Barache}, {Barata},
  {Barbier}, {Barblan}, {Barrado y Navascu{\'e}s}, {Barros}, {Barstow},
  {Becciani}, {Bellazzini}, {Bello Garc{\'\i}a}, {Belokurov}, {Bendjoya},
  {Berihuete}, {Bianchi}, {Bienaym{\'e}}, {Billebaud}, {Blagorodnova},
  {Blanco-Cuaresma}, {Boch}, {Bombrun}, {Borrachero}, {Bouquillon}, {Bourda},
  {Bouy}, {Bragaglia}, {Breddels}, {Brouillet}, {Br{\"u}semeister},
  {Bucciarelli}, {Burgess}, {Burgon}, {Burlacu}, {Busonero}, {Buzzi}, {Caffau},
  {Cambras}, {Campbell}, {Cancelliere}, {Cantat-Gaudin}, {Carlucci},
  {Carrasco}, {Castellani}, {Charlot}, {Charnas}, {Chiavassa}, {Clotet},
  {Cocozza}, {Collins}, {Costigan}, {Crifo}, {Cross}, {Crosta}, {Crowley},
  {Dafonte}, {Damerdji}, {Dapergolas}, {David}, {David}, {De Cat}, {de Felice},
  {de Laverny}, {De Luise}, {De March}, {de Martino}, {de Souza}, {Debosscher},
  {del Pozo}, {Delbo}, {Delgado}, {Delgado}, {Di Matteo}, {Diakite},
  {Distefano}, {Dolding}, {Dos Anjos}, {Drazinos}, {Duran}, {Dzigan},
  {Edvardsson}, {Enke}, {Evans}, {Eynard Bontemps}, {Fabre}, {Fabrizio},
  {Faigler}, {Falc{\~a}o}, {Farr{\`a}s Casas}, {Federici}, {Fedorets},
  {Fern{\'a}ndez-Hern{\'a}ndez}, {Fernique}, {Fienga}, {Figueras}, {Filippi},
  {Findeisen}, {Fonti}, {Fouesneau}, {Fraile}, {Fraser}, {Fuchs}, {Gai},
  {Galleti}, {Galluccio}, {Garabato}, {Garc{\'\i}a-Sedano}, {Garofalo},
  {Garralda}, {Gavras}, {Gerssen}, {Geyer}, {Gilmore}, {Girona}, {Giuffrida},
  {Gomes}, {Gonz{\'a}lez-Marcos}, {Gonz{\'a}lez-N{\'u}{\~n}ez},
  {Gonz{\'a}lez-Vidal}, {Granvik}, {Guerrier}, {Guillout}, {Guiraud},
  {G{\'u}rpide}, {Guti{\'e}rrez-S{\'a}nchez}, {Guy}, {Haigron},
  {Hatzidimitriou}, {Haywood}, {Heiter}, {Helmi}, {Hobbs}, {Hofmann}, {Holl},
  {Holland}, {Hunt}, {Hypki}, {Icardi}, {Irwin}, {Jevardat de Fombelle},
  {Jofr{\'e}}, {Jonker}, {Jorissen}, {Julbe}, {Karampelas}, {Kochoska},
  {Kohley}, {Kolenberg}, {Kontizas}, {Koposov}, {Kordopatis}, {Koubsky},
  {Krone-Martins}, {Kudryashova}, {Kull}, {Bachchan}, {Lacoste-Seris}, {Lanza},
  {Lavigne}, {Le Poncin-Lafitte}, {Lebreton}, {Lebzelter}, {Leccia}, {Leclerc},
  {Lecoeur-Taibi}, {Lemaitre}, {Lenhardt}, {Leroux}, {Liao}, {Licata},
  {Lindstr{\o}m}, {Lister}, {Livanou}, {Lobel}, {L{\"o}ffler}, {L{\'o}pez},
  {Lorenz}, {MacDonald}, {Magalh{\~a}es Fernandes}, {Managau}, {Mann},
  {Mantelet}, {Marchal}, {Marchant}, {Marconi}, {Marinoni}, {Marrese},
  {Marschalk{\'o}}, {Marshall}, {Mart{\'\i}n-Fleitas}, {Martino}, {Mary},
  {Matijevi{\v{c}}}, {Mazeh}, {McMillan}, {Messina}, {Michalik}, {Millar},
  {Miranda}, {Molina}, {Molinaro}, {Molinaro}, {Moln{\'a}r}, {Moniez},
  {Montegriffo}, {Mor}, {Mora}, {Morbidelli}, {Morel}, {Morgenthaler},
  {Morris}, {Mulone}, {Muraveva}, {Musella}, {Narbonne}, {Nelemans},
  {Nicastro}, {Noval}, {Ord{\'e}novic}, {Ordieres-Mer{\'e}}, {Osborne},
  {Pagani}, {Pagano}, {Pailler}, {Palacin}, {Palaversa}, {Parsons}, {Pecoraro},
  {Pedrosa}, {Pentik{\"a}inen}, {Pichon}, {Piersimoni}, {Pineau}, {Plachy},
  {Plum}, {Poujoulet}, {Pr{\v{s}}a}, {Pulone}, {Ragaini}, {Rago}, {Rambaux},
  {Ramos-Lerate}, {Ranalli}, {Rauw}, {Read}, {Regibo}, {Reyl{\'e}}, {Ribeiro},
  {Rimoldini}, {Ripepi}, {Riva}, {Rixon}, {Roelens}, {Romero-G{\'o}mez},
  {Rowell}, {Royer}, {Ruiz-Dern}, {Sadowski}, {Sagrist{\`a} Sell{\'e}s},
  {Sahlmann}, {Salgado}, {Salguero}, {Sarasso}, {Savietto}, {Schultheis},
  {Sciacca}, {Segol}, {Segovia}, {Segransan}, {Shih}, {Smareglia}, {Smart},
  {Solano}, {Solitro}, {Sordo}, {Soria Nieto}, {Souchay}, {Spagna}, {Spoto},
  {Stampa}, {Steele}, {Steidelm{\"u}ller}, {Stephenson}, {Stoev}, {Suess},
  {S{\"u}veges}, {Surdej}, {Szabados}, {Szegedi-Elek}, {Tapiador}, {Taris},
  {Tauran}, {Taylor}, {Teixeira}, {Terrett}, {Tingley}, {Trager}, {Turon},
  {Ulla}, {Utrilla}, {Valentini}, {van Elteren}, {Van Hemelryck}, {van
  Leeuwen}, {Varadi}, {Vecchiato}, {Veljanoski}, {Via}, {Vicente}, {Vogt},
  {Voss}, {Votruba}, {Voutsinas}, {Walmsley}, {Weiler}, {Weingrill}, {Wevers},
  {Wyrzykowski}, {Yoldas}, {{\v{Z}}erjal}, {Zucker}, {Zurbach}, {Zwitter},
  {Alecu}, {Allen}, {Allende Prieto}, {Amorim}, {Anglada-Escud{\'e}},
  {Arsenijevic}, {Azaz}, {Balm}, {Beck}, {Bernstein}, {Bigot}, {Bijaoui},
  {Blasco}, {Bonfigli}, {Bono}, {Boudreault}, {Bressan}, {Brown}, {Brunet},
  {Bunclark}, {Buonanno}, {Butkevich}, {Carret}, {Carrion}, {Chemin},
  {Ch{\'e}reau}, {Corcione}, {Darmigny}, {de Boer}, {de Teodoro}, {de Zeeuw},
  {Delle Luche}, {Domingues}, {Dubath}, {Fodor}, {Fr{\'e}zouls}, {Fries},
  {Fustes}, {Fyfe}, {Gallardo}, {Gallegos}, {Gardiol}, {Gebran}, {Gomboc},
  {G{\'o}mez}, {Grux}, {Gueguen}, {Heyrovsky}, {Hoar}, {Iannicola}, {Isasi
  Parache}, {Janotto}, {Joliet}, {Jonckheere}, {Keil}, {Kim}, {Klagyivik},
  {Klar}, {Knude}, {Kochukhov}, {Kolka}, {Kos}, {Kutka}, {Lainey}, {LeBouquin},
  {Liu}, {Loreggia}, {Makarov}, {Marseille}, {Martayan}, {Martinez-Rubi},
  {Massart}, {Meynadier}, {Mignot}, {Munari}, {Nguyen}, {Nordlander}, {Ocvirk},
  {O'Flaherty}, {Olias Sanz}, {Ortiz}, {Osorio}, {Oszkiewicz}, {Ouzounis},
  {Palmer}, {Park}, {Pasquato}, {Peltzer}, {Peralta}, {P{\'e}turaud},
  {Pieniluoma}, {Pigozzi}, {Poels}, {Prat}, {Prod'homme}, {Raison}, {Rebordao},
  {Risquez}, {Rocca-Volmerange}, {Rosen}, {Ruiz-Fuertes}, {Russo}, {Sembay},
  {Serraller Vizcaino}, {Short}, {Siebert}, {Silva}, {Sinachopoulos}, {Slezak},
  {Soffel}, {Sosnowska}, {Strai{\v{z}}ys}, {ter Linden}, {Terrell}, {Theil},
  {Tiede}, {Troisi}, {Tsalmantza}, {Tur}, {Vaccari}, {Vachier}, {Valles}, {Van
  Hamme}, {Veltz}, {Virtanen}, {Wallut}, {Wichmann}, {Wilkinson}, {Ziaeepour},
  \& {Zschocke}}]{Gaia2016Brown}
{Gaia Collaboration}, {Brown}, A.~G.~A., {Vallenari}, A., {et~al.}
  2016{\natexlab{a}}, \aap, 595, A2

\bibitem[{{Gaia Collaboration} {et~al.}(2019){Gaia Collaboration}, {Eyer},
  {Rimoldini}, {Audard}, {Anderson}, {Nienartowicz}, {Glass}, {Marchal},
  {Grenon}, {Mowlavi}, {Holl}, {Clementini}, {Aerts}, {Mazeh}, {Evans},
  {Szabados}, {Brown}, {Vallenari}, {Prusti}, {de Bruijne}, {Babusiaux},
  {Bailer-Jones}, {Biermann}, {Jansen}, {Jordi}, {Klioner}, {Lammers},
  {Lindegren}, {Luri}, {Mignard}, {Panem}, {Pourbaix}, {Randich}, {Sartoretti},
  {Siddiqui}, {Soubiran}, {van Leeuwen}, {Walton}, {Arenou}, {Bastian},
  {Cropper}, {Drimmel}, {Katz}, {Lattanzi}, {Bakker}, {Cacciari},
  {Casta{\~n}eda}, {Chaoul}, {Cheek}, {De Angeli}, {Fabricius}, {Guerra},
  {Masana}, {Messineo}, {Panuzzo}, {Portell}, {Riello}, {Seabroke}, {Tanga},
  {Th{\'e}venin}, {Gracia-Abril}, {Comoretto}, {Garcia-Reinaldos}, {Teyssier},
  {Altmann}, {Andrae}, {Bellas-Velidis}, {Benson}, {Berthier}, {Blomme},
  {Burgess}, {Busso}, {Carry}, {Cellino}, {Clotet}, {Creevey}, {Davidson}, {De
  Ridder}, {Delchambre}, {Dell'Oro}, {Ducourant},
  {Fern{\'a}ndez-Hern{\'a}ndez}, {Fouesneau}, {Fr{\'e}mat}, {Galluccio},
  {Garc{\'\i}a-Torres}, {Gonz{\'a}lez-N{\'u}{\~n}ez}, {Gonz{\'a}lez-Vidal},
  {Gosset}, {Guy}, {Halbwachs}, {Hambly}, {Harrison}, {Hern{\'a}ndez},
  {Hestroffer}, {Hodgkin}, {Hutton}, {Jasniewicz}, {Jean-Antoine-Piccolo},
  {Jordan}, {Korn}, {Krone-Martins}, {Lanzafame}, {Lebzelter}, {L{\"o}ffler},
  {Manteiga}, {Marrese}, {Mart{\'\i}n-Fleitas}, {Moitinho}, {Mora}, {Muinonen},
  {Osinde}, {Pancino}, {Pauwels}, {Petit}, {Recio-Blanco}, {Richards}, {Robin},
  {Sarro}, {Siopis}, {Smith}, {Sozzetti}, {S{\"u}veges}, {Torra}, {van Reeven},
  {Abbas}, {Abreu Aramburu}, {Accart}, {Altavilla}, {{\'A}lvarez}, {Alvarez},
  {Alves}, {Andrei}, {Anglada Varela}, {Antiche}, {Antoja}, {Arcay},
  {Astraatmadja}, {Bach}, {Baker}, {Balaguer-N{\'u}{\~n}ez}, {Balm}, {Barache},
  {Barata}, {Barbato}, {Barblan}, {Barklem}, {Barrado}, {Barros}, {Barstow},
  {Bartholom{\'e} Mu{\~n}oz}, {Bassilana}, {Becciani}, {Bellazzini},
  {Berihuete}, {Bertone}, {Bianchi}, {Bienaym{\'e}}, {Blanco-Cuaresma}, {Boch},
  {Boeche}, {Bombrun}, {Borrachero}, {Bossini}, {Bouquillon}, {Bourda},
  {Bragaglia}, {Bramante}, {Breddels}, {Bressan}, {Brouillet},
  {Br{\"u}semeister}, {Brugaletta}, {Bucciarelli}, {Burlacu}, {Busonero},
  {Butkevich}, {Buzzi}, {Caffau}, {Cancelliere}, {Cannizzaro}, {Cantat-Gaudin},
  {Carballo}, {Carlucci}, {Carrasco}, {Casamiquela}, {Castellani},
  {Castro-Ginard}, {Charlot}, {Chemin}, {Chiavassa}, {Cocozza}, {Costigan},
  {Cowell}, {Crifo}, {Crosta}, {Crowley}, {Cuypers}, {Dafonte}, {Damerdji},
  {Dapergolas}, {David}, {David}, {de Laverny}, {De Luise}, {De March}, {de
  Martino}, {de Souza}, {de Torres}, {Debosscher}, {del Pozo}, {Delbo},
  {Delgado}, {Delgado}, {Diakite}, {Diener}, {Distefano}, {Dolding},
  {Drazinos}, {Dur{\'a}n}, {Edvardsson}, {Enke}, {Eriksson}, {Esquej}, {Eynard
  Bontemps}, {Fabre}, {Fabrizio}, {Faigler}, {Falc{\~a}o}, {Farr{\`a}s Casas},
  {Federici}, {Fedorets}, {Fernique}, {Figueras}, {Filippi}, {Findeisen},
  {Fonti}, {Fraile}, {Fraser}, {Fr{\'e}zouls}, {Gai}, {Galleti}, {Garabato},
  {Garc{\'\i}a-Sedano}, {Garofalo}, {Garralda}, {Gavel}, {Gavras}, {Gerssen},
  {Geyer}, {Giacobbe}, {Gilmore}, {Girona}, {Giuffrida}, {Gomes}, {Granvik},
  {Gueguen}, {Guerrier}, {Guiraud}, {Guti{\'e}rrez-S{\'a}nchez}, {Haigron},
  {Hatzidimitriou}, {Hauser}, {Haywood}, {Heiter}, {Helmi}, {Heu}, {Hilger},
  {Hobbs}, {Hofmann}, {Holland}, {Huckle}, {Hypki}, {Icardi}, {Jan{\ss}en},
  {Jevardat de Fombelle}, {Jonker}, {Juh{\'a}sz}, {Julbe}, {Karampelas},
  {Kewley}, {Klar}, {Kochoska}, {Kohley}, {Kolenberg}, {Kontizas}, {Kontizas},
  {Koposov}, {Kordopatis}, {Kostrzewa-Rutkowska}, {Koubsky}, {Lambert},
  {Lanza}, {Lasne}, {Lavigne}, {Le Fustec}, {Le Poncin-Lafitte}, {Lebreton},
  {Leccia}, {Leclerc}, {Lecoeur-Taibi}, {Lenhardt}, {Leroux}, {Liao}, {Licata},
  {Lindstr{\o}m}, {Lister}, {Livanou}, {Lobel}, {L{\'o}pez}, {Lorenz},
  {Managau}, {Mann}, {Mantelet}, {Marchant}, {Marconi}, {Marinoni},
  {Marschalk{\'o}}, {Marshall}, {Martino}, {Marton}, {Mary}, {Massari},
  {Matijevi{\v{c}}}, {McMillan}, {Messina}, {Michalik}, {Millar}, {Molina},
  {Molinaro}, {Moln{\'a}r}, {Montegriffo}, {Mor}, {Morbidelli}, {Morel},
  {Morgenthaler}, {Morris}, {Mulone}, {Muraveva}, {Musella}, {Nelemans},
  {Nicastro}, {Noval}, {O'Mullane}, {Ord{\'e}novic}, {Ord{\'o}{\~n}ez-Blanco},
  {Osborne}, {Pagani}, {Pagano}, {Pailler}, {Palacin}, {Palaversa}, {Panahi},
  {Pawlak}, {Piersimoni}, {Pineau}, {Plachy}, {Plum}, {Poggio}, {Poujoulet},
  {Pr{\v{s}}a}, {Pulone}, {Racero}, {Ragaini}, {Rambaux}, {Ramos-Lerate},
  {Regibo}, {Reyl{\'e}}, {Riclet}, {Ripepi}, {Riva}, {Rivard}, {Rixon},
  {Roegiers}, {Roelens}, {Romero-G{\'o}mez}, {Rowell}, {Royer}, {Ruiz-Dern},
  {Sadowski}, {Sagrist{\`a} Sell{\'e}s}, {Sahlmann}, {Salgado}, {Salguero},
  {Sanna}, {Santana-Ros}, {Sarasso}, {Savietto}, {Schultheis}, {Sciacca},
  {Segol}, {Segovia}, {S{\'e}gransan}, {Shih}, {Siltala}, {Silva}, {Smart},
  {Smith}, {Solano}, {Solitro}, {Sordo}, {Soria Nieto}, {Souchay}, {Spagna},
  {Spoto}, {Stampa}, {Steele}, {Steidelm{\"u}ller}, {Stephenson}, {Stoev},
  {Suess}, {Surdej}, {Szegedi-Elek}, {Tapiador}, {Taris}, {Tauran}, {Taylor},
  {Teixeira}, {Terrett}, {Teyssandier}, {Thuillot}, {Titarenko}, {Torra
  Clotet}, {Turon}, {Ulla}, {Utrilla}, {Uzzi}, {Vaillant}, {Valentini},
  {Valette}, {van Elteren}, {Van Hemelryck}, {van Leeuwen}, {Vaschetto},
  {Vecchiato}, {Veljanoski}, {Viala}, {Vicente}, {Vogt}, {von Essen}, {Voss},
  {Votruba}, {Voutsinas}, {Walmsley}, {Weiler}, {Wertz}, {Wevers},
  {Wyrzykowski}, {Yoldas}, {{\v{Z}}erjal}, {Ziaeepour}, {Zorec}, {Zschocke},
  {Zucker}, {Zurbach}, \& {Zwitter}}]{Eyer2019}
{Gaia Collaboration}, {Eyer}, L., {Rimoldini}, L., {et~al.} 2019, \aap, 623,
  A110

\bibitem[{{Gaia Collaboration} {et~al.}(2021{\natexlab{b}}){Gaia
  Collaboration}, {Luri}, {Chemin}, {Clementini}, {Delgado}, {McMillan},
  {Romero-G{\'o}mez}, {Balbinot}, {Castro-Ginard}, {Mor}, {Ripepi}, {Sarro},
  {Cioni}, {Fabricius}, {Garofalo}, {Helmi}, {Muraveva}, {Brown}, {Vallenari},
  {Prusti}, {de Bruijne}, {Babusiaux}, {Biermann}, {Creevey}, {Evans}, {Eyer},
  {Hutton}, {Jansen}, {Jordi}, {Klioner}, {Lammers}, {Lindegren}, {Mignard},
  {Panem}, {Pourbaix}, {Randich}, {Sartoretti}, {Soubiran}, {Walton}, {Arenou},
  {Bailer-Jones}, {Bastian}, {Cropper}, {Drimmel}, {Katz}, {Lattanzi}, {van
  Leeuwen}, {Bakker}, {Casta{\~n}eda}, {De Angeli}, {Ducourant}, {Fouesneau},
  {Fr{\'e}mat}, {Guerra}, {Guerrier}, {Guiraud}, {Jean-Antoine Piccolo},
  {Masana}, {Messineo}, {Mowlavi}, {Nicolas}, {Nienartowicz}, {Pailler},
  {Panuzzo}, {Riclet}, {Roux}, {Seabroke}, {Sordo}, {Tanga}, {Th{\'e}venin},
  {Gracia-Abril}, {Portell}, {Teyssier}, {Altmann}, {Andrae}, {Bellas-Velidis},
  {Benson}, {Berthier}, {Blomme}, {Brugaletta}, {Burgess}, {Busso}, {Carry},
  {Cellino}, {Cheek}, {Damerdji}, {Davidson}, {Delchambre}, {Dell'Oro},
  {Fern{\'a}ndez-Hern{\'a}ndez}, {Galluccio}, {Garc{\'\i}a-Lario},
  {Garcia-Reinaldos}, {Gonz{\'a}lez-N{\'u}{\~n}ez}, {Gosset}, {Haigron},
  {Halbwachs}, {Hambly}, {Harrison}, {Hatzidimitriou}, {Heiter},
  {Hern{\'a}ndez}, {Hestroffer}, {Hodgkin}, {Holl}, {Jan{\ss}en}, {Jevardat de
  Fombelle}, {Jordan}, {Krone-Martins}, {Lanzafame}, {L{\"o}ffler}, {Lorca},
  {Manteiga}, {Marchal}, {Marrese}, {Moitinho}, {Mora}, {Muinonen}, {Osborne},
  {Pancino}, {Pauwels}, {Recio-Blanco}, {Richards}, {Riello}, {Rimoldini},
  {Robin}, {Roegiers}, {Rybizki}, {Siopis}, {Smith}, {Sozzetti}, {Ulla},
  {Utrilla}, {van Leeuwen}, {van Reeven}, {Abbas}, {Abreu Aramburu}, {Accart},
  {Aerts}, {Aguado}, {Ajaj}, {Altavilla}, {{\'A}lvarez}, {{\'A}lvarez
  Cid-Fuentes}, {Alves}, {Anderson}, {Anglada Varela}, {Antoja}, {Audard},
  {Baines}, {Baker}, {Balaguer-N{\'u}{\~n}ez}, {Balog}, {Barache}, {Barbato},
  {Barros}, {Barstow}, {Bartolom{\'e}}, {Bassilana}, {Bauchet},
  {Baudesson-Stella}, {Becciani}, {Bellazzini}, {Bernet}, {Bertone}, {Bianchi},
  {Blanco-Cuaresma}, {Boch}, {Bombrun}, {Bossini}, {Bouquillon}, {Bragaglia},
  {Bramante}, {Breedt}, {Bressan}, {Brouillet}, {Bucciarelli}, {Burlacu},
  {Busonero}, {Butkevich}, {Buzzi}, {Caffau}, {Cancelliere}, {C{\'a}novas},
  {Cantat-Gaudin}, {Carballo}, {Carlucci}, {Carnerero}, {Carrasco},
  {Casamiquela}, {Castellani}, {Castro Sampol}, {Chaoul}, {Charlot},
  {Chiavassa}, {Comoretto}, {Cooper}, {Cornez}, {Cowell}, {Crifo}, {Crosta},
  {Crowley}, {Dafonte}, {Dapergolas}, {David}, {David}, {de Laverny}, {De
  Luise}, {De March}, {De Ridder}, {de Souza}, {de Teodoro}, {de Torres}, {del
  Peloso}, {del Pozo}, {Delgado}, {Delisle}, {Di Matteo}, {Diakite}, {Diener},
  {Distefano}, {Dolding}, {Eappachen}, {Enke}, {Esquej}, {Fabre}, {Fabrizio},
  {Faigler}, {Fedorets}, {Fernique}, {Fienga}, {Figueras}, {Fouron},
  {Fragkoudi}, {Fraile}, {Franke}, {Gai}, {Garabato}, {Garcia-Gutierrez},
  {Garc{\'\i}a-Torres}, {Gavras}, {Gerlach}, {Geyer}, {Giacobbe}, {Gilmore},
  {Girona}, {Giuffrida}, {Gomez}, {Gonzalez-Santamaria}, {Gonz{\'a}lez-Vidal},
  {Granvik}, {Guti{\'e}rrez-S{\'a}nchez}, {Guy}, {Hauser}, {Haywood},
  {Hidalgo}, {Hilger}, {H{\l}adczuk}, {Hobbs}, {Holland}, {Huckle},
  {Jasniewicz}, {Jonker}, {Juaristi Campillo}, {Julbe}, {Karbevska},
  {Kervella}, {Khanna}, {Kochoska}, {Kontizas}, {Kordopatis}, {Korn},
  {Kostrzewa-Rutkowska}, {Kruszy{\'n}ska}, {Lambert}, {Lanza}, {Lasne}, {Le
  Campion}, {Le Fustec}, {Lebreton}, {Lebzelter}, {Leccia}, {Leclerc},
  {Lecoeur-Taibi}, {Liao}, {Licata}, {Lindstr{\o}m}, {Lister}, {Livanou},
  {Lobel}, {Madrero Pardo}, {Managau}, {Mann}, {Marchant}, {Marconi}, {Marcos
  Santos}, {Marinoni}, {Marocco}, {Marshall}, {Martin Polo},
  {Mart{\'\i}n-Fleitas}, {Masip}, {Massari}, {Mastrobuono-Battisti}, {Mazeh},
  {Messina}, {Michalik}, {Millar}, {Mints}, {Molina}, {Molinaro}, {Moln{\'a}r},
  {Montegriffo}, {Morbidelli}, {Morel}, {Morris}, {Mulone}, {Munoz}, {Murphy},
  {Musella}, {Noval}, {Ord{\'e}novic}, {Orr{\`u}}, {Osinde}, {Pagani},
  {Pagano}, {Palaversa}, {Palicio}, {Panahi}, {Pawlak}, {Pe{\~n}alosa
  Esteller}, {Penttil{\"a}}, {Piersimoni}, {Pineau}, {Plachy}, {Plum},
  {Poggio}, {Poretti}, {Poujoulet}, {Pr{\v{s}}a}, {Pulone}, {Racero},
  {Ragaini}, {Rainer}, {Raiteri}, {Rambaux}, {Ramos}, {Ramos-Lerate}, {Re
  Fiorentin}, {Regibo}, {Reyl{\'e}}, {Riva}, {Rixon}, {Robichon}, {Robin},
  {Roelens}, {Rohrbasser}, {Rowell}, {Royer}, {Rybicki}, {Sadowski},
  {Sagrist{\`a} Sell{\'e}s}, {Sahlmann}, {Salgado}, {Salguero}, {Samaras},
  {Gimenez}, {Sanna}, {Santove{\~n}a}, {Sarasso}, {Schultheis}, {Sciacca},
  {Segol}, {Segovia}, {S{\'e}gransan}, {Semeux}, {Siddiqui}, {Siebert},
  {Siltala}, {Slezak}, {Smart}, {Solano}, {Solitro}, {Souami}, {Souchay},
  {Spagna}, {Spoto}, {Steele}, {Steidelm{\"u}ller}, {Stephenson},
  {S{\"u}veges}, {Szabados}, {Szegedi-Elek}, {Taris}, {Tauran}, {Taylor},
  {Teixeira}, {Thuillot}, {Tonello}, {Torra}, {Torra}, {Turon}, {Unger},
  {Vaillant}, {van Dillen}, {Vanel}, {Vecchiato}, {Viala}, {Vicente},
  {Voutsinas}, {Weiler}, {Wevers}, {Wyrzykowski}, {Yoldas}, {Yvard}, {Zhao},
  {Zorec}, {Zucker}, {Zurbach}, \& {Zwitter}}]{GaiaLuri2021}
{Gaia Collaboration}, {Luri}, X., {Chemin}, L., {et~al.} 2021{\natexlab{b}},
  \aap, 649, A7

\bibitem[{{Gaia Collaboration} {et~al.}(2016{\natexlab{b}}){Gaia
  Collaboration}, {Prusti}, {de Bruijne}, {Brown}, {Vallenari}, {Babusiaux},
  {Bailer-Jones}, {Bastian}, {Biermann}, {Evans}, {Eyer}, {Jansen}, {Jordi},
  {Klioner}, {Lammers}, {Lindegren}, {Luri}, {Mignard}, {Milligan}, {Panem},
  {Poinsignon}, {Pourbaix}, {Randich}, {Sarri}, {Sartoretti}, {Siddiqui},
  {Soubiran}, {Valette}, {van Leeuwen}, {Walton}, {Aerts}, {Arenou}, {Cropper},
  {Drimmel}, {H{\o}g}, {Katz}, {Lattanzi}, {O'Mullane}, {Grebel}, {Holland},
  {Huc}, {Passot}, {Bramante}, {Cacciari}, {Casta{\~n}eda}, {Chaoul}, {Cheek},
  {De Angeli}, {Fabricius}, {Guerra}, {Hern{\'a}ndez}, {Jean-Antoine-Piccolo},
  {Masana}, {Messineo}, {Mowlavi}, {Nienartowicz}, {Ord{\'o}{\~n}ez-Blanco},
  {Panuzzo}, {Portell}, {Richards}, {Riello}, {Seabroke}, {Tanga},
  {Th{\'e}venin}, {Torra}, {Els}, {Gracia-Abril}, {Comoretto},
  {Garcia-Reinaldos}, {Lock}, {Mercier}, {Altmann}, {Andrae}, {Astraatmadja},
  {Bellas-Velidis}, {Benson}, {Berthier}, {Blomme}, {Busso}, {Carry},
  {Cellino}, {Clementini}, {Cowell}, {Creevey}, {Cuypers}, {Davidson}, {De
  Ridder}, {de Torres}, {Delchambre}, {Dell'Oro}, {Ducourant}, {Fr{\'e}mat},
  {Garc{\'\i}a-Torres}, {Gosset}, {Halbwachs}, {Hambly}, {Harrison}, {Hauser},
  {Hestroffer}, {Hodgkin}, {Huckle}, {Hutton}, {Jasniewicz}, {Jordan},
  {Kontizas}, {Korn}, {Lanzafame}, {Manteiga}, {Moitinho}, {Muinonen},
  {Osinde}, {Pancino}, {Pauwels}, {Petit}, {Recio-Blanco}, {Robin}, {Sarro},
  {Siopis}, {Smith}, {Smith}, {Sozzetti}, {Thuillot}, {van Reeven}, {Viala},
  {Abbas}, {Abreu Aramburu}, {Accart}, {Aguado}, {Allan}, {Allasia},
  {Altavilla}, {{\'A}lvarez}, {Alves}, {Anderson}, {Andrei}, {Anglada Varela},
  {Antiche}, {Antoja}, {Ant{\'o}n}, {Arcay}, {Atzei}, {Ayache}, {Bach},
  {Baker}, {Balaguer-N{\'u}{\~n}ez}, {Barache}, {Barata}, {Barbier}, {Barblan},
  {Baroni}, {Barrado y Navascu{\'e}s}, {Barros}, {Barstow}, {Becciani},
  {Bellazzini}, {Bellei}, {Bello Garc{\'\i}a}, {Belokurov}, {Bendjoya},
  {Berihuete}, {Bianchi}, {Bienaym{\'e}}, {Billebaud}, {Blagorodnova},
  {Blanco-Cuaresma}, {Boch}, {Bombrun}, {Borrachero}, {Bouquillon}, {Bourda},
  {Bouy}, {Bragaglia}, {Breddels}, {Brouillet}, {Br{\"u}semeister},
  {Bucciarelli}, {Budnik}, {Burgess}, {Burgon}, {Burlacu}, {Busonero}, {Buzzi},
  {Caffau}, {Cambras}, {Campbell}, {Cancelliere}, {Cantat-Gaudin}, {Carlucci},
  {Carrasco}, {Castellani}, {Charlot}, {Charnas}, {Charvet}, {Chassat},
  {Chiavassa}, {Clotet}, {Cocozza}, {Collins}, {Collins}, {Costigan}, {Crifo},
  {Cross}, {Crosta}, {Crowley}, {Dafonte}, {Damerdji}, {Dapergolas}, {David},
  {David}, {De Cat}, {de Felice}, {de Laverny}, {De Luise}, {De March}, {de
  Martino}, {de Souza}, {Debosscher}, {del Pozo}, {Delbo}, {Delgado},
  {Delgado}, {di Marco}, {Di Matteo}, {Diakite}, {Distefano}, {Dolding}, {Dos
  Anjos}, {Drazinos}, {Dur{\'a}n}, {Dzigan}, {Ecale}, {Edvardsson}, {Enke},
  {Erdmann}, {Escolar}, {Espina}, {Evans}, {Eynard Bontemps}, {Fabre},
  {Fabrizio}, {Faigler}, {Falc{\~a}o}, {Farr{\`a}s Casas}, {Faye}, {Federici},
  {Fedorets}, {Fern{\'a}ndez-Hern{\'a}ndez}, {Fernique}, {Fienga}, {Figueras},
  {Filippi}, {Findeisen}, {Fonti}, {Fouesneau}, {Fraile}, {Fraser}, {Fuchs},
  {Furnell}, {Gai}, {Galleti}, {Galluccio}, {Garabato}, {Garc{\'\i}a-Sedano},
  {Gar{\'e}}, {Garofalo}, {Garralda}, {Gavras}, {Gerssen}, {Geyer}, {Gilmore},
  {Girona}, {Giuffrida}, {Gomes}, {Gonz{\'a}lez-Marcos},
  {Gonz{\'a}lez-N{\'u}{\~n}ez}, {Gonz{\'a}lez-Vidal}, {Granvik}, {Guerrier},
  {Guillout}, {Guiraud}, {G{\'u}rpide}, {Guti{\'e}rrez-S{\'a}nchez}, {Guy},
  {Haigron}, {Hatzidimitriou}, {Haywood}, {Heiter}, {Helmi}, {Hobbs},
  {Hofmann}, {Holl}, {Holland}, {Hunt}, {Hypki}, {Icardi}, {Irwin}, {Jevardat
  de Fombelle}, {Jofr{\'e}}, {Jonker}, {Jorissen}, {Julbe}, {Karampelas},
  {Kochoska}, {Kohley}, {Kolenberg}, {Kontizas}, {Koposov}, {Kordopatis},
  {Koubsky}, {Kowalczyk}, {Krone-Martins}, {Kudryashova}, {Kull}, {Bachchan},
  {Lacoste-Seris}, {Lanza}, {Lavigne}, {Le Poncin-Lafitte}, {Lebreton},
  {Lebzelter}, {Leccia}, {Leclerc}, {Lecoeur-Taibi}, {Lemaitre}, {Lenhardt},
  {Leroux}, {Liao}, {Licata}, {Lindstr{\o}m}, {Lister}, {Livanou}, {Lobel},
  {L{\"o}ffler}, {L{\'o}pez}, {Lopez-Lozano}, {Lorenz}, {Loureiro},
  {MacDonald}, {Magalh{\~a}es Fernandes}, {Managau}, {Mann}, {Mantelet},
  {Marchal}, {Marchant}, {Marconi}, {Marie}, {Marinoni}, {Marrese},
  {Marschalk{\'o}}, {Marshall}, {Mart{\'\i}n-Fleitas}, {Martino}, {Mary},
  {Matijevi{\v{c}}}, {Mazeh}, {McMillan}, {Messina}, {Mestre}, {Michalik},
  {Millar}, {Miranda}, {Molina}, {Molinaro}, {Molinaro}, {Moln{\'a}r},
  {Moniez}, {Montegriffo}, {Monteiro}, {Mor}, {Mora}, {Morbidelli}, {Morel},
  {Morgenthaler}, {Morley}, {Morris}, {Mulone}, {Muraveva}, {Musella},
  {Narbonne}, {Nelemans}, {Nicastro}, {Noval}, {Ord{\'e}novic},
  {Ordieres-Mer{\'e}}, {Osborne}, {Pagani}, {Pagano}, {Pailler}, {Palacin},
  {Palaversa}, {Parsons}, {Paulsen}, {Pecoraro}, {Pedrosa}, {Pentik{\"a}inen},
  {Pereira}, {Pichon}, {Piersimoni}, {Pineau}, {Plachy}, {Plum}, {Poujoulet},
  {Pr{\v{s}}a}, {Pulone}, {Ragaini}, {Rago}, {Rambaux}, {Ramos-Lerate},
  {Ranalli}, {Rauw}, {Read}, {Regibo}, {Renk}, {Reyl{\'e}}, {Ribeiro},
  {Rimoldini}, {Ripepi}, {Riva}, {Rixon}, {Roelens}, {Romero-G{\'o}mez},
  {Rowell}, {Royer}, {Rudolph}, {Ruiz-Dern}, {Sadowski}, {Sagrist{\`a}
  Sell{\'e}s}, {Sahlmann}, {Salgado}, {Salguero}, {Sarasso}, {Savietto},
  {Schnorhk}, {Schultheis}, {Sciacca}, {Segol}, {Segovia}, {Segransan},
  {Serpell}, {Shih}, {Smareglia}, {Smart}, {Smith}, {Solano}, {Solitro},
  {Sordo}, {Soria Nieto}, {Souchay}, {Spagna}, {Spoto}, {Stampa}, {Steele},
  {Steidelm{\"u}ller}, {Stephenson}, {Stoev}, {Suess}, {S{\"u}veges}, {Surdej},
  {Szabados}, {Szegedi-Elek}, {Tapiador}, {Taris}, {Tauran}, {Taylor},
  {Teixeira}, {Terrett}, {Tingley}, {Trager}, {Turon}, {Ulla}, {Utrilla},
  {Valentini}, {van Elteren}, {Van Hemelryck}, {van Leeuwen}, {Varadi},
  {Vecchiato}, {Veljanoski}, {Via}, {Vicente}, {Vogt}, {Voss}, {Votruba},
  {Voutsinas}, {Walmsley}, {Weiler}, {Weingrill}, {Werner}, {Wevers},
  {Whitehead}, {Wyrzykowski}, {Yoldas}, {{\v{Z}}erjal}, {Zucker}, {Zurbach},
  {Zwitter}, {Alecu}, {Allen}, {Allende Prieto}, {Amorim},
  {Anglada-Escud{\'e}}, {Arsenijevic}, {Azaz}, {Balm}, {Beck}, {Bernstein},
  {Bigot}, {Bijaoui}, {Blasco}, {Bonfigli}, {Bono}, {Boudreault}, {Bressan},
  {Brown}, {Brunet}, {Bunclark}, {Buonanno}, {Butkevich}, {Carret}, {Carrion},
  {Chemin}, {Ch{\'e}reau}, {Corcione}, {Darmigny}, {de Boer}, {de Teodoro}, {de
  Zeeuw}, {Delle Luche}, {Domingues}, {Dubath}, {Fodor}, {Fr{\'e}zouls},
  {Fries}, {Fustes}, {Fyfe}, {Gallardo}, {Gallegos}, {Gardiol}, {Gebran},
  {Gomboc}, {G{\'o}mez}, {Grux}, {Gueguen}, {Heyrovsky}, {Hoar}, {Iannicola},
  {Isasi Parache}, {Janotto}, {Joliet}, {Jonckheere}, {Keil}, {Kim},
  {Klagyivik}, {Klar}, {Knude}, {Kochukhov}, {Kolka}, {Kos}, {Kutka}, {Lainey},
  {LeBouquin}, {Liu}, {Loreggia}, {Makarov}, {Marseille}, {Martayan},
  {Martinez-Rubi}, {Massart}, {Meynadier}, {Mignot}, {Munari}, {Nguyen},
  {Nordlander}, {Ocvirk}, {O'Flaherty}, {Olias Sanz}, {Ortiz}, {Osorio},
  {Oszkiewicz}, {Ouzounis}, {Palmer}, {Park}, {Pasquato}, {Peltzer}, {Peralta},
  {P{\'e}turaud}, {Pieniluoma}, {Pigozzi}, {Poels}, {Prat}, {Prod'homme},
  {Raison}, {Rebordao}, {Risquez}, {Rocca-Volmerange}, {Rosen}, {Ruiz-Fuertes},
  {Russo}, {Sembay}, {Serraller Vizcaino}, {Short}, {Siebert}, {Silva},
  {Sinachopoulos}, {Slezak}, {Soffel}, {Sosnowska}, {Strai{\v{z}}ys}, {ter
  Linden}, {Terrell}, {Theil}, {Tiede}, {Troisi}, {Tsalmantza}, {Tur},
  {Vaccari}, {Vachier}, {Valles}, {Van Hamme}, {Veltz}, {Virtanen}, {Wallut},
  {Wichmann}, {Wilkinson}, {Ziaeepour}, \& {Zschocke}}]{Gaia2016}
{Gaia Collaboration}, {Prusti}, T., {de Bruijne}, J.~H.~J., {et~al.}
  2016{\natexlab{b}}, \aap, 595, A1

\bibitem[{{Gaia Collaboration, Drimmel et al.}(2022)}]{DR3-DPACP-075}
{Gaia Collaboration, Drimmel et al.} 2022, \aap\ submitted

\bibitem[{{Gaia Collaboration, Vallenari et al.}(2022)}]{GaiaVallenari}
{Gaia Collaboration, Vallenari et al.} 2022, \aap, in prep.

\bibitem[{{Gallenne} {et~al.}(2019){Gallenne}, {Kervella}, {Borgniet},
  {M{\'e}rand}, {Pietrzy{\'n}ski}, {Gieren}, {Monnier}, {Schaefer}, {Evans},
  {Anderson}, {Baron}, {Roettenbacher}, \& {Karczmarek}}]{Gallenne2019}
{Gallenne}, A., {Kervella}, P., {Borgniet}, S., {et~al.} 2019, \aap, 622, A164

\bibitem[{{Gautschy}(1987)}]{Gautschy1987}
{Gautschy}, A. 1987, Vistas in Astronomy, 30, 197

\bibitem[{{Genovali} {et~al.}(2014){Genovali}, {Lemasle}, {Bono}, {Romaniello},
  {Fabrizio}, {Ferraro}, {Iannicola}, {Laney}, {Nonino}, {Bergemann},
  {Buonanno}, {Fran{\c{c}}ois}, {Inno}, {Kudritzki}, {Matsunaga}, {Pedicelli},
  {Primas}, \& {Th{\'e}venin}}]{Genovali2014}
{Genovali}, K., {Lemasle}, B., {Bono}, G., {et~al.} 2014, \aap, 566, A37

\bibitem[{{Gieren}(1977)}]{Gieren1977}
{Gieren}, W. 1977, \aaps, 28, 193

\bibitem[{{Gieren} {et~al.}(2018){Gieren}, {Storm}, {Konorski}, {G{\'o}rski},
  {Pilecki}, {Thompson}, {Pietrzy{\'n}ski}, {Graczyk}, {Barnes}, {Fouqu{\'e}},
  {Nardetto}, {Gallenne}, {Karczmarek}, {Suchomska}, {Wielg{\'o}rski},
  {Taormina}, \& {Zgirski}}]{Gieren2018}
{Gieren}, W., {Storm}, J., {Konorski}, P., {et~al.} 2018, \aap, 620, A99

\bibitem[{{Gorynya} {et~al.}(1996){Gorynya}, {Samus'}, {Rastorguev}, \&
  {Sachkov}}]{Gorynya1996}
{Gorynya}, N.~A., {Samus'}, N.~N., {Rastorguev}, A.~S., \& {Sachkov}, M.~E.
  1996, Astronomy Letters, 22, 175

\bibitem[{{Heinze} {et~al.}(2018){Heinze}, {Tonry}, {Denneau}, {Flewelling},
  {Stalder}, {Rest}, {Smith}, {Smartt}, \& {Weiland}}]{Heinze2018}
{Heinze}, A.~N., {Tonry}, J.~L., {Denneau}, L., {et~al.} 2018, \aj, 156, 241

\bibitem[{{Holl} {et~al.}(2018){Holl}, {Audard}, {Nienartowicz}, {Jevardat de
  Fombelle}, {Marchal}, {Mowlavi}, {Clementini}, {De Ridder}, {Evans}, {Guy},
  {Lanzafame}, {Lebzelter}, {Rimoldini}, {Roelens}, {Zucker}, {Distefano},
  {Garofalo}, {Lecoeur-Ta{\"\i}bi}, {Lopez}, {Molinaro}, {Muraveva}, {Panahi},
  {Regibo}, {Ripepi}, {Sarro}, {Aerts}, {Anderson}, {Charnas}, {Barblan},
  {Blanco-Cuaresma}, {Busso}, {Cuypers}, {De Angeli}, {Glass}, {Grenon},
  {Juh{\'a}sz}, {Kochoska}, {Koubsky}, {Lanza}, {Leccia}, {Lorenz}, {Marconi},
  {Marschalk{\'o}}, {Mazeh}, {Messina}, {Mignard}, {Moitinho}, {Moln{\'a}r},
  {Morgenthaler}, {Musella}, {Ordenovic}, {Ord{\'o}{\~n}ez}, {Pagano},
  {Palaversa}, {Pawlak}, {Plachy}, {Pr{\v{s}}a}, {Riello}, {S{\"u}veges},
  {Szabados}, {Szegedi-Elek}, {Votruba}, \& {Eyer}}]{Holl2018}
{Holl}, B., {Audard}, M., {Nienartowicz}, K., {et~al.} 2018, \aap, 618, A30

\bibitem[{{Holl et al.}(2022)}]{DR3-DPACP-164}
{Holl et al.} 2022, \aap\ in prep.

\bibitem[{{Huang} {et~al.}(2020{\natexlab{a}}){Huang}, {Vanderburg}, {P{\'a}l},
  {Sha}, {Yu}, {Fong}, {Fausnaugh}, {Shporer}, {Guerrero}, {Vanderspek}, \&
  {Ricker}}]{QLP-I-2020RNAAS...4..204H}
{Huang}, C.~X., {Vanderburg}, A., {P{\'a}l}, A., {et~al.} 2020{\natexlab{a}},
  Research Notes of the American Astronomical Society, 4, 204

\bibitem[{{Huang} {et~al.}(2020{\natexlab{b}}){Huang}, {Vanderburg}, {P{\'a}l},
  {Sha}, {Yu}, {Fong}, {Fausnaugh}, {Shporer}, {Guerrero}, {Vanderspek}, \&
  {Ricker}}]{QLP-II-2020RNAAS...4..206H}
{Huang}, C.~X., {Vanderburg}, A., {P{\'a}l}, A., {et~al.} 2020{\natexlab{b}},
  Research Notes of the American Astronomical Society, 4, 206

\bibitem[{{Jayasinghe} {et~al.}(2019){Jayasinghe}, {Stanek}, {Kochanek},
  {Shappee}, {Holoien}, {Thompson}, {Prieto}, {Dong}, {Pawlak}, {Pejcha},
  {Shields}, {Pojmanski}, {Otero}, {Britt}, \& {Will}}]{Jayasinghe2019}
{Jayasinghe}, T., {Stanek}, K.~Z., {Kochanek}, C.~S., {et~al.} 2019, \mnras,
  486, 1907

\bibitem[{{Kienzle} {et~al.}(1999){Kienzle}, {Moskalik}, {Bersier}, \&
  {Pont}}]{Kienzle1999}
{Kienzle}, F., {Moskalik}, P., {Bersier}, D., \& {Pont}, F. 1999, \aap, 341,
  818

\bibitem[{{Kinemuchi} {et~al.}(2008){Kinemuchi}, {Harris}, {Smith},
  {Silbermann}, {Snyder}, {La Cluyz{\'e}}, \& {Clark}}]{Kinemuchi2008}
{Kinemuchi}, K., {Harris}, H.~C., {Smith}, H.~A., {et~al.} 2008, \aj, 136, 1921

\bibitem[{{Klagyivik} {et~al.}(2013){Klagyivik}, {Szabados}, {Szing}, {Leccia},
  \& {Mowlavi}}]{Klagyivik2013}
{Klagyivik}, P., {Szabados}, L., {Szing}, A., {Leccia}, S., \& {Mowlavi}, N.
  2013, \mnras, 434, 2418

\bibitem[{{Kodric} {et~al.}(2018){Kodric}, {Riffeser}, {Hopp}, {Goessl},
  {Seitz}, {Bender}, {Koppenhoefer}, {Obermeier}, {Snigula}, {Lee}, {Burgett},
  {Draper}, {Hodapp}, {Kaiser}, {Kudritzki}, {Metcalfe}, {Tonry}, \&
  {Wainscoat}}]{Kodric2018}
{Kodric}, M., {Riffeser}, A., {Hopp}, U., {et~al.} 2018, \aj, 156, 130

\bibitem[{{Kunimoto} {et~al.}(2021){Kunimoto}, {Huang}, {Tey}, {Fong}, {Hesse},
  {Shporer}, {Guerrero}, {Fausnaugh}, {Vanderspek}, \&
  {Ricker}}]{QLP-0-2021RNAAS...5..234K}
{Kunimoto}, M., {Huang}, C., {Tey}, E., {et~al.} 2021, Research Notes of the
  American Astronomical Society, 5, 234

\bibitem[{{Leavitt} \& {Pickering}(1912)}]{Leavitt1912}
{Leavitt}, H.~S. \& {Pickering}, E.~C. 1912, Harvard College Observatory
  Circular, 173, 1

\bibitem[{{Lenz} \& {Breger}(2005)}]{Lenz2005}
{Lenz}, P. \& {Breger}, M. 2005, Communications in Asteroseismology, 146, 53

\bibitem[{{Luck}(2018)}]{Luck2018}
{Luck}, R.~E. 2018, \aj, 156, 171

\bibitem[{{Luck} \& {Lambert}(2011)}]{Luck2011}
{Luck}, R.~E. \& {Lambert}, D.~L. 2011, \aj, 142, 136

\bibitem[{{Madore}(1982)}]{Madore1982}
{Madore}, B.~F. 1982, \apj, 253, 575

\bibitem[{{Marconi} {et~al.}(2004){Marconi}, {Fiorentino}, \&
  {Caputo}}]{Marconi2004}
{Marconi}, M., {Fiorentino}, G., \& {Caputo}, F. 2004, \aap, 417, 1101

\bibitem[{{Matsunaga} {et~al.}(2011){Matsunaga}, {Feast}, \&
  {Soszy{\'n}ski}}]{Matsunaga2011}
{Matsunaga}, N., {Feast}, M.~W., \& {Soszy{\'n}ski}, I. 2011, \mnras, 413, 223

\bibitem[{{Pellerin} \& {Macri}(2011)}]{Pellerin2011}
{Pellerin}, A. \& {Macri}, L.~M. 2011, \apjs, 193, 26

\bibitem[{{Perina} {et~al.}(2009){Perina}, {Federici}, {Bellazzini},
  {Cacciari}, {Fusi Pecci}, \& {Galleti}}]{Perina2009}
{Perina}, S., {Federici}, L., {Bellazzini}, M., {et~al.} 2009, \aap, 507, 1375

\bibitem[{{Petterson} {et~al.}(2004){Petterson}, {Cottrell}, \&
  {Albrow}}]{Petterson2004}
{Petterson}, O.~K.~L., {Cottrell}, P.~L., \& {Albrow}, M.~D. 2004, \mnras, 350,
  95

\bibitem[{{Petterson} {et~al.}(2005){Petterson}, {Cottrell}, {Albrow}, \&
  {Fokin}}]{Petterson2005}
{Petterson}, O.~K.~L., {Cottrell}, P.~L., {Albrow}, M.~D., \& {Fokin}, A. 2005,
  \mnras, 362, 1167

\bibitem[{{Pietrukowicz} {et~al.}(2021){Pietrukowicz}, {Soszy{\'n}ski}, \&
  {Udalski}}]{Piet2021}
{Pietrukowicz}, P., {Soszy{\'n}ski}, I., \& {Udalski}, A. 2021, \actaa, 71, 205

\bibitem[{{Plachy} {et~al.}(2021){Plachy}, {P{\'a}l}, {B{\'o}di}, {Szab{\'o}},
  {Moln{\'a}r}, {Szabados}, {Benk{\H{o}}}, {Anderson}, {Bellinger}, {Bhardwaj},
  {Ebadi}, {Gazeas}, {Hambsch}, {Hasanzadeh}, {Jurkovic}, {Kalaee}, {Kervella},
  {Kolenberg}, {Miko{\l}ajczyk}, {Nardetto}, {Nemec}, {Netzel}, {Ngeow},
  {Ozuyar}, {Pascual-Granado}, {Pilecki}, {Ripepi}, {Skarka}, {Smolec},
  {S{\'o}dor}, {Szab{\'o}}, {Christensen-Dalsgaard}, {Jenkins}, {Kjeldsen},
  {Ricker}, \& {Vanderspek}}]{Plachy-2021ApJS..253...11P}
{Plachy}, E., {P{\'a}l}, A., {B{\'o}di}, A., {et~al.} 2021, \apjs, 253, 11

\bibitem[{{Poggio} {et~al.}(2021){Poggio}, {Drimmel}, {Cantat-Gaudin}, {Ramos},
  {Ripepi}, {Zari}, {Andrae}, {Blomme}, {Chemin}, {Clementini}, {Figueras},
  {Fouesneau}, {Fr{\'e}mat}, {Lobel}, {Marshall}, {Muraveva}, \&
  {Romero-G{\'o}mez}}]{Poggio2021}
{Poggio}, E., {Drimmel}, R., {Cantat-Gaudin}, T., {et~al.} 2021, \aap, 651,
  A104

\bibitem[{{Ricker} {et~al.}(2015){Ricker}, {Winn}, {Vanderspek}, {Latham},
  {Bakos}, {Bean}, {Berta-Thompson}, {Brown}, {Buchhave}, {Butler}, {Butler},
  {Chaplin}, {Charbonneau}, {Christensen-Dalsgaard}, {Clampin}, {Deming},
  {Doty}, {De Lee}, {Dressing}, {Dunham}, {Endl}, {Fressin}, {Ge}, {Henning},
  {Holman}, {Howard}, {Ida}, {Jenkins}, {Jernigan}, {Johnson}, {Kaltenegger},
  {Kawai}, {Kjeldsen}, {Laughlin}, {Levine}, {Lin}, {Lissauer}, {MacQueen},
  {Marcy}, {McCullough}, {Morton}, {Narita}, {Paegert}, {Palle}, {Pepe},
  {Pepper}, {Quirrenbach}, {Rinehart}, {Sasselov}, {Sato}, {Seager},
  {Sozzetti}, {Stassun}, {Sullivan}, {Szentgyorgyi}, {Torres}, {Udry}, \&
  {Villasenor}}]{TESS-2015JATIS...1a4003R}
{Ricker}, G.~R., {Winn}, J.~N., {Vanderspek}, R., {et~al.} 2015, Journal of
  Astronomical Telescopes, Instruments, and Systems, 1, 014003

\bibitem[{{Riello, M.} {et~al.}(2021){Riello, M.}, {De Angeli, F.}, {Evans, D.
  W.}, {Montegriffo, P.}, {Carrasco, J. M.}, {Busso, G.}, {Palaversa, L.},
  {Burgess, P. W.}, {Diener, C.}, {Davidson, M.}, {Rowell, N.}, {Fabricius,
  C.}, {Jordi, C.}, {Bellazzini, M.}, {Pancino, E.}, {Harrison, D. L.},
  {Cacciari, C.}, {van Leeuwen, F.}, {Hambly, N. C.}, {Hodgkin, S. T.},
  {Osborne, P. J.}, {Altavilla, G.}, {Barstow, M. A.}, {Brown, A. G. A.},
  {Castellani, M.}, {Cowell, S.}, {De Luise, F.}, {Gilmore, G.}, {Giuffrida,
  G.}, {Hidalgo, S.}, {Holland, G.}, {Marinoni, S.}, {Pagani, C.}, {Piersimoni,
  A. M.}, {Pulone, L.}, {Ragaini, S.}, {Rainer, M.}, {Richards, P. J.}, {Sanna,
  N.}, {Walton, N. A.}, {Weiler, M.}, \& {Yoldas, A.}}]{Riello2021}
{Riello, M.}, {De Angeli, F.}, {Evans, D. W.}, {et~al.} 2021, A\&A, 649, A3

\bibitem[{{Riess} {et~al.}(2016){Riess}, {Macri}, {Hoffmann}, {Scolnic},
  {Casertano}, {Filippenko}, {Tucker}, {Reid}, {Jones}, {Silverman},
  {Chornock}, {Challis}, {Yuan}, {Brown}, \& {Foley}}]{Riess2016}
{Riess}, A.~G., {Macri}, L.~M., {Hoffmann}, S.~L., {et~al.} 2016, \apj, 826, 56

\bibitem[{{Rimoldini et al.}(2022)}]{DR3-DPACP-165}
{Rimoldini et al.} 2022, \aap\ in prep.

\bibitem[{{Ripepi} {et~al.}(1997){Ripepi}, {Barone}, {Milano}, \&
  {Russo}}]{Ripepi1997}
{Ripepi}, V., {Barone}, F., {Milano}, L., \& {Russo}, G. 1997, \aap, 318, 797

\bibitem[{{Ripepi} {et~al.}(2022{\natexlab{a}}){Ripepi}, {Catanzaro},
  {Clementini}, {De Somma}, {Drimmel}, {Leccia}, {Marconi}, {Molinaro},
  {Musella}, \& {Poggio}}]{Ripepi2022a}
{Ripepi}, V., {Catanzaro}, G., {Clementini}, G., {et~al.} 2022{\natexlab{a}},
  \aap, 659, A167

\bibitem[{{Ripepi} {et~al.}(2022{\natexlab{b}}){Ripepi}, {Chemin}, {Molinaro},
  {Cioni}, {Bekki}, {Clementini}, {de Grijs}, {De Somma}, {El Youssoufi},
  {Girardi}, {Groenewegen}, {Ivanov}, {Marconi}, {McMillan}, \& {van
  Loon}}]{Ripepi2022}
{Ripepi}, V., {Chemin}, L., {Molinaro}, R., {et~al.} 2022{\natexlab{b}},
  \mnras, 512, 563

\bibitem[{{Ripepi} {et~al.}(2017){Ripepi}, {Cioni}, {Moretti}, {Marconi},
  {Bekki}, {Clementini}, {de Grijs}, {Emerson}, {Groenewegen}, {Ivanov},
  {Molinaro}, {Muraveva}, {Oliveira}, {Piatti}, {Subramanian}, \& {van
  Loon}}]{Ripepi2017}
{Ripepi}, V., {Cioni}, M.-R.~L., {Moretti}, M.~I., {et~al.} 2017, \mnras, 472,
  808

\bibitem[{{Ripepi} {et~al.}(2014){Ripepi}, {Marconi}, {Moretti}, {Clementini},
  {Cioni}, {de Grijs}, {Emerson}, {Groenewegen}, {Ivanov}, \&
  {Oliveira}}]{Ripepi2014}
{Ripepi}, V., {Marconi}, M., {Moretti}, M.~I., {et~al.} 2014, \mnras, 437, 2307

\bibitem[{{Ripepi} {et~al.}(2019){Ripepi}, {Molinaro}, {Musella}, {Marconi},
  {Leccia}, \& {Eyer}}]{Ripepi2019}
{Ripepi}, V., {Molinaro}, R., {Musella}, I., {et~al.} 2019, \aap, 625, A14

\bibitem[{{Ripepi} {et~al.}(2015){Ripepi}, {Moretti}, {Marconi}, {Clementini},
  {Cioni}, {de Grijs}, {Emerson}, {Groenewegen}, {Ivanov}, {Muraveva},
  {Piatti}, \& {Subramanian}}]{Ripepi2015}
{Ripepi}, V., {Moretti}, M.~I., {Marconi}, M., {et~al.} 2015, \mnras, 446, 3034

\bibitem[{{Romaniello} {et~al.}(2008){Romaniello}, {Primas}, {Mottini},
  {Pedicelli}, {Lemasle}, {Bono}, {Fran{\c{c}}ois}, {Groenewegen}, \&
  {Laney}}]{Romaniello2008}
{Romaniello}, M., {Primas}, F., {Mottini}, M., {et~al.} 2008, \aap, 488, 731

\bibitem[{{Romaniello} {et~al.}(2022){Romaniello}, {Riess}, {Mancino},
  {Anderson}, {Freudling}, {Kudritzki}, {Macr{\`\i}}, {Mucciarelli}, \&
  {Yuan}}]{Romaniello2022}
{Romaniello}, M., {Riess}, A., {Mancino}, S., {et~al.} 2022, \aap, 658, A29

\bibitem[{{Sandage} \& {Tammann}(2006)}]{Sandage2006}
{Sandage}, A. \& {Tammann}, G.~A. 2006, \araa, 44, 93

\bibitem[{{Sartoretti et al.}(2022)}]{Sartoretti}
{Sartoretti et al.} 2022, {Gaia DR3 documentation Chapter 6: Spectroscopy},
  Gaia DR3 documentation

\bibitem[{{Shappee} {et~al.}(2014){Shappee}, {Prieto}, {Grupe}, {Kochanek},
  {Stanek}, {De Rosa}, {Mathur}, {Zu}, {Peterson}, {Pogge}, {Komossa}, {Im},
  {Jencson}, {Holoien}, {Basu}, {Beacom}, {Szczygie{\l}}, {Brimacombe},
  {Adams}, {Campillay}, {Choi}, {Contreras}, {Dietrich}, {Dubberley},
  {Elphick}, {Foale}, {Giustini}, {Gonzalez}, {Hawkins}, {Howell}, {Hsiao},
  {Koss}, {Leighly}, {Morrell}, {Mudd}, {Mullins}, {Nugent}, {Parrent},
  {Phillips}, {Pojmanski}, {Rosing}, {Ross}, {Sand}, {Terndrup}, {Valenti},
  {Walker}, \& {Yoon}}]{Shappee2014}
{Shappee}, B.~J., {Prieto}, J.~L., {Grupe}, D., {et~al.} 2014, \apj, 788, 48

\bibitem[{{Skowron} {et~al.}(2019){Skowron}, {Skowron}, {Mr{\'o}z}, {Udalski},
  {Pietrukowicz}, {Soszy{\'n}ski}, {Szyma{\'n}ski}, {Poleski}, {Koz{\l}owski},
  {Ulaczyk}, {Rybicki}, \& {Iwanek}}]{Skowron2019}
{Skowron}, D.~M., {Skowron}, J., {Mr{\'o}z}, P., {et~al.} 2019, Science, 365,
  478

\bibitem[{{Soszy{\'n}ski} {et~al.}(2016{\natexlab{a}}){Soszy{\'n}ski},
  {Pawlak}, {Pietrukowicz}, {Udalski}, {Szyma{\'n}ski}, {Wyrzykowski},
  {Ulaczyk}, {Poleski}, {Koz{\l}owski}, {Skowron}, {Skowron}, {Mr{\'o}z}, \&
  {Hamanowicz}}]{Sos2016}
{Soszy{\'n}ski}, I., {Pawlak}, M., {Pietrukowicz}, P., {et~al.}
  2016{\natexlab{a}}, \actaa, 66, 405

\bibitem[{{Soszy{\'n}ski} {et~al.}(2021){Soszy{\'n}ski}, {Pietrukowicz},
  {Skowron}, {Udalski}, {Szyma{\'n}ski}, {Skowron}, {Poleski}, {Koz{\l}owski},
  {Mr{\'o}z}, {Ulaczyk}, {Rybicki}, {Iwanek}, {Wrona}, \&
  {Gromadzki}}]{Sos2021}
{Soszy{\'n}ski}, I., {Pietrukowicz}, P., {Skowron}, J., {et~al.} 2021, \actaa,
  71, 189

\bibitem[{{Soszy{\'n}ski} {et~al.}(2016{\natexlab{b}}){Soszy{\'n}ski},
  {Smolec}, {Dziembowski}, {Udalski}, {Szyma{\'n}ski}, {Wyrzykowski},
  {Ulaczyk}, {Poleski}, {Pietrukowicz}, {Koz{\l}owski}, {Skowron}, {Skowron},
  {Mr{\'o}z}, \& {Pawlak}}]{Sos2016arrd}
{Soszy{\'n}ski}, I., {Smolec}, R., {Dziembowski}, W.~A., {et~al.}
  2016{\natexlab{b}}, \mnras, 463, 1332

\bibitem[{{Soszy{\'n}ski} {et~al.}(2019{\natexlab{a}}){Soszy{\'n}ski},
  {Udalski}, {Szyma{\'n}ski}, {Pietrukowicz}, {Skowron}, {Skowron}, {Poleski},
  {Koz{\l}owski}, {Mr{\'o}z}, {Ulaczyk}, {Rybicki}, {Iwanek}, \&
  {Wrona}}]{Sos2019_MCs}
{Soszy{\'n}ski}, I., {Udalski}, A., {Szyma{\'n}ski}, M.~K., {et~al.}
  2019{\natexlab{a}}, \actaa, 69, 87

\bibitem[{{Soszy{\'n}ski} {et~al.}(2020){Soszy{\'n}ski}, {Udalski},
  {Szyma{\'n}ski}, {Pietrukowicz}, {Skowron}, {Skowron}, {Poleski},
  {Koz{\l}owski}, {Mr{\'o}z}, {Ulaczyk}, {Rybicki}, {Iwanek}, {Wrona}, \&
  {Gromadzki}}]{Sos2020}
{Soszy{\'n}ski}, I., {Udalski}, A., {Szyma{\'n}ski}, M.~K., {et~al.} 2020,
  \actaa, 70, 101

\bibitem[{{Soszy{\'n}ski} {et~al.}(2015{\natexlab{a}}){Soszy{\'n}ski},
  {Udalski}, {Szyma{\'n}ski}, {Pietrzy{\'n}ski}, {Wyrzykowski}, {Ulaczyk},
  {Poleski}, {Pietrukowicz}, {Koz{\l}owski}, {Skowron}, {Skowron}, {Mr{\'o}z},
  \& {Pawlak}}]{Sos2015}
{Soszy{\'n}ski}, I., {Udalski}, A., {Szyma{\'n}ski}, M.~K., {et~al.}
  2015{\natexlab{a}}, \actaa, 65, 233

\bibitem[{{Soszy{\'n}ski} {et~al.}(2015{\natexlab{b}}){Soszy{\'n}ski},
  {Udalski}, {Szyma{\'n}ski}, {Skowron}, {Pietrzy{\'n}ski}, {Poleski},
  {Pietrukowicz}, {Skowron}, {Mr{\'o}z}, {Koz{\l}owski}, {Wyrzykowski},
  {Ulaczyk}, \& {Pawlak}}]{Sos2015b}
{Soszy{\'n}ski}, I., {Udalski}, A., {Szyma{\'n}ski}, M.~K., {et~al.}
  2015{\natexlab{b}}, \actaa, 65, 297

\bibitem[{{Soszy{\'n}ski} {et~al.}(2018){Soszy{\'n}ski}, {Udalski},
  {Szyma{\'n}ski}, {Wyrzykowski}, {Ulaczyk}, {Poleski}, {Pietrukowicz},
  {Koz{\l}owski}, {Skowron}, {Skowron}, {Mr{\'o}z}, {Rybicki}, \&
  {Iwanek}}]{Sos2018}
{Soszy{\'n}ski}, I., {Udalski}, A., {Szyma{\'n}ski}, M.~K., {et~al.} 2018,
  \actaa, 68, 89

\bibitem[{{Soszy{\'n}ski} {et~al.}(2017){Soszy{\'n}ski}, {Udalski},
  {Szyma{\'n}ski}, {Wyrzykowski}, {Ulaczyk}, {Poleski}, {Pietrukowicz},
  {Koz{\l}owski}, {Skowron}, {Skowron}, {Mr{\'o}z}, \& {Pawlak}}]{Sos2017}
{Soszy{\'n}ski}, I., {Udalski}, A., {Szyma{\'n}ski}, M.~K., {et~al.} 2017,
  \actaa, 67, 103

\bibitem[{{Soszy{\'n}ski} {et~al.}(2019{\natexlab{b}}){Soszy{\'n}ski},
  {Udalski}, {Wrona}, {Szyma{\'n}ski}, {Pietrukowicz}, {Skowron}, {Skowron},
  {Poleski}, {Koz{\l}owski}, {Mr{\'o}z}, {Ulaczyk}, {Rybicki}, {Iwanek}, \&
  {Gromadzki}}]{Sos2019}
{Soszy{\'n}ski}, I., {Udalski}, A., {Wrona}, M., {et~al.} 2019{\natexlab{b}},
  \actaa, 69, 321

\bibitem[{{Storm} {et~al.}(2004){Storm}, {Carney}, {Gieren}, {Fouqu{\'e}},
  {Freedman}, {Madore}, \& {Habgood}}]{Storm2004}
{Storm}, J., {Carney}, B.~W., {Gieren}, W.~P., {et~al.} 2004, \aap, 415, 521

\bibitem[{{Storm} {et~al.}(2011){Storm}, {Gieren}, {Fouqu{\'e}}, {Barnes},
  {Pietrzy{\'n}ski}, {Nardetto}, {Weber}, {Granzer}, \&
  {Strassmeier}}]{Storm2011}
{Storm}, J., {Gieren}, W., {Fouqu{\'e}}, P., {et~al.} 2011, \aap, 534, A94

\bibitem[{{Szabados}(1989)}]{Szabados1989}
{Szabados}, L. 1989, Communications of the Konkoly Observatory Hungary, 94, 1

\bibitem[{{Tarricq} {et~al.}(2022){Tarricq}, {Soubiran}, {Casamiquela},
  {Castro-Ginard}, {Olivares}, {Miret-Roig}, \& {Galli}}]{Tarricq2022}
{Tarricq}, Y., {Soubiran}, C., {Casamiquela}, L., {et~al.} 2022, \aap, 659, A59

\bibitem[{{Taylor}(2005)}]{Taylor2005}
{Taylor}, M.~B. 2005, in Astronomical Society of the Pacific Conference Series,
  Vol. 347, Astronomical Data Analysis Software and Systems XIV, ed.
  P.~{Shopbell}, M.~{Britton}, \& R.~{Ebert}, 29

\bibitem[{{Torrealba} {et~al.}(2015){Torrealba}, {Catelan}, {Drake},
  {Djorgovski}, {McNaught}, {Belokurov}, {Koposov}, {Graham}, {Mahabal},
  {Larson}, \& {Christensen}}]{Torrealba2015}
{Torrealba}, G., {Catelan}, M., {Drake}, A.~J., {et~al.} 2015, \mnras, 446,
  2251

\bibitem[{{Udalski} {et~al.}(2018){Udalski}, {Soszy{\'n}ski}, {Pietrukowicz},
  {Szyma{\'n}ski}, {Skowron}, {Skowron}, {Mr{\'o}z}, {Poleski}, {Koz{\l}owski},
  {Ulaczyk}, {Rybicki}, {Iwanek}, \& {Wrona}}]{Udalski2018}
{Udalski}, A., {Soszy{\'n}ski}, I., {Pietrukowicz}, P., {et~al.} 2018, \actaa,
  68, 315

\bibitem[{{Usenko} {et~al.}(2014){Usenko}, {Kniazev}, {Berdnikov}, \&
  {Kravtsov}}]{Usenko2014}
{Usenko}, I.~A., {Kniazev}, A.~Y., {Berdnikov}, L.~N., \& {Kravtsov}, V.~V.
  2014, Astronomy Letters, 40, 800

\bibitem[{{Wenger} {et~al.}(2000){Wenger}, {Ochsenbein}, {Egret}, {Dubois},
  {Bonnarel}, {Borde}, {Genova}, {Jasniewicz}, {Lalo{\"e}}, {Lesteven}, \&
  {Monier}}]{Wenger2000}
{Wenger}, M., {Ochsenbein}, F., {Egret}, D., {et~al.} 2000, \aaps, 143, 9

\bibitem[{{Wesselink}(1946)}]{Wesselink1946}
{Wesselink}, A.~J. 1946, \bain, 10, 91

\end{thebibliography}

\begin{appendix}

\section{Light curve examples}

\label{appendix}
   \begin{figure*}
   \centering
   \vbox{
   \hbox{
   \includegraphics[width=6cm]{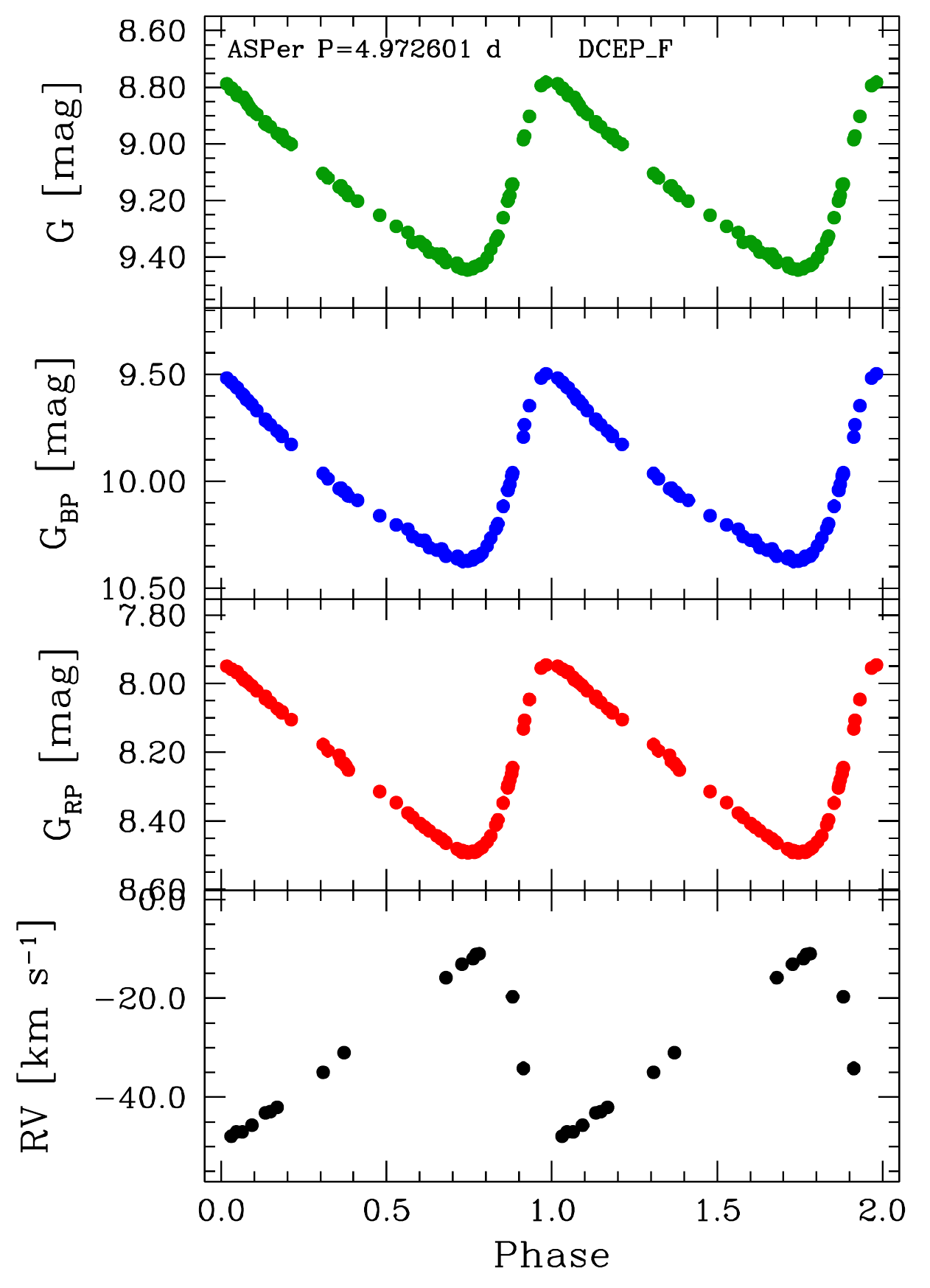}
   \includegraphics[width=6.1cm]{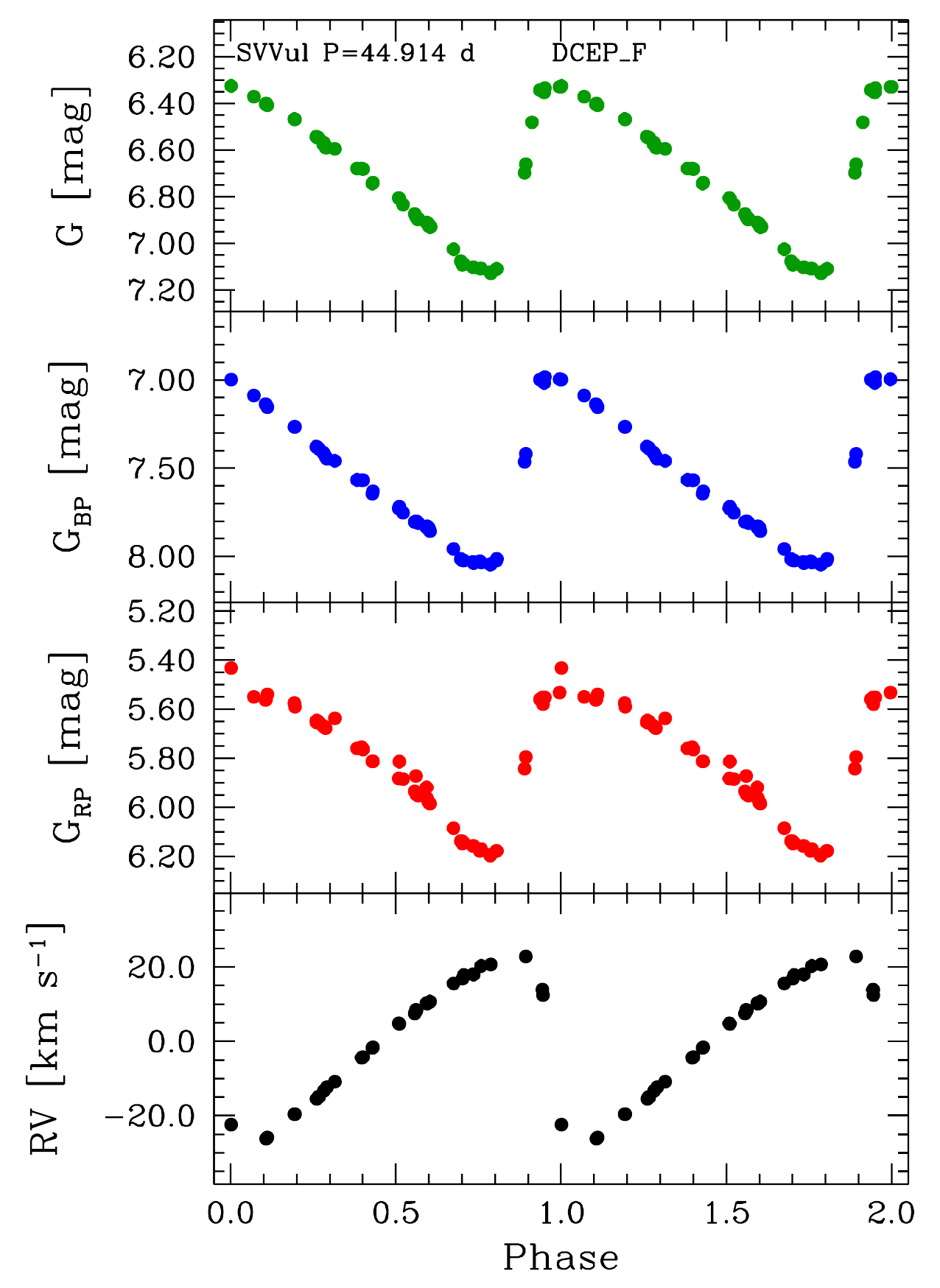}
   \includegraphics[width=6cm]{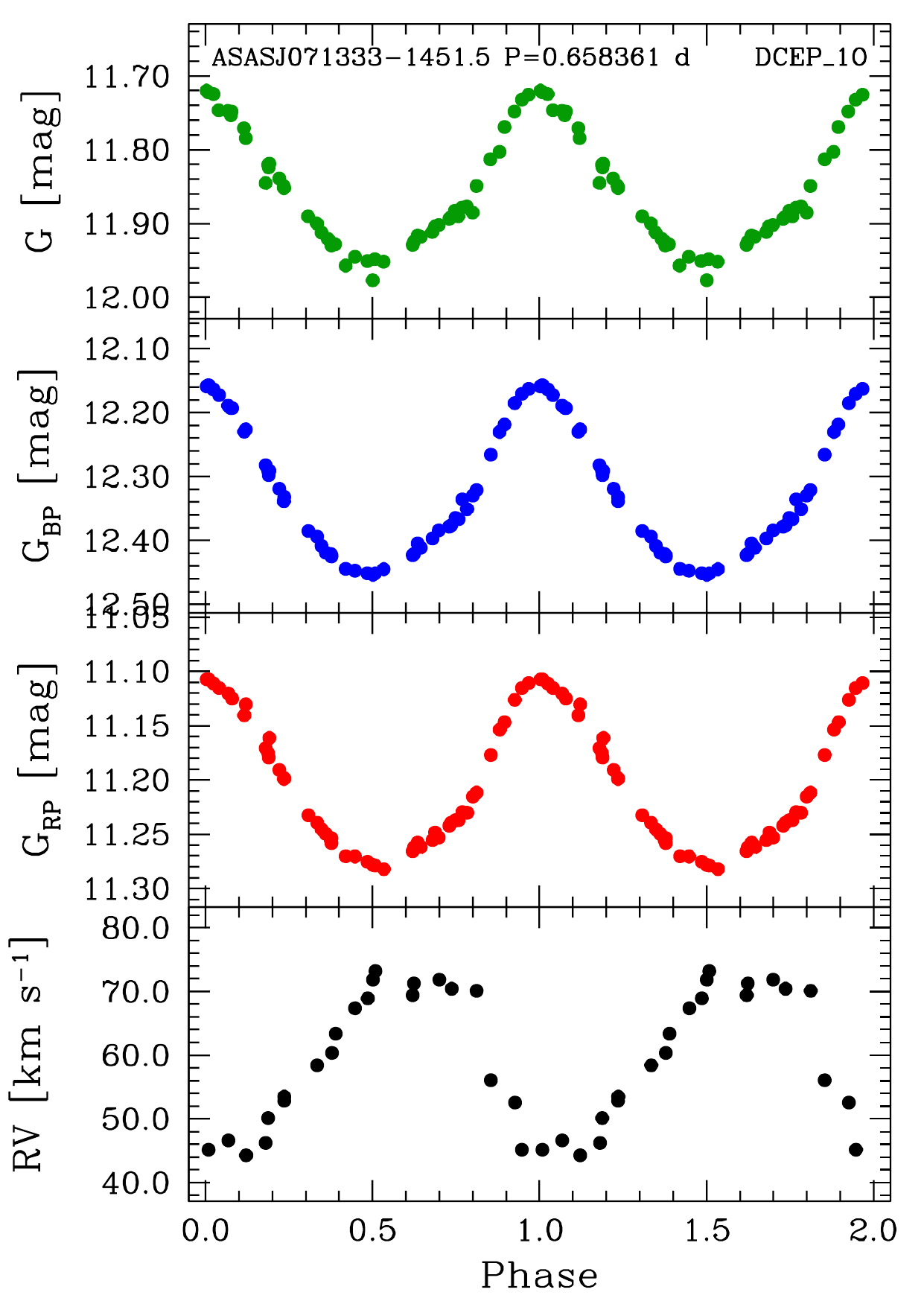}    
   }
   \hbox{
 \includegraphics[width=6cm]{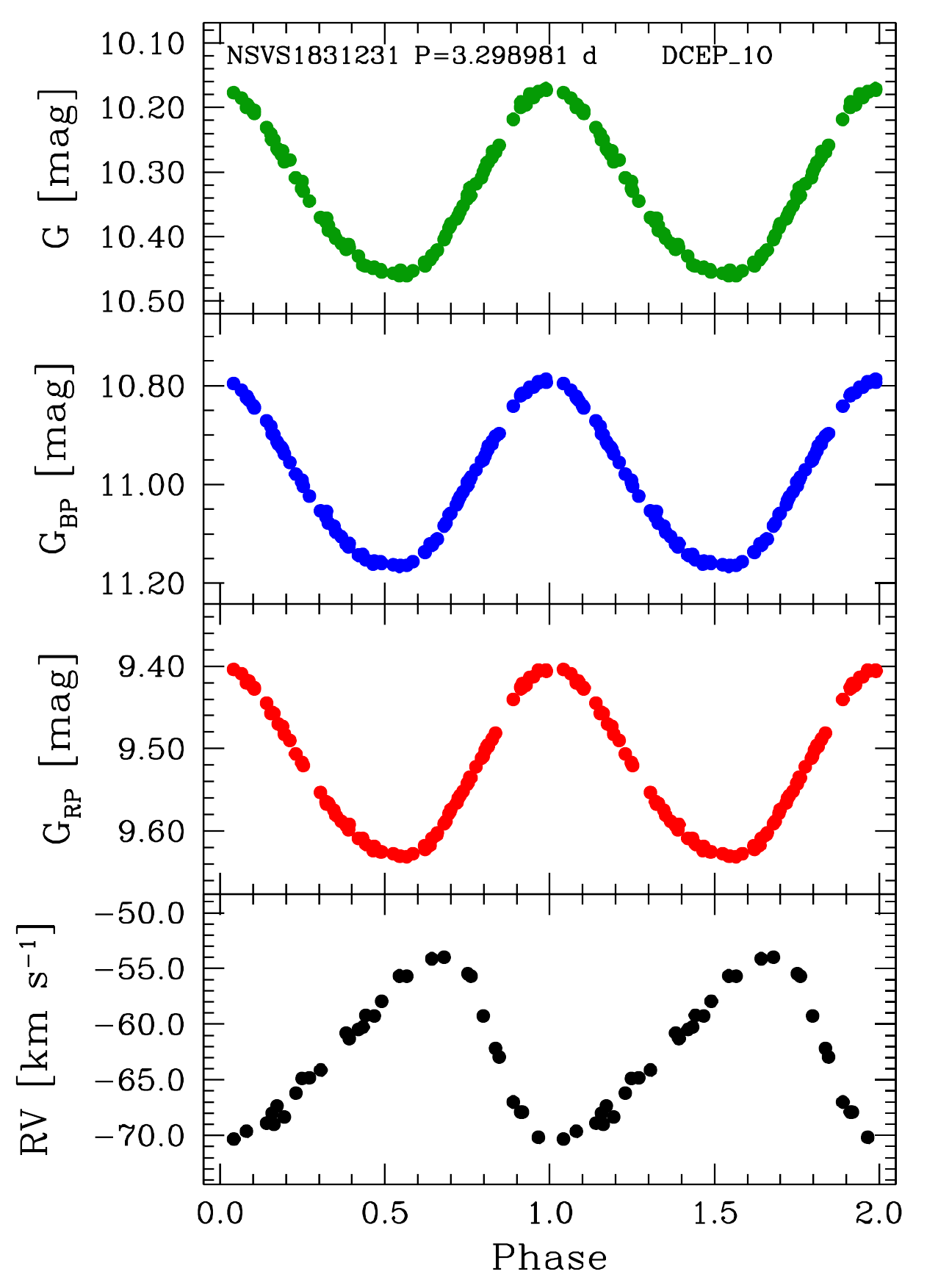}
   \includegraphics[width=6cm]{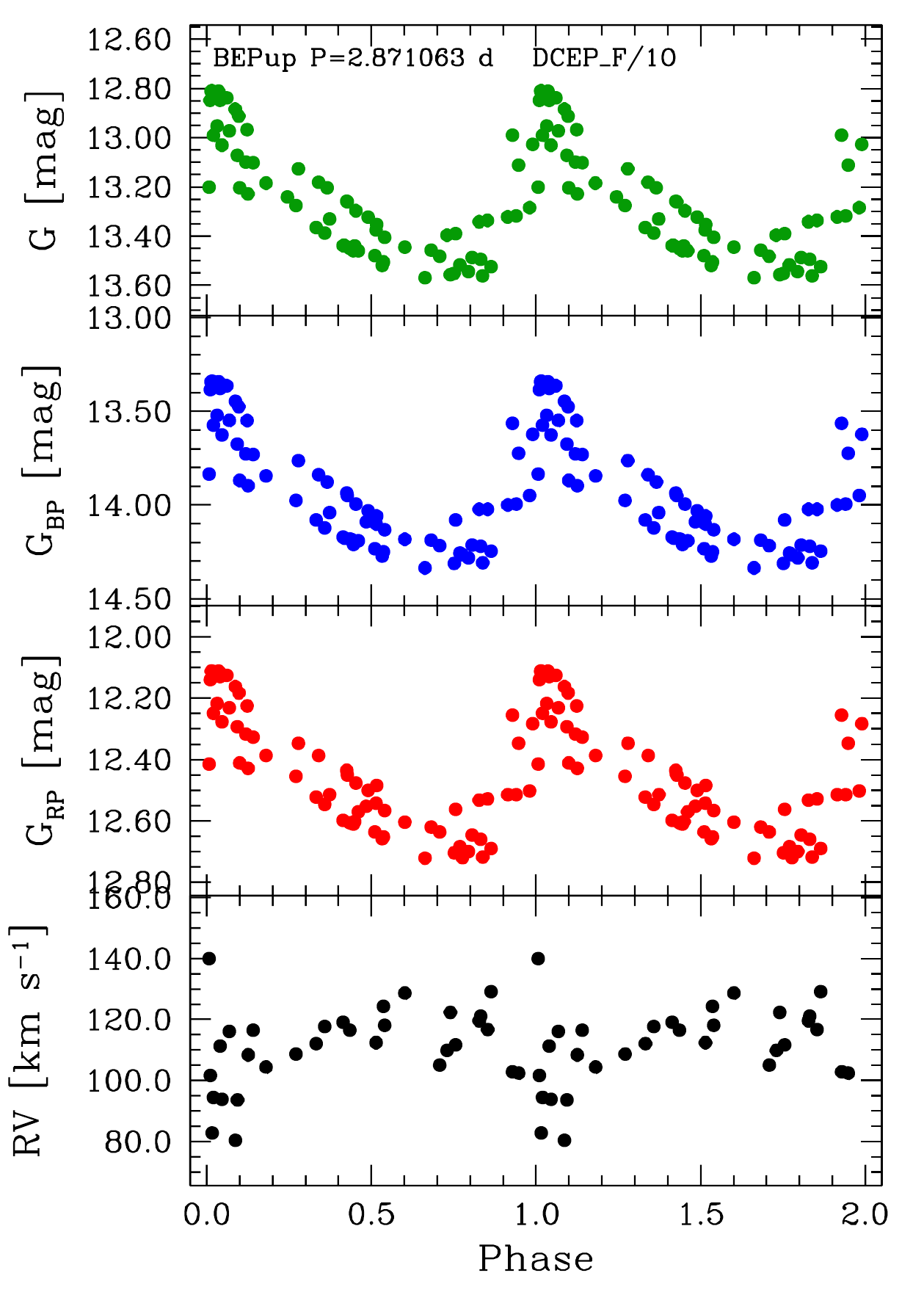}
    \includegraphics[width=6.1cm]{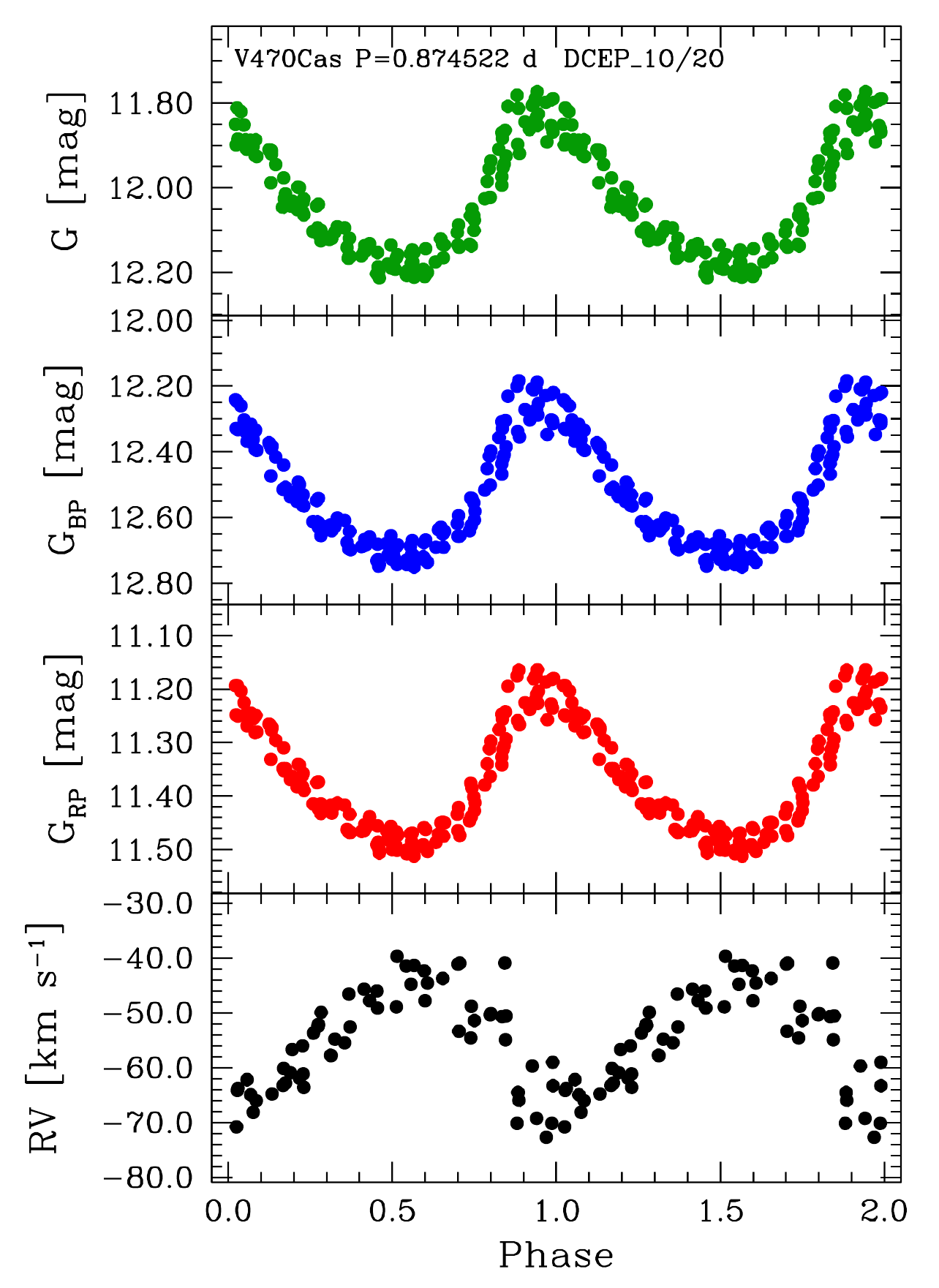}
   }
      }
      \caption{Light and RV curves for a selected sample of DCEPs of different modes.
              }
         \label{fig:lc_dceps}
   \end{figure*}

   \begin{figure*}
   \centering
   \hbox{
   \includegraphics[width=6cm]{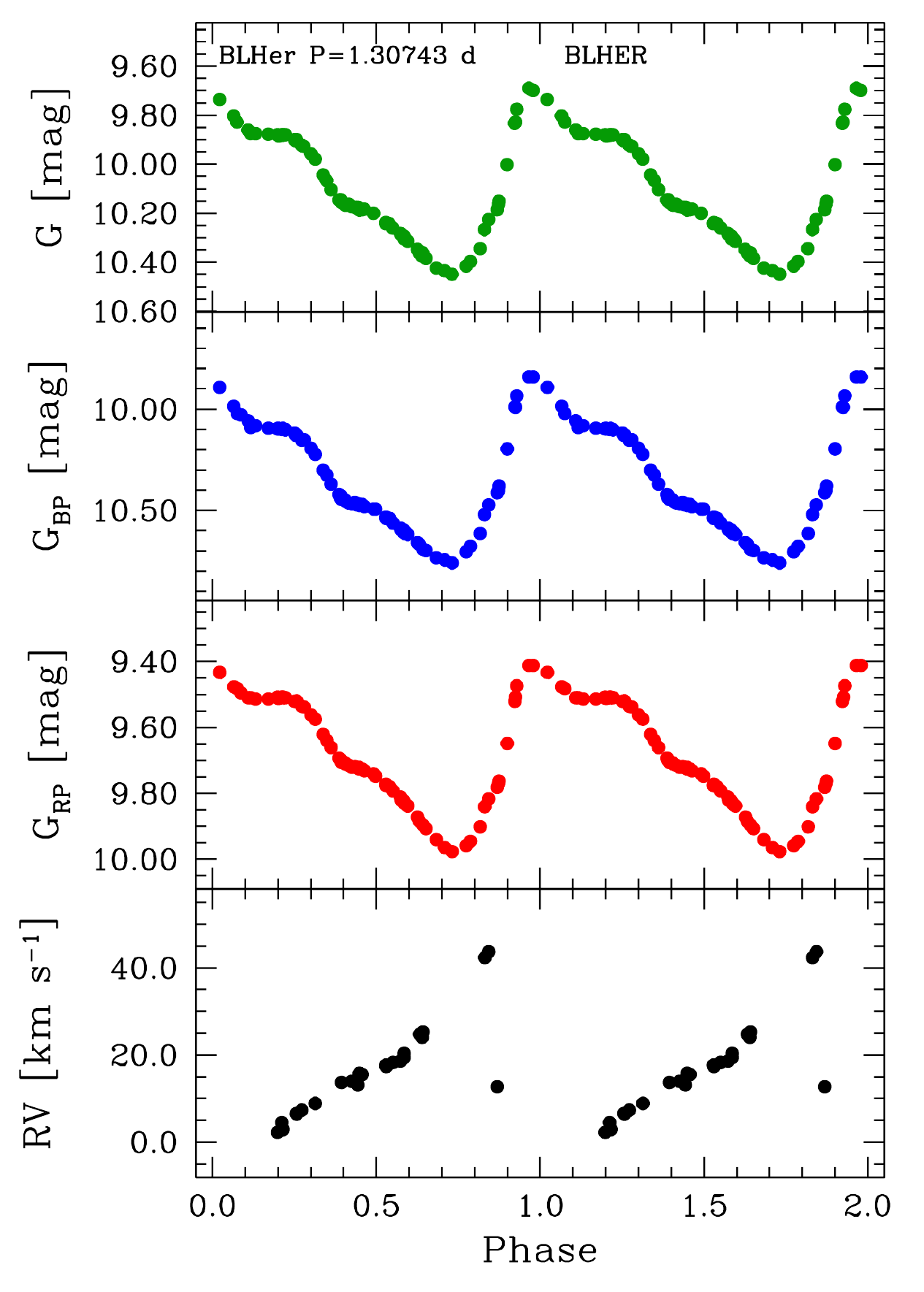}
   \includegraphics[width=6cm]{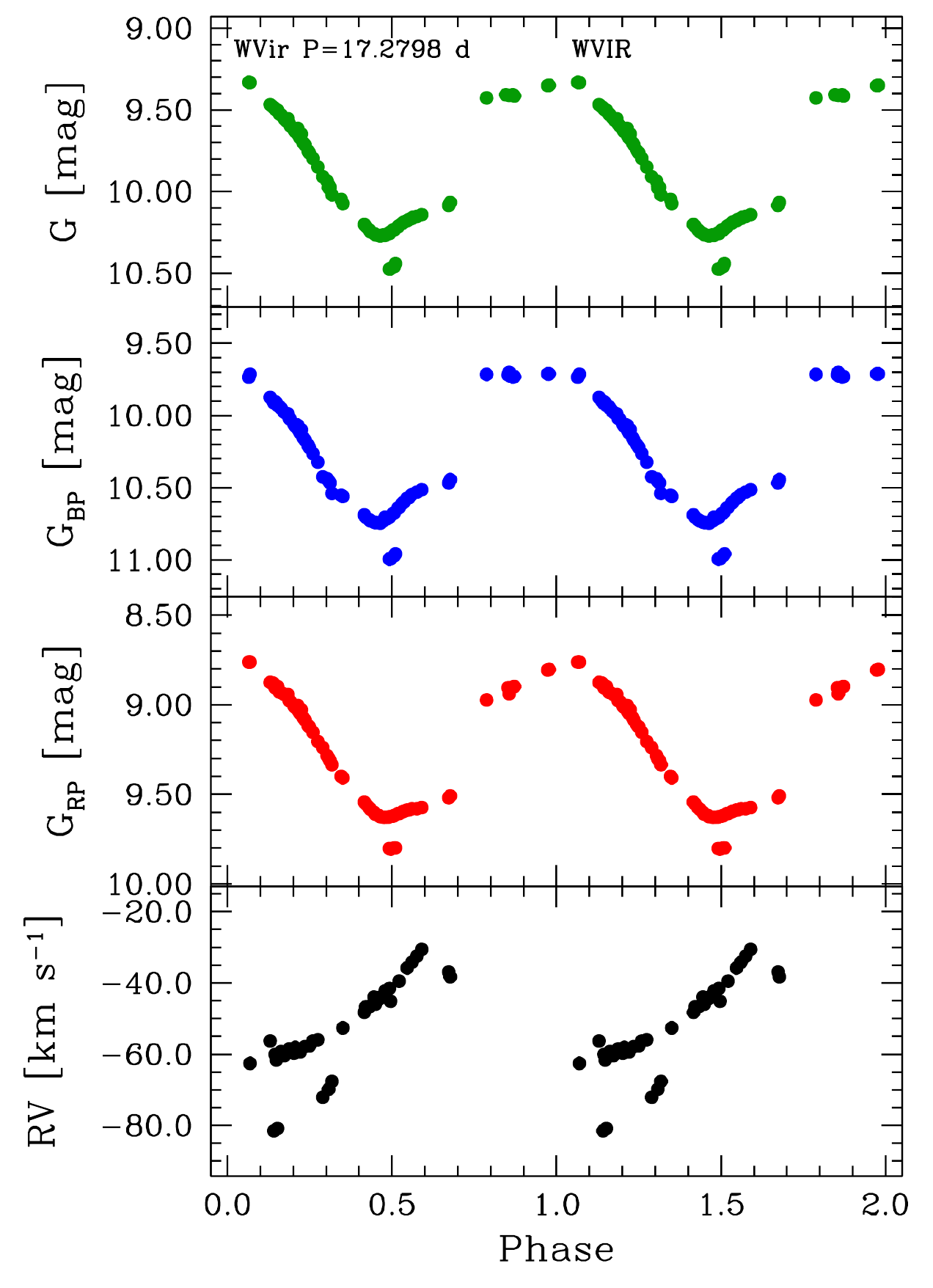}
   \includegraphics[width=6cm]{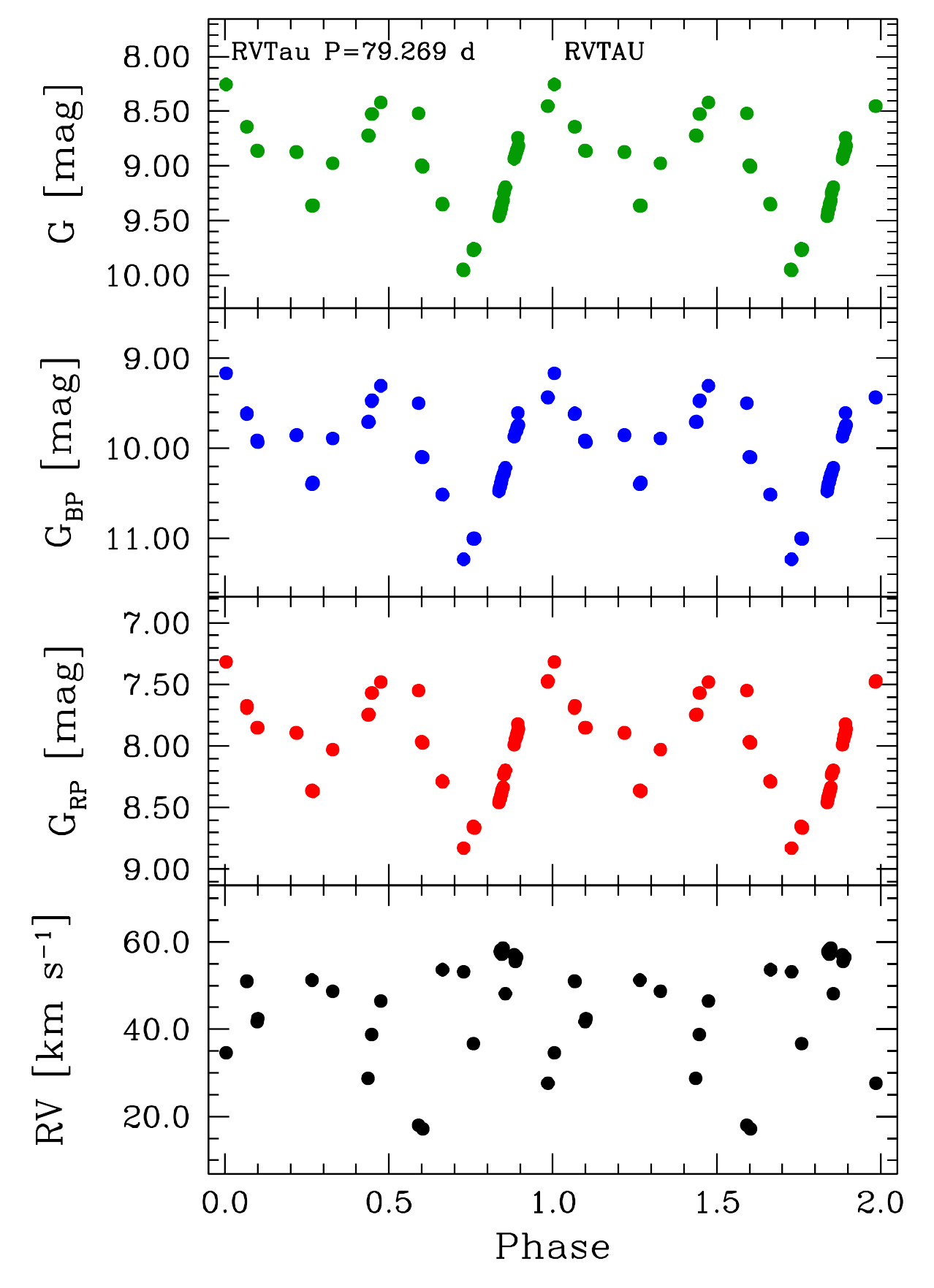}    
   }
      \caption{Light and RV curves for the prototypes of the BLHER (left), WVIR (centre) and RVTAU (right) classes.  }
         \label{fig:lc_t2ceps}
   \end{figure*}

   \begin{figure*}
   \begin{center}
       
   \hbox{
   \includegraphics[width=6cm]{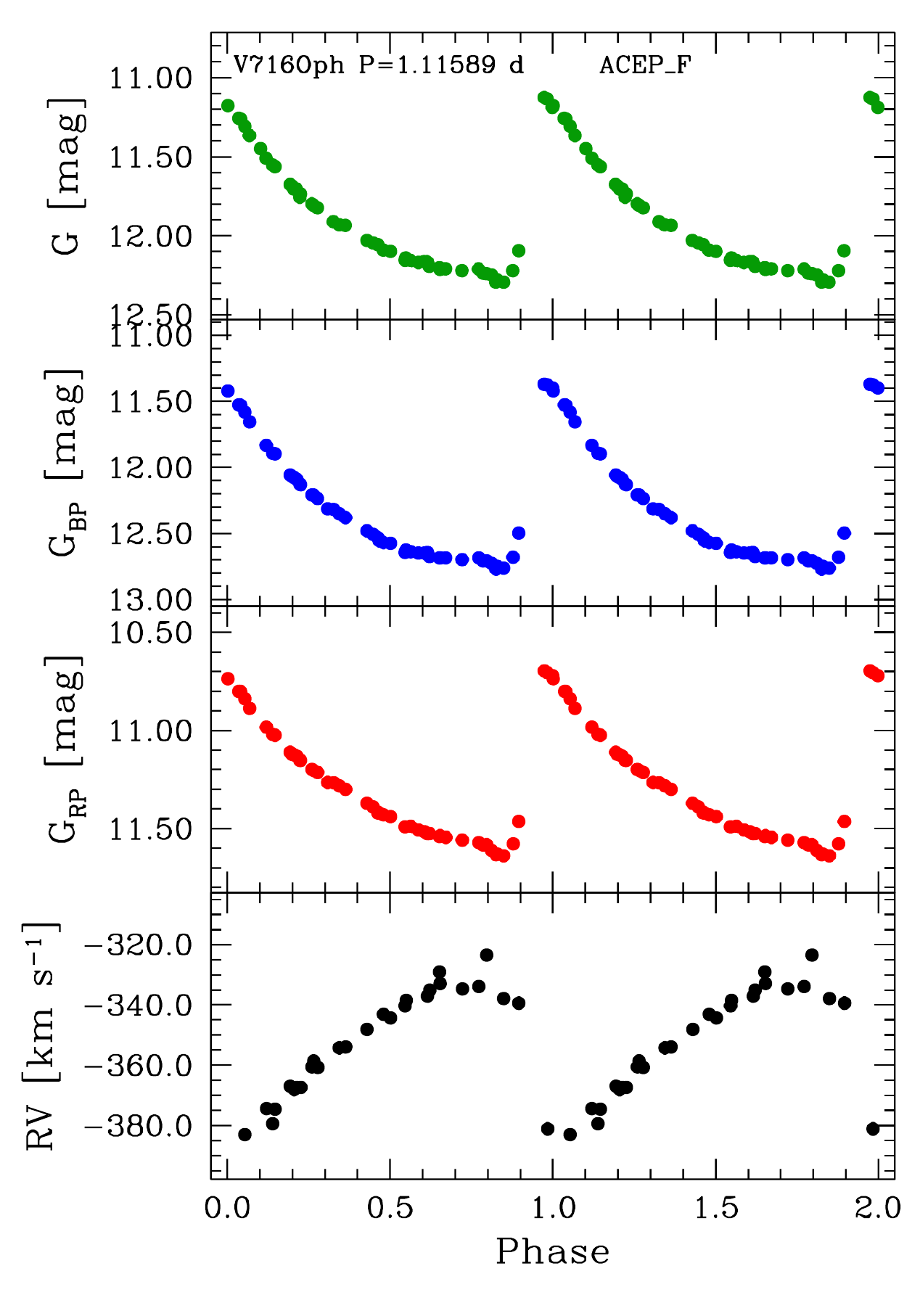}
   \includegraphics[width=6.15cm]{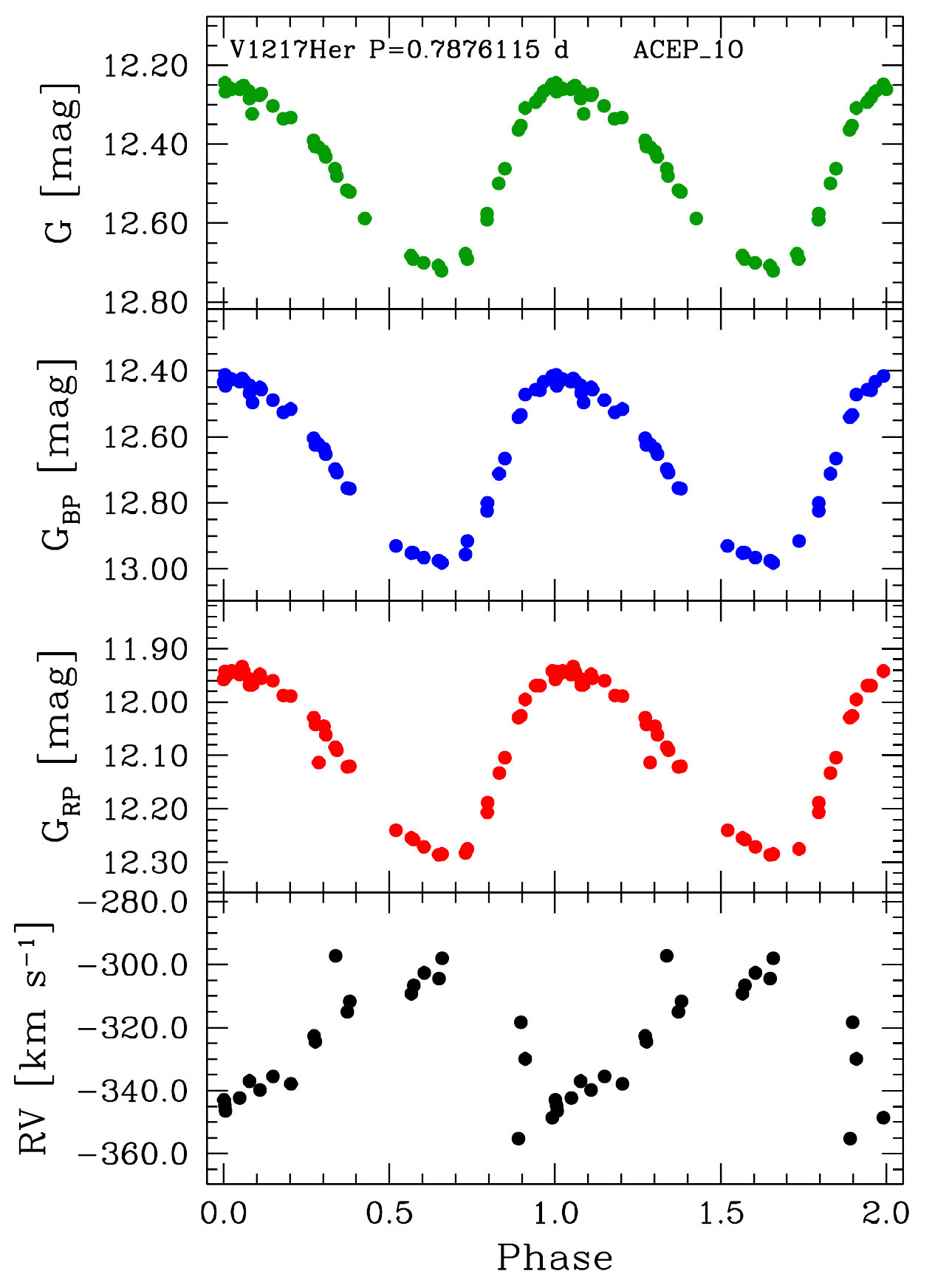}
   }
      \caption{Light and RV curves for an ACEP\_F (left) and an ACEP\_1O (right) variables. }
         \label{fig:lc_aceps}
    \end{center}

   \end{figure*}

\end{appendix}

\end{document}